\documentclass[twocolumn,nopacs,floatfix,amsmath,nofootinbib,amssymb,floatfix]{revtex4}
\usepackage{graphicx,color,dcolumn,booktabs,bm}
\usepackage{longtable,lscape}
\usepackage{txfonts}
\usepackage{overpic}
\usepackage{amssymb}
\usepackage{indentfirst}
\usepackage{feynmf}   
\usepackage{slashed}  
\usepackage{cases}
\usepackage{color}
\usepackage{multirow}
\usepackage{longtable}
\usepackage{ulem}
\usepackage{enumerate}
\usepackage{graphicx,color,dcolumn,booktabs,bm}
\usepackage[colorlinks,
            citecolor=blue,
            anchorcolor=red,
            menucolor=red,
            linkcolor=red,
            filecolor=red,
            runcolor=red,
            urlcolor=blue,
            frenchlinks=red]{hyperref}
\usepackage{epstopdf}
\usepackage{array}
\begin{document}

\title{Revisiting semileptonic decays of $\Lambda_{b(c)}$ supported by baryon spectroscopy}
\author{Yu-Shuai Li$^{1,2}$}\email{liysh20@lzu.edu.cn}
\author{Xiang Liu$^{1,2,3}$}\email{xiangliu@lzu.edu.cn}
\author{Fu-Sheng Yu$^{3,4,5}$}\email{yufsh@lzu.edu.cn}
\affiliation{$^1$School of Physical Science and Technology, Lanzhou University, Lanzhou 730000, China\\
$^2$Research Center for Hadron and CSR Physics, Lanzhou University and Institute of Modern Physics of CAS, Lanzhou 730000, China\\
$^3$Lanzhou Center for Theoretical Physics, Key Laboratory of Theoretical Physics of Gansu Province, and Frontiers Science Center for Rare Isotopes, Lanzhou University, Lanzhou 730000, China\\
$^4$School of Nuclear Science and Technology, Lanzhou University, Lanzhou 730000, China\\
$^5$Center for High Energy Physics, Peking University, Beijing 100871, China}

\begin{abstract}

The semileptonic decays of $\Lambda_{b}\to\Lambda_{c}^{(*)}\ell\nu_\ell$ and $\Lambda_{c}\to\Lambda^{(*)}\ell\nu_\ell$ are studied in the light-front quark model in this work. Instead of the quark-diquark approximation, we use the three-body wave functions obtained by baryon spectroscopy. The ground states $(1/2^{+})$, the $\lambda$-mode orbital excited states  $(1/2^{-})$, and the first radial excited states $(1/2^{+})$ of $\Lambda_{(c)}^{(*)}$ are considered. The discussions are given for the form factors, partial widths, branching fractions, leptonic forward-backward asymmetries, hadron polarizations, lepton polarizations, and the lepton flavor universalities. Our results are useful for the inputs of heavy baryon decays and understanding the baryon structures, and helpful for experimental measurements.

\end{abstract}


\maketitle

\section{introduction}
\label{sec1}

In recent years, a lot of progress has been made in understanding of the weak decays of heavy-flavor baryons, such as the observation of the double-charm baryon $\Xi_{cc}^{++}$ \cite{Aaij:2017ueg,Aaij:2018gfl}, the first measurements of the absolute branching fractions of $\Lambda_c^+$ and $\Xi_c^{+,0}$ decays \cite{Zupanc:2013iki,Ablikim:2015prg,Ablikim:2015flg,Ablikim:2016vqd,Ablikim:2018woi,Li:2018qak}, and the evidence of $CP$ violation in $\Lambda_b^0\to p\pi^+\pi^-\pi^-$ \cite{Aaij:2016cla}. The semileptonic decays play an important role in the understanding of the dynamics of beauty and charm baryon decays. The simpler dynamics of semileptonic decays compared to the nonleptonic ones, can help us to search for the new physics beyond the Standard Model (SM) and study the structures of excited states of baryons.

The lepton flavor universality violation (LFUV) was tested in the recent years using the ratio of $R(D^{(*)})=\mathcal{B}(B\to D^{(*)}\tau\nu_\tau)/\mathcal{B}(B\to D^{(*)}e(\mu)\nu_{e(\mu)})$. The experimental measurements deviate from the Standard Model predictions by 3.1$\sigma$ \cite{Amhis:2019ckw}, indicating a hint for new physics. It was pointed out that the $\Lambda_b\to \Lambda_c\ell\nu$ decays provide a theoretical cleaner place to test the LFUV \cite{Bernlochner:2018kxh,Bernlochner:2018bfn}. Up to the order of $\mathcal{O}(\Lambda_{\rm QCD}^2/m_c^2)$ in the heavy-quark effective theory, there are only three Isgur-Wise functions in the $\Lambda_b\to \Lambda_c$ transitions, compared to the 10 parameters in $B\to D^{(*)}$ transitions. Due to the large data collected by LHCb in the recent past, some excited states involving processes of $\Lambda_b\to \Lambda_c^{*}\ell\nu_\ell$ can also be measured \cite{Bernlochner:2021vlv}. Thus, the $\Lambda_b\to \Lambda_c^{(*)}\ell\nu_\ell$ decays in the same theoretical framework deserve to be systematically studied.

The comparison between the semileptonic exclusive decay of $\Lambda_c^+\to \Lambda e^+\nu_e$ and the inclusive $\Lambda_c^+\to e^+ X$ decay provides an interesting result. The BESIII measurements show that the branching fraction of the exclusive process of $\mathcal{B}(\Lambda_c^+\to \Lambda e^+\nu_e)=(3.63\pm0.43)\%$ \cite{Ablikim:2015prg} dominates the inclusive decay of $\mathcal{B}(\Lambda_c^+\to e^+ X)=(3.95\pm0.35)\%$ \cite{Ablikim:2018woi}, with a fraction of $(91.9 \pm 13.6) \%$. It implies a little room for other semileptonic processes involving excited final states. This phenomenon is quite different from the charmed meson decays. For example, $\mathcal{B}(D^+\to \overline K^0 e^+\nu_e)=(8.73 \pm 0.10) \%$ is much smaller than $\mathcal{B}(D^+\to e^+ X)=(16.07\pm0.30) \%$ \cite{Zyla:2020zbs}. Therefore, it is necessary to explore the excited-state involving processes of $\Lambda_c^+\to \Lambda^{(*)} e^+\nu_e$, to solve this problem.

The key issue in the theoretical study of semileptonic decays is to calculate the weak transition form factors. They are also the important inputs in the nonleptonic decays, such as in the prediction of the discovery channels of the doubly charmed baryons \cite{Yu:2017zst,Wang:2017mqp,Han:2021azw,Yu:2019lxw}.
The current research status on the form factors is reviewed in the next section. The main difficulty in the calculations is from the three-body problem in the baryons. A lot of theoretical works are based on the approximation of the quark-diquark scheme \cite{Guo:2005qa,Zhao:2018zcb,Ke:2007tg,Zhu:2018jet,Chua:2018lfa,Chua:2019yqh}.

In this work, we calculate the form factors of $\Lambda_{b}\to\Lambda_{c}^{(*)}\left(1/2^{\pm}\right)$ and $\Lambda_{c}\to\Lambda^{(*)}\left(1/2^{\pm}\right)$ transitions in the light-front quark model with a triquark picture.
There are some important improvements. Firstly, the triquark picture is closer to the conventional baryons' structures compared to the diquark approximation. Secondly, the baryons' wave functions are obtained by the Gaussian expansion method (GEM), avoiding the effective parameters $\beta$'s which are the major uncertainties in the previous works \cite{Ke:2007tg,Wang:2017mqp,Zhu:2018jet,Ke:2019smy,Chua:2019yqh,Geng:2020fng,Geng:2020gjh}.
Finally, the parameters in the wave functions are fixed by the baryon spectra, since a series of baryons are involved in this work.

The organization of this paper is as follows. In Sec. \ref{sec22} the current research status on the theoretical calculations of form factors are given. In Sec. \ref{sec2}, the formalisms for the form factors of the related processes are derived in the framework of light-front quark model. The formulas of the Godfrey-Isgure (GI) model and the GEM, which are used to obtain the spatial wave functions of baryons, are illustrated. In Sec. \ref{sec3} we give our numerical results; either the spatial wave functions for involved baryons or the numerical results of form factors. Besides, the relevant semileptonic differential decay rates and the branching fractions are also obtained by using these form factors. We also made a comparison with other theoretical predictions and experimental data. The conclusion and discussion are given in the last section.

\section{The present research status}
\label{sec22}

The semileptonic decays of $\Lambda_Q$ have been widely studied with various approaches, which include flavor symmetry, various quark models, the light-front approach, QCD sum rules, light-cone sum rules, lattice QCD (LQCD) and so on. In this section, we briefly introduce the present status of the semileptonic decays of  $\Lambda_Q$.

In the early 1980s, the evidence of $\Lambda_c^+$ semileptonic decay \cite{Ballagh:1981yh,Vella:1982ei} was found. However, the precisions on the branching ratios were very low in the experiments \cite{Klein:1989pu,Albrecht:1991bu,Bergfeld:1994gt}. The recent progress on this issue was made by the BESIII Collaboration \cite{Ablikim:2015prg,Ablikim:2016vqd}, where  the first measurement of the absolute branching ratio for $\Lambda_c^+\to\Lambda\ell^+\nu_{\ell}$ was given
\begin{equation*}
\begin{split}
\mathcal{B}(\Lambda_c^+\to\Lambda e^{+}\nu_{e})&=(3.63\pm0.38\pm0.20)\%,\\
\mathcal{B}(\Lambda_c^+\to\Lambda \mu^{+}\nu_{\mu})&=(3.49\pm0.46\pm0.27)\%.
\end{split}
\end{equation*}
Obviously, the new measurement can be applied to test the different theoretical approaches describing the $\Lambda_c^+$ semileptonic decays.

In 1989, Marcial {\it et al.} \cite{PerezMarcial:1989yh} predicted the widths and branching fractions of $\Lambda_c^+\to\Lambda e^+\nu_e$ by using both the nonrelativistic quark model and the MIT bag model (MBM). Here, the predicted branching ratio of $\mathcal{B}(\Lambda_c^+\to\Lambda e^+\nu_e)$ is $1.4\sim 4.2$ \%. Later, Hussain and Korner \cite{Hussain:1990ai} studied the same topic with a relativistic spectator quark model where the interaction between the spectator quark and the acting quark is  ignored. This treatment was widely used in the study of heavy to heavy transitions, and was expanded to $c\to s$ sector by them. The estimated branching fraction is $4.45\%$ \cite{Hussain:1990ai}. Efimov {\it et al.} also focused on $\Lambda_c^+\to\Lambda e^+\nu_e$ decay at the same time. Using the quark confinement model, they predicted $\mathcal{B}(\Lambda_c^+\to\Lambda e^+\nu_e)=5.72\%$ \cite{Efimov:1991ex}. Cheng and Tseng \cite{Cheng:1995fe} applied the nonrelativistic quark model and considered the flavor suppression factor, and obtained $\mathcal{B}(\Lambda_c^+\to\Lambda e^+\nu_e)=1.44\%$.
In fact, QCDSR is also an effective approach to study the semileptonic decay of $\Lambda_c$. For example, Carvalho {\it et al.} estimated $\mathcal{B}(\Lambda_c^+\to\Lambda e^+\nu_e)=(2.69\pm0.37)\%$ \cite{MarquesdeCarvalho:1999bqs}. By including the higher twist contributions in the light cone sum rule calculation, Liu {\it et al.} \cite{Liu:2009sn} calculated $\mathcal{B}(\Lambda_c^+\to\Lambda l^+\nu_l)$, which is $3.0\%$ or $2.0\%$ if adopting the Chernyak-Zhitnitsky-type or the Ioffe-type interpolating current, respectively. In addition, Zhao {\it et al.} \cite{Zhao:2020mod} also presented the involved form factors and decay rates in QCDSR, where the results are  consistent with the experimental data within errors. Pervin {\it et al.} \cite{Pervin:2005ve} performed the calculation in the framework of the consistent quark model with both nonrelativistic and semirelativistic Hamiltonians. The obtained decay rate from the semirelativistic Hamiltonian can describe the experimental data well, while the one from the nonrelativistic Hamiltonian shows apparent divergence \cite{Pervin:2005ve}. Migura {\it et al.} used a relativistically covariant constituent quark model with the Bethe-Salpeter equation and found $\mathcal{B}(\Lambda_c^+\to\Lambda e^{+}\nu_{e})=3.19\%$ and $\mathcal{B}(\Lambda_c^+\to\Lambda\mu^{+}\nu_{\mu})=2.97\%$ \cite{Migura:2006en}.

After measuring the absolute branching fractions of $\mathcal{B}(\Lambda_c^+\to\Lambda\ell^{+}\nu_{\ell})$ by BESIII \cite{Ablikim:2015prg,Ablikim:2016vqd}, Faustov and Galkin \cite{Faustov:2020thr} investigated the semileptonic decays of $\Lambda_c$ within a relativistic quark model (RQM) based on the quasipotential approach. By taking into account the relativistic effects, their estimations for $\mathcal{B}(\Lambda_c^+\to\Lambda e^+\nu_{e})$ and $\mathcal{B}(\Lambda_c^+\to\Lambda\mu^+\nu_{\mu})$ are $3.25\%$ and $3.14\%$, respectively, which are good comparison with the current measurements. Besides, Gutsche {\it et al.} \cite{Gutsche:2015rrt} applied the covariant confined quark model (CCQM). Apart from the decay rates, they also presented some detailed results for other physical observables, which are also important to the study of semileptonic decay. The light-front approach was also widely used to analyze the semileptonic decays of $\Lambda_c$. Zhao \cite{Zhao:2018zcb} studied the $\Lambda_c^+\to\Lambda e^{+}\nu_{e}$ process in the light-front approach by treating the spectator quarks as a diquark system and used an effective parameter to simulate the baryon wave function. With this approximation, he obtained $\mathcal{B}(\Lambda_c^+\to\Lambda e^{+}\nu_{e})=1.63\%$, which is smaller than the BESIII data. {On the contrary, without considering diquark approximation, Geng {\it et al.} \cite{Geng:2020fng,Geng:2020gjh} estimated $\mathcal{B}(\Lambda_c^+\to\Lambda e^+\nu_{e})=(3.36\pm0.87)\%$ and $\mathcal{B}(\Lambda_c^+\to\Lambda\mu^+\nu_{\mu})=(3.21\pm0.85)\%$ in the MIT bag model with the spacial wave functions determined by the baryon spectroscopy \cite{Geng:2020fng} and $\mathcal{B}(\Lambda_c^+\to\Lambda e^+\nu_{e})=(3.55\pm1.04)\%$ and $\mathcal{B}(\Lambda_c^+\to\Lambda\mu^+\nu_{\mu})=(3.40\pm1.02)\%$ in the light-front formalism \cite{Geng:2020gjh} , which are all consistent with the BESIII data.}

In the following, we continue to introduce the research status of the semileptonic decays of $\Lambda_b$. Firstly, we should show the current experimental data in the Particle Data Group (PDG) \cite{Zyla:2020zbs} \begin{equation}
\begin{split}
\mathcal{B}(\Lambda_b\to\Lambda_c^+\ell^{-}\nu_{\ell})&=(6.4^{+1.4}_{-1.3})\%,\\
\mathcal{B}(\Lambda_b\to\Lambda_c(2595)^+\ell^{-}\nu_{\ell})&=(0.79^{+0.40}_{-0.35})\%
\end{split}
\end{equation}
with $\ell^{-}=e^{-}$ or $\mu^{-}$.
Obviously, there still needs to be some improvement in these measurements.

The $\Lambda_b\to\Lambda_c^{+}\ell^{-}\nu_{\ell}$ semileptonic decays have been studied for a long time \cite{Mannel:1991ii,Korner:1991ph,Mannel:1991bs,Singleton:1990ye,Dai:1996xv, Lee:1998bj,Ivanov:1998ya,Cardarelli:1998tq,MarquesdeCarvalho:1999bqs,Guo:1999ss}. Different from the $\Lambda_c^+$ semileptonic decays, the $\tau^-$-mode of the $\Lambda_b\to\Lambda_c^{+}\ell^{-}\nu_{\ell}$ semi-leptonic decay is allowed kinematically.
Similar to the discussion of the $B\to D^{(*)}$ semileptonic decays \cite{Lees:2012xj,Belle:2019rba}, investigating the ratio $R(\Lambda_c)=\mathcal{B}(\Lambda_b\to\Lambda_c^+\ell^-\nu_{\ell})/\mathcal{B}(\Lambda_b\to\Lambda_c^+\tau^-\nu_{\tau})$ with $\ell^-=e^-$ or $\mu^-$ is also an interesting research topic. Thus, in this work, we will focus on this issue.
There were some theoretical investigations of the $\Lambda_b$ semileptonic decays.
Ke {\it et al.} \cite{Ke:2007tg,Zhu:2018jet,Ke:2019smy} calculated the branching fractions in standard and covariant light-front quark models within the diquark picture. Their results show that the quark-diquark picture works well for heavy baryons. Additionally, they also reinvestigated the same topic by the same approach without introducing diquark approximation.
A RQM \cite{Ebert:2006rp,Faustov:2016pal} with quark-diquark approximation was used by Ebert, Faustov and Galkin. The calculated branching fraction is consistent with the experimental data; other physical observables were also given. Gutsche {\it et al.} \cite{Gutsche:2015mxa,Gutsche:2018nks} calculated the $\Lambda_b\to\Lambda_c^+\ell^-\nu_{\ell}$$(\ell^-=e^-,\mu^-,\tau^-)$ observables by using the covariant confined quark model. Ranmani {\it et al.} \cite{Rahmani:2020kjd} calculated them in a potential model with a modified QCD Cornell interaction. Thakkar \cite{Thakkar:2020vpv} studied the semileptonic decays by using the hypercentral constituent quark model, where the six-dimensional hypercentral Schr\"odinger equation was solved for extracting the wave functions of heavy baryons.
Of course, the QCDSR was also applied to the study of the $\Lambda_b$ semileptonic decays \cite{Wang:2003it,Huang:2005mea,Wang:2009yma,Azizi:2018axf}.

Until now, theorists have paid more attentions to the semileptonic decays of $\Lambda_Q$ to a ground state of $\Lambda$ or $\Lambda_c$.
Since 2005, the semileptonic decays of $\Lambda_Q$ to an excited state of $\Lambda$ or $\Lambda_c$ have been studied.
Pervin {\it et al.} \cite{Pervin:2005ve} studied the semileptonic decays of $\Lambda_{Q}$ to $J^P=\frac{1}{2}^{\pm}, \frac{3}{2}^-$ final states in a constituent quark model with both nonrelativistic and semirelativistic Hamiltonians.
Gussain and Roberts \cite{Hussain:2017lir} studied the semileptonic decays of $\Lambda_c^+$ to excited $\Lambda$ states with a nonrelativistic quark model.
Gutsche {\it et al.} \cite{Gutsche:2018nks} analyzed  $\Lambda_{b}\to\Lambda_{c}\left(\frac{1}{2}^{\pm},\frac{3}{2}^{-}\right)\ell^-\nu_{\ell}$ with CCQM.
Nieves et al. \cite{Nieves:2019kdh} studied the $\Lambda_{b}\to\Lambda_{c}(2595, 2625)\ell^-\nu_{\ell}$ with heavy quark spin symmetry.
Be\v{c}irevi\'c {\it et al.} \cite{Becirevic:2020nmb} studied $\Lambda_{b}\to\Lambda_{c}\left(\frac{1}{2}^{\pm}\right)\ell^-\nu_{\ell}$ by using the Bakamjian-Thomas approach with a spectroscopic model.
Recently, the BESIII Collaboration \cite{Ablikim:2019hff} released a white paper of future plans, where measuring $\Lambda_{c}\to\Lambda^{*}\ell^+\nu_{\ell}$ was mentioned. To enlarge our knowledge on the semileptonic decays of $\Lambda_Q$ to excited $\Lambda$ or $\Lambda_c$ state, joint effort from experimentalists and theorists are needed.

In fact, the LQCD \cite{Bowler:1997ej,Gottlieb:2003yb,Detmold:2015aaa,Meinel:2016dqj,Meinel:2021rbm} is an effective approach to study the $\Lambda_Q$ semileptonic decays. Meinal {\it et al.} \cite{Detmold:2015aaa,Meinel:2016dqj} calculated the form factors and decay rates of $\Lambda_b\to\Lambda_c\ell^-\nu_{\ell}$ and $\Lambda_c\to\Lambda\ell^+\nu_{\ell}$ processes with all possible leptonic channels. Very recently, they presented the first lattice QCD calculation of the form factors describing the $\Lambda_b\to\Lambda_c\left(\frac{1}{2}^-,\frac{3}{2}^-\right)\ell^-\nu_{\ell}$ decays \cite{Meinel:2021rbm}. We are looking forward to more progress  on the $\Lambda_Q$ semileptonic decays by LQCD, which may provide valuable information to theoretical studies.

The numerical results from literature are summarized in the table in Sec.\ref{sec4}, with a comparison to our predictions.

\section{semileptonic decays of $\Lambda_Q$}
\label{sec2}

\subsection{The form factors relevant to semileptonic decays of $\Lambda_Q$}
\label{semi-leptonic decay and LFQM}

In this subsection, we briefly present how to calculate the form factors involved in these semi-leptonic decays of $\Lambda_b$ and $\Lambda_c$. For illustration, we take
$\Lambda_{b}\to\Lambda_{c}^{(*)}\left(1/2^{\pm}\right)$ as an example, where $\Lambda_{c}\left(1/2^{+}\right)$ and $\Lambda_{c}\left(1/2^{-}\right)$ denote $\Lambda_{c}\left(2286\right)$ and $\Lambda_{c}\left(2595\right)$, respectively. In addition, we use subscript $*$ to mark the radial excitation.

\begin{figure}[htbp]\centering
  \includegraphics[width=6cm]{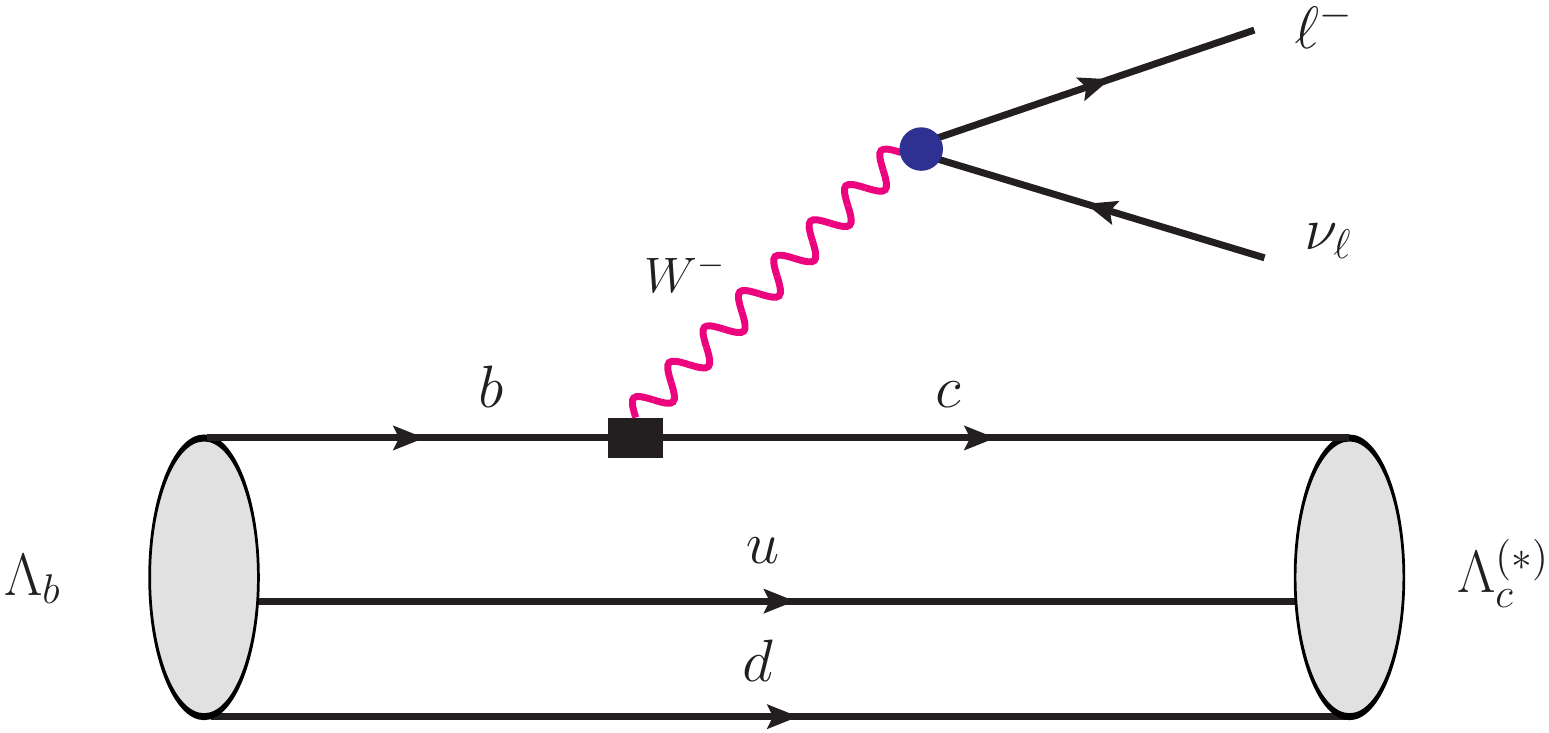}\\
  \caption{The diagram for depicting the $\Lambda_{b}\to\Lambda_{c}^{(*)}\ell^{-}\nu_{\ell}$ decays in tree level.}  \label{transition}
\end{figure}

The Hamiltonian depicted corresponding semileptonic decays is written as
\begin{equation}
\mathcal{H}_{eff}=\frac{G_F}{\sqrt{2}}V_{cb}\bar{\ell}\gamma_{\mu}(1-\gamma_5)\nu_\ell\bar{c}\gamma^{\mu}(1-\gamma_5)b,
\end{equation}
where $G_F$ is the Fermi constant and $V_{cb}$ is the CKM matrix element. In Fig. \ref{transition}, the Feynman diagram for the weak transition is given.

Given that the quarks are confined in hadron, the weak transition matrix element cannot be calculated by QCD since these weak transitions are involved in low-energy aspects of QCD. Generally, the matrix elements of hadron section $\langle\Lambda_{c}^{(*)}(1/2^{\pm})|\bar{c}\gamma^{\mu}(1-\gamma_5)b|\Lambda_b(1/2^+)\rangle$ can be parametrized in terms of several dimensionless form factors as

\begin{equation}
\begin{split}
\langle&\Lambda_{c}^{(*)}(1/2^{+})(P',J'_z)|\bar{c}\gamma^{\mu}(1-\gamma_5)b|\Lambda_b(P,J_z)\rangle\\
&=\bar{u}(P',J'_z)\left[f^V_1(q^2)\gamma^{\mu}+i\frac{f^V_2(q^2)}{M}\sigma^{\mu\nu}q_{\nu}+\frac{f^V_3(q^2)}{M}q^{\mu}\right.\\
&\quad\left.-\left(g^A_1(q^2)\gamma^{\mu}+i\frac{g^A_2(q^2)}{M}\sigma^{\mu\nu}q_{\nu}+\frac{g^A_3(q^2)}{M}q^{\mu}\right)\gamma_5\right]u(P,J_z),
\label{general expression of parameter 05+}
\end{split}
\end{equation}
\begin{equation}
\begin{split}
\langle&\Lambda_{c}(1/2^{-})(P',J'_z)|\bar{c}\gamma^{\mu}(1-\gamma_5)b|\Lambda_b(P,J_z)\rangle\\
&=\bar{u}(P',J'_z)\left[\left(g^V_1(q^2)\gamma^{\mu}+i\frac{g^V_2(q^2)}{M}\sigma^{\mu\nu}q_{\nu}+\frac{g^V_3(q^2)}{M}q^{\mu}\right)\gamma_5\right.\\
&\quad\left.-\left(f^A_1(q^2)\gamma^{\mu}+i\frac{f^A_2(q^2)}{M}\sigma^{\mu\nu}q_{\nu}+\frac{f^A_3(q^2)}{M}q^{\mu}\right)\right]u(P,J_z),
\label{general expression of parameter 05-}
\end{split}
\end{equation}
for $1/2^+\to1/2^+$ and $1/2^+\to1/2^-$ transitions, respectively. Here, $M(P)$, $M'(P')$ denote the masses (four momentums) of the initial and final hadrons respectively, and $q=P-P'$.

Following the standard procedure, the angular distribution for the decay $\Lambda_{b}\to\Lambda_{c}^{(*)}W^{-}(\to \ell^{-}\bar{\nu}_{\ell})$ reads as
\begin{equation}
\begin{split}
\frac{d\Gamma}{dq^2d\cos\theta_{l}}=&
\frac{G_F^2}{(2\pi)^3}|V_{cb}|^2\frac{\sqrt{\lambda(M^2,M'^2,q^2)}(q^2-m_l^2)^2}{128M^3q^2}\\
&\times\left[A_{1}+\frac{m_l^2}{q^2}A_{2}\right],
\label{width_q2theta}
\end{split}
\end{equation}
where $\lambda(x,y,z)=x^2+y^2+z^2-2xy-2xz-2yz$ is the kinematical triangle Kallen function and
\begin{equation}
\begin{split}
A_{1}=&2\sin^2\theta_{l}(H^2_{1/2,0}+H^2_{-1/2,0})+(1-\cos\theta_{l})^2H^2_{1/2,1}\\
&+(1+\cos\theta_{l})^2H^2_{-1/2,-1},\\
A_{2}=&2\cos^2\theta_{l}(H^2_{1/2,0}+H^2_{-1/2,0})+\sin^2\theta_{l}(H^2_{1/2,1}+H^2_{-1/2,-1})\\
&-4\cos\theta_{l}(H_{1/2,t}H_{1/2,0}+H_{-1/2,t}H_{-1/2,0})\\
&+2(H^2_{1/2,t}+H^2_{-1/2,t}).\label{A1A2}
\end{split}
\end{equation}
$\theta_{l}$ is the angle between momenta of the final hadron and the lepton in the $q^2$ rest frame. The helicity amplitudes in Eq. (\ref{A1A2}), $H^{V,A}_{\lambda_{\Lambda_{c}^{(*)}},\lambda_{W^{-}}}$, can be expressed as a series functions of the form factors defined in Eqs. (\ref{general expression of parameter 05+}) and (\ref{general expression of parameter 05-}),
\begin{equation}
\begin{split}
&H^{V}_{\frac{1}{2},0}(1/2^+\to1/2^{\pm})=\frac{\sqrt{Q_{\mp}}}{\sqrt{q^2}}\left(M_{\pm}f(g)^V_1(q^2)\mp\frac{q^2}{M}f(g)^V_2(q^2)\right),\\
&H^{A}_{\frac{1}{2},0}(1/2^+\to1/2^{\pm})=\frac{\sqrt{Q_{\pm}}}{\sqrt{q^2}}\left(M_{\mp}g(f)^A_1(q^2)\pm\frac{q^2}{M}g(f)^A_2(q^2)\right),\\
&H^{V}_{\frac{1}{2},1}(1/2^+\to1/2^{\pm})=\sqrt{2Q_{\mp}}\left(f(g)^V_1(q^2)\mp\frac{M_{\pm}}{M}f(g)^V_2(q^2)\right),\\
&H^{A}_{\frac{1}{2},1}(1/2^+\to1/2^{\pm})=\sqrt{2Q_{\pm}}\left(g(f)^A_1(q^2)\pm\frac{M_{\mp}}{M}g(f)^A_2(q^2)\right),\\
&H^{V}_{\frac{1}{2},t}(1/2^+\to1/2^{\pm})=\frac{\sqrt{Q_{\pm}}}{\sqrt{q^2}}\left(M_{\mp}f(g)^V_1(q^2)\pm\frac{q^2}{M}f(g)^V_3(q^2)\right),\\
&H^{A}_{\frac{1}{2},t}(1/2^+\to1/2^{\pm})=\frac{\sqrt{Q_{\mp}}}{\sqrt{q^2}}\left(M_{\pm}g(f)^A_1(q^2)\mp\frac{q^2}{M}g(f)^A_3(q^2)\right),\\
\end{split}
\end{equation}
where $Q_{\pm}$ is defined as $Q_{\pm}=(M \pm M')^2-q^2$ and $M_{\pm}=M\pm M'$. The negative helicities for transitions involved final hadrons with $J^{P}=$ $1/2^{+}$ and $1/2^{-}$ can be obtained by
\begin{equation}
H^V_{-\lambda',-\lambda_W}=\mathcal{P}^{V}H^V_{\lambda',\lambda_W},~H^A_{-\lambda',-\lambda_W}=\mathcal{P}^{A}H^A_{\lambda',\lambda_W}
\end{equation}
with $(\mathcal{P}^{V}, \mathcal{P}^{A})=$ $(+,-)$ and $(-,+)$, respectively. The total helicity amplitudes can be obtained by
\begin{equation}
H_{\lambda',\lambda_W}=H^V_{\lambda',\lambda_W}-H^A_{\lambda',\lambda_W}.
\end{equation}

It is obvious that the form factors play important roles in the calculation of the semileptonic decays, which can be calculated by concrete models. In this work, we adopt the light-front quark model, where the general expressions of $\Lambda_{b}\to\Lambda_{c}^{(*)}\left(1/2^{\pm}\right)$ weak transitions are written as \cite{Ke:2019smy}
\begin{widetext}
\begin{equation}
\begin{split}
\langle\Lambda_{c}&^{(*)}\left(1/2^+\right)(\bar{P}',J'_z)|\bar{c}\gamma^{\mu}(1-\gamma_5)b|\Lambda_b\left(1/2^+\right)(\bar{P},J_z)\rangle\\
=&\int\left(\frac{dx_1d^{2}\vec{k}_{1\bot}}{2(2\pi)^3}\right)\left(\frac{dx_2d^{2}\vec{k}_{2\bot}}{{2(2\pi)^3}}\right)
\frac{\phi_{\Lambda_{b}}(x_i,\vec{k}_{i\bot})\phi^{*}_{\Lambda_{c}^{(*)}}(x_i',\vec{k}_{i\bot}')}{16\sqrt{x_3 x_3'{M_0}^3{M_0'}^3}}
\frac{\text{Tr}[(\slashed{\bar{P}}'-M_0')\gamma_5(\slashed{p}_1+m_1)(\slashed{\bar{P}}+M_0)\gamma_5(\slashed{p}_2-m_2)]}{\sqrt{(e_1+m_1)(e_2+m_2)(e_3+m_3)(e_1'+m_1')(e_2'+m_2')(e_3'+m_3')}}\\
&\times\bar{u}(\bar{P}',J_z')(\slashed{p}_3'+m_3')\gamma^{\mu}(1-\gamma_5)(\slashed{p}_3+m_3)u(\bar{P},J_z),
\label{general expression of 05+}
\end{split}
\end{equation}
\begin{equation}
\begin{split}
\langle\Lambda_{c}&^{*}\left(1/2^-\right)(\bar{P}',J'_z)|\bar{c}\gamma^{\mu}(1-\gamma_5)b|\Lambda_b\left(1/2^+\right)(\bar{P},J_z)\rangle\\
=&\int\left(\frac{dx_1 d^{2}\vec{k}_{1\bot}}{2(2\pi)^3}\right)\left(\frac{dx_2d^{2}\vec{k}_{2\bot}}{{2(2\pi)^3}}\right)
\frac{\phi_{\Lambda_{b}}(x_i,\vec{k}_{i\bot})\phi^{*}_{\Lambda_{c}^{*}}(x_i',\vec{k}_{i\bot}')}{16\sqrt{3x_3x_3'{M_0}^3{M_0'}^3}}
\frac{\text{Tr}[(\slashed{\bar{P}}'-M_0')\gamma_5(\slashed{p}_1+m_1)(\slashed{\bar{P}}+M_0)\gamma_5(\slashed{p}_2-m_2)]}{\sqrt{(e_1+m_1)(e_2+m_2)(e_3+m_3)(e_1'+m_1')(e_2'+m_2')(e_3'-m_3')}}\\
&\times\frac{\sum_{i}m_i'/m_3'}{\sum_{i}m_i'+M_0'}\bar{u}(\bar{P}',J_z')\left(\slashed{K}'-\frac{m_{3}'^{2}-\mu'^2}{2M_0'}\right)\gamma_{5}(\slashed{p}_3'+m_3')\gamma^{\mu}(1-\gamma_5)(\slashed{p}_3+m_3)u(\bar{P},J_z),
\label{general expression of 05-}
\end{split}
\end{equation}
where $K'=\frac{(m_1'+m_2')p_3'-m_3'(p_1+p_2)}{m_1'+m_2'+m_3'}$ is the $\lambda-$mode momentum of the final hadron. And, $\mu'=m_1'm_2'/m_1'+m_2'$; $\bar{P}=p_1+p_2+p_3$ and $\bar{P}'=p_1+p_2+p_3'$ are the light-front momentum for initial or final hadrons respectively, considering $p_1=p_1'$ and $p_2=p_2'$ in the spectator model. $\phi(x_i,\vec{k}_{i\bot})$ and $\phi^{*}(x_i',\vec{k}_{i\bot}')$ represent the wave functions for initial or final hadrons, respectively.

We follow Ref. \cite{Chua:2019yqh}, where they chose a special gauge $q^{+}=0$, to get the form factors with $V^{+}$, $A^{+}$, $\vec{q}_{\bot}\cdot\vec{V}$, $\vec{q}_{\bot}\cdot\vec{A}$, $\vec{n}_{\bot}\cdot\vec{V}$, and $\vec{n}_{\bot}\cdot\vec{A}$ in the way shown below. Firstly, multiply $\sum_{J_z,J_z'}\delta_{J_z,J_z'}$, $\sum_{J_z,J_z'}(\sigma^3)_{J_z,J_z'}$, $\sum_{J_z,J_z'}(\vec{\sigma}\cdot \vec{q}_{\bot})_{J_z,J_z'}$, $\sum_{J_z,J_z'}(\vec{\sigma}\cdot \vec{q}_{\bot}\sigma^3)_{J_z,J_z'}$ to both sides of Eq. (\ref{general expression of parameter 05+}) and Eq. (\ref{general expression of parameter 05-}), and then use Eq. (11) of Ref. \cite{Chua:2019yqh} in the right side, use Eq. (\ref{general expression of 05+}), Eq. (\ref{general expression of 05-}) and Eq. (10) of Ref. \cite{Chua:2019yqh} in the left side. Following the above steps we can finally obtain the concrete expressions for the involved form factors.

The expressions for the form factors describing $\Lambda_{b}\to\Lambda_{c}^{(*)}(1/2^{+})$ transitions are
\begin{equation}
\begin{split}
f^V_1(q^2)=&\int\mathcal{DS}\frac{1}{8P^+P'^+}
\text{Tr}[(\slashed{\bar{P}}+M_0)\gamma^+(\slashed{\bar{P}}'+M_0')(\slashed{p}_3'+m_3')\gamma^+(\slashed{p}_3+m_3)],\\
f^V_2(q^2)=&\int\mathcal{DS}\frac{iM}{8P^+P'^+q^2}
\text{Tr}[(\slashed{\bar{P}}+M_0)\sigma^{+\mu}q_{\mu}(\slashed{\bar{P}}'+M_0')(\slashed{p}_3'+m_3')\gamma^+(\slashed{p}_3+m_3)],\\
f^V_3(q^2)=&\frac{M}{M+M'}\left(f_1^V(q^2)(1-\frac{2\bar{P}\cdot q}{q^2})+\int\mathcal{DS}\frac{1}{4\sqrt{P^+P'^+}q^2}
\text{Tr}[(\slashed{\bar{P}}+M_0)\gamma^+(\slashed{\bar{P}}'+M_0')(\slashed{p}_3'+m_3')\slashed{q}(\slashed{p}_3+m_3)]\right),\\
g^A_1(q^2)=&\int\mathcal{DS}\frac{1}{8P^+P'^+}
\text{Tr}[(\slashed{\bar{P}}+M_0)\gamma^+\gamma_5(\slashed{\bar{P}}'+M_0')(\slashed{p}_3'+m_3')\gamma^+\gamma_5(\slashed{p}_3+m_3)],\\
g^A_2(q^2)=&\int\mathcal{DS}\frac{-iM}{8P^+P'^+q^2}
\text{Tr}[(\slashed{\bar{P}}+M_0)\sigma^{+\mu}q_{\mu}\gamma_5(\slashed{\bar{P}}'+M_0')(\slashed{p}_3'+m_3')\gamma^+\gamma_5(\slashed{p}_3+m_3)],\\
g^A_3(q^2)=&\frac{M}{M-M'}\left(g^A_1(q^2)(-1+\frac{2\bar{P}\cdot q}{q^2})+\int\mathcal{DS}\frac{-1}{4\sqrt{P^+P'^+}q^2}
\text{Tr}[(\slashed{\bar{P}}+M_0)\gamma^+\gamma_5(\slashed{\bar{P}}'+M_0')(\slashed{p}_3'+m_3')~\slashed{q}~\gamma_5~(\slashed{p}_3+m_3)]\right)
\label{formfactorS}
\end{split}
\end{equation}
with
\begin{equation}
\mathcal{DS}=\frac{dx_1d^2\vec{k}_{1\bot}dx_2d^2\vec{k}_{2\bot}}{2(2\pi)^32(2\pi)^3}
\frac{\phi^*(x_i',\vec{k}_{i\bot}')\phi(x_i,\vec{k}_{i\bot})}{16\sqrt{x_3x_3'M_0^3M_0'^3}}
\frac{\text{Tr}[(\slashed{\bar{P}}'-M_0')\gamma_5(\slashed{p}_1+m_1)(\slashed{\bar{P}}+M_0)\gamma_5(\slashed{p}_2-m_2)]}{\sqrt{(e_1+m_1)(e_2+m_2)(e_3+m_3)(e_1'+m_1')(e_2+m_2')(e_3'+m_3')}},
\end{equation}
while the expressions for the form factors describing $\Lambda_{b}\to\Lambda_{c}(1/2^{-})$ transition are
\begin{equation}
\begin{split}
g^V_1(q^2)=&\int\mathcal{DP}\frac{1}{8P^+P'^+}
\text{Tr}[(\slashed{\bar{P}}+M_0)\gamma^+\gamma_5(\slashed{\bar{P}}'+M_0')\left(\slashed{K}'-\frac{m_{3}'^{2}-\mu'^2}{2M_0'}\right)\gamma_{5}(\slashed{p}_3'+m_3')\gamma^+(\slashed{p}_3+m_3)],\\
g^V_2(q^2)=&\int\mathcal{DP}\frac{-iM}{8P^+P'^+q^2}
\text{Tr}[(\slashed{\bar{P}}+M_0)\sigma^{+\mu}q_{\mu}\gamma_5(\slashed{\bar{P}}'+M_0')\left(\slashed{K}'-\frac{m_{3}'^{2}-\mu'^2}{2M_0'}\right)\gamma_{5}(\slashed{p}_3'+m_3')\gamma^+(\slashed{p}_3+m_3)],\\
g^V_3(q^2)=&\frac{M}{M+M'}\left(f_1^V(q^2)(1-\frac{2\bar{P}\cdot q}{q^2})\right.\\
&\left.+\int\mathcal{DP}\frac{1}{4\sqrt{P^+P'^+}q^2}
\text{Tr}[(\slashed{\bar{P}}+M_0)\gamma^+\gamma_5(\slashed{\bar{P}}'+M_0')\left(\slashed{K}'-\frac{m_{3}'^{2}-\mu'^2}{2M_0'}\right)\gamma_{5}(\slashed{p}_3'+m_3')\slashed{q}(\slashed{p}_3+m_3)]\right),\\
f^A_1(q^2)=&\int\mathcal{DP}\frac{1}{8P^+P'^+}
\text{Tr}[(\slashed{\bar{P}}+M_0)\gamma^+(\slashed{\bar{P}}'+M_0')\left(\slashed{K}'-\frac{m_{3}'^{2}-\mu'^2}{2M_0'}\right)\gamma_{5}(\slashed{p}_3'+m_3')\gamma^+\gamma_5(\slashed{p}_3+m_3)],\\
f^A_2(q^2)=&\int\mathcal{DP}\frac{iM}{8P^+P'^+q^2}
\text{Tr}[(\slashed{\bar{P}}+M_0)\sigma^{+\mu}q_{\mu}(\slashed{\bar{P}}'+M_0')\left(\slashed{K}'-\frac{m_{3}'^{2}-\mu'^2}{2M_0'}\right)\gamma_{5}(\slashed{p}_3'+m_3')\gamma^+\gamma_5(\slashed{p}_3+m_3)],\\
f^A_3(q^2)=&\frac{M}{M-M'}\left(g^A_1(q^2)(-1+\frac{2\bar{P}\cdot q}{q^2})\right.\\
&\left.+\int\mathcal{DP}\frac{-1}{4\sqrt{P^+P'^+}q^2}
\text{Tr}[(\slashed{\bar{P}}+M_0)\gamma^+(\slashed{\bar{P}}'+M_0')\left(\slashed{K}'-\frac{m_{3}'^{2}-\mu'^2}{2M_0'}\right)\gamma_{5}(\slashed{p}_3'+m_3')\slashed{q}\gamma_5(\slashed{p}_3+m_3)]\right)
\end{split}
\end{equation}
with
\begin{equation}
\mathcal{DP}=\frac{dx_1d^2\vec{k}_{1\bot}dx_2d^2\vec{k}_{2\bot}}{2(2\pi)^32(2\pi)^3}
\frac{\phi^*(x_i',\vec{k}_{i\bot}')~\phi(x_i,\vec{k}_{i\bot})}{16\sqrt{3x_3x_3'M_0^3M_0'^3}}\frac{(\sum_{i}m_i')/m_3'}{(M_0'+\sum_{i}m_i')}
\frac{\text{Tr}[(\slashed{\bar{P}}'-M_0')\gamma_5(\slashed{p}_1+m_1)(\slashed{\bar{P}}+M_0)\gamma_5(\slashed{p}_2-m_2)]}{\sqrt{(e_1+m_1)(e_2+m_2)(e_3+m_3)(e_1'+m_1')(e_2+m_2')(e_3'-m_3')}}.
\label{formfactorP}
\end{equation}

\end{widetext}

The ones for $\Lambda_{c}\to\Lambda^{(*)}\left(1/2^{\pm}\right)$ modes have the same expressions. Since the light-front quark model has been widely used for studying the weak transition form factors, we will not repeatedly introduce the explicit forms for this model here. One can turn to Refs. \cite{Ke:2019smy,Chua:2019yqh,Chua:2018lfa,Ke:2007tg, Cheng:2004cc,Wei:2009np,Chang:2019obq,Chang:2020wvs,Ke:2019lcf,Ke:2012wa,Wang:2017mqp} for detailed introductions.

The spatial wave functions of these baryons involved in the discussed semileptonic decay are crucial input when calculating the form factors. In the previous works, the spatial wave functions were generally used in simple harmonic oscillator form with the oscillator parameter $\beta$ \cite{Ke:2019smy,Chua:2019yqh,Ke:2007tg,Chua:2018lfa}. In this work, we adopt the exact spatial wave functions for the involved baryons with the help of the GEM, which will be introduced in the next subsection. This approach should be supported by the study of the mass spectrum of the baryon.

\subsection{The spatial wave function of these involved baryons}
\label{The wave function}

Differing from former treatments \cite{Ke:2019smy,Chua:2019yqh,Ke:2007tg,Chua:2018lfa}, in this work we adopt GEM to get the wave functions of baryons. As is known, baryons are formed by three constituent quarks in the traditional constituent quark model. Different from mesons, one baryon bound state is the typical three-body system, and its wave function can be extracted by solving the three-body Schr\"odinger equation. In this work, we use the semirelativistic potentials introduced by Godfrey and Isgur \cite{Godfrey:1985xj}, which was applied by Capstick and Isgur to study baryon spectroscopy \cite{Capstick:1986bm}.

In the light quark sector, the relativistic effects in effective potential should be considered. Those effects are introduced by two ways in the GI model. At the first step, the GI model introduces a smeared function, and thus the Hamiltonian depicted three-body bound system can be written as
\begin{equation}
\mathcal{H}=K+\sum_{i=1<j}^{3}(S_{ij}+G_{ij}+V^{\text{so}(s)}_{ij}+V^{\text{so}(v)}_{ij}+V^{\text{tens}}_{ij}+V^{\text{con}}_{ij}),
\end{equation}
where $K$, $S$, $G$, $V^{\text{so}(s)}$, $V^{\text{so}(v)}$,$V^{\text{tens}}$ and $V^{\text{con}}$ are the kinetic energy, the spin-independent linear confinement piece, the Coulomb-like potential, the scalar type-spin-orbit interaction, the vector type-spin-orbit interaction, the tensor potential, and the spin-dependent contact potential, respectively. They are written as \cite{Godfrey:1985xj,Capstick:1986bm,Song:2015nia,Pang:2017dlw}
\begin{eqnarray}
K&=&\sum_{i=1,2,3}\sqrt{m_i^2+p_i^2},\\
S_{ij}&=&-\frac{3}{4}\left(br_{ij}\left[\frac{e^{-\sigma^2r_{ij}^2}}{\sqrt{\pi}\sigma r_{ij}}+\left(1+\frac{1}{2\sigma^2r_{ij}^2}\right)\frac{2}{\sqrt{\pi}}\right.\right.\nonumber\\&&\left.\left.
\times\int_{0}^{\sigma r_{ij}}e^{-x^2}dx\right]\right)\mathbf{F_i}\cdot\mathbf{F_j}+\frac{c}{3},\\
G_{ij}&=&\sum_{k}\frac{\alpha_k}{r_{ij}}\left[\frac{2}{\sqrt{\pi}}\int_{0}^{\tau_k r_{ij}}e^{-x^2}dx\right]\mathbf{F_i}\cdot\mathbf{F_j},
\end{eqnarray}
for spin-independent terms with
\begin{equation} \sigma^2=\sigma_0^2\left[\frac{1}{2}+\frac{1}{2}\left(\frac{4m_im_j}{(m_i+m_j)^2}\right)^4+s^2\left(\frac{2m_im_j}{m_i+m_j}\right)^2\right],
\end{equation}
and
\begin{equation}
\begin{split}
V^{\text{so}(s)}_{ij}=&-\frac{\mathbf{r_{ij}}\times\mathbf{p_{i}}\cdot\mathbf{S_i}}{2m_i^2}\frac{1}{r_{ij}}
\frac{\partial S_{ij}}{r_{ij}}+\frac{\mathbf{r_{ij}}\times\mathbf{p_{j}}\cdot\mathbf{S_j}}{2m_j^2}\frac{1}{r_{ij}}\frac{\partial S_{ij}}{\partial r_{ij}},\\
V^{\text{so}(v)}_{ij}=&\frac{\mathbf{r_{ij}}\times\mathbf{p_{i}}\cdot\mathbf{S_i}}{2m_i^2}\frac{1}{r_{ij}}\frac{\partial G_{ij}}{r_{ij}}
-\frac{\mathbf{r_{ij}}\times\mathbf{p_{j}}\cdot\mathbf{S_j}}{2m_j^2}\frac{1}{r_{ij}}\frac{\partial G_{ij}}{r_{ij}}
\nonumber\\&-\frac{\mathbf{r_{ij}}\times\mathbf{p_{j}}\cdot\mathbf{S_i}-\mathbf{r_{ij}}\times\mathbf{p_{i}}\cdot\mathbf{S_j}}{m_i~m_j}\frac{1}{r_{ij}}\frac{\partial G_{ij}}{\partial r_{ij}},\\
V^{\text{tens}}_{ij}=&-\frac{1}{m_im_j}\left[\left(\mathbf{S_i}\cdot\mathbf{\hat r_{ij}}\right)\left(\mathbf{S_j}\cdot \mathbf{\hat r_{ij}}\right)-\frac{\mathbf{S_i}\cdot\mathbf{S_j}}{3}\right]\left(\frac{\partial^2G_{ij}}{\partial r^2}-\frac{\partial G_{ij}}{r_{ij}\partial r_{ij}}\right),\\
V^{\text{con}}_{ij}=&\frac{2\mathbf{S_i}\cdot\mathbf{S_j}}{3m_i m_j}\nabla^2G_{ij},
\end{split}
\end{equation}
for the spin-dependent terms, where, $m_i$ and $m_j$ are the mass of quark i and j, respectively. $\langle\mathbf{F_i}\cdot\mathbf{F_j}\rangle=-2/3$ for quark-quark interaction.

At the second step, the general potential (which relies on the center-of-mass of interacting quarks and momentum) is made up for the loss of relativistic effects in the nonrelativistic limit \cite{Godfrey:1985xj,Capstick:1986bm,Song:2015nia,Pang:2017dlw,Wang:2018rjg},
\begin{equation}
\begin{split}
&G_{ij}\to\left(1+\frac{p^2}{E_iE_j}\right)^{1/2} G_{ij}\left(1+\frac{p^2}{E_iE_j}\right)^{1/2},\\
&\frac{V^{k}_{ij}}{m_im_j}\to\left(\frac{m_im_j}{E_iE_j}\right)^{1/2+\epsilon_k}\frac{V^k_{ij}}{m_im_j}\left(\frac{m_im_j}{E_iE_j}\right)^{1/2+\epsilon_k}
\end{split}
\end{equation}
with $E_i=\sqrt{p^2+m_i^2}$, {where subscript $k$ was applied to distinguish the contributions from the contact, the tensor, the vector spin-orbit, and the scalar spin-orbit terms. In addition, $\epsilon_k$ represents the relevant modification parameters, which are listed in Table \ref{parametersofGI}}.

\begin{table}
\centering
\caption{The parameters used in the semirelativistic potential model.}
\label{parametersofGI}
\renewcommand\arraystretch{1.2}
\begin{tabular*}{86mm}{c@{\extracolsep{\fill}}cccc}
\toprule[1pt]
\toprule[0.7pt]
Parameters     & Values   & Parameters   & Values\\
\toprule[0.7pt]
$m_u~(\text{GeV})$     & $0.220$   & $\epsilon^{\text{so}(s)}$   & $0.448$\\
$m_d~(\text{GeV})$     & $0.220$   & $\epsilon^{\text{so}(v)}$   & $-0.062$\\
$m_s~(\text{GeV})$     & $0.419$   & $\epsilon^{\text{tens}}$   & $0.379$\\
$m_c~(\text{GeV})$     & $1.628$   & $\epsilon^{\text{con}}$   & $-0.142$\\
$m_b~(\text{GeV})$     & $4.977$   & $\sigma_0~(\text{GeV})$   & $2.242$\\
$b~(\text{GeV}^2)$     & $0.142$   & $s$   & $0.805$\\
$c~(\text{GeV})  $     & $-0.302$  & $$    & $$\\
\bottomrule[0.7pt]
\bottomrule[1pt]
\end{tabular*}
\end{table}

\begin{table*}[htbp]
\centering
\caption{The spatial wave functions of $\Lambda_b$, $\Lambda_{c}^{(*)}\left(1/2^{\pm}\right)$ as well as $\Lambda^{(*)}\left(1/2^{\pm}\right)$ with the GI model by the GEM. The Gaussian bases $(n_{\rho},n_{\lambda})$ listed in the third column are arranged as: $[(1,1),(1,2),\cdots,(1,n_{\lambda_{max}}),(2,1),(2,2),\cdots,(2,n_{\lambda_{max}}),\cdots,(n_{\rho_{max}},1),(n_{\rho_{max}},2),\cdots,(n_{\rho_{max}},n_{\lambda_{max}})]$. Here, we also listed the names of states, which are given in PDG, in the second column.}
\label{Wave functions}
\renewcommand\arraystretch{1.35}
{\footnotesize
\begin{tabular*}{165mm}{c@{\extracolsep{\fill}}ccc}
\toprule[1pt]
\toprule[0.7pt]
States  & Name in PDG  &Eigenvectors\\
\toprule[0.7pt]
\multirow{3}*{\shortstack{$\Lambda_b\left(\frac{1}{2}^{+}\right)$}}        &\multirow{3}*{$\Lambda_b$}
& $[-0.0177,-0.0561,-0.0930,-0.0020,-0.0022,0.0007,-0.0382,0.0018,-0.0009,-0.0024,0.0020,-0.0006,$\\
& &$0.0172,-0.3036,-0.2450,-0.0090,-0.0030,0.0010,0.0070,0.0590,-0.3754,-0.0091,-0.0024,0.0008,$\\
& &$-0.0090,-0.0043,-0.0265,-0.1238,0.0256,-0.0057,0.0026,0.0022,0.0169,0.0149,-0.0088,0.0018]$\\
\toprule[0.7pt]
\multirow{3}*{\shortstack{$\Lambda_c\left(\frac{1}{2}^{+}\right)$}}  &\multirow{3}*{$\Lambda_c^+$}
& $[-0.0162,-0.0404,-0.1010,-0.0174,0.0003,0.0001,-0.0321,-0.0094,-0.0024,-0.0027,0.0019,-0.0005,$\\
& &$0.0163,-0.2232,-0.2985,-0.0527,0.0045,-0.0007,0.0002,0.0701,-0.3035,-0.0831,0.0127,-0.0027,$\\
& &$-0.0054,-0.0095,-0.0129,-0.1277,0.0183,-0.0042,0.0018,0.0024,0.0135,0.0186,-0.0085,0.0016]$\\
\toprule[0.7pt]
\multirow{3}*{\shortstack{$\Lambda_c\left(\frac{1}{2}^{-}\right)$}}        &\multirow{3}*{$\Lambda_c(2595)^+$}
& $[0.0044,0.0227,0.0983,0.0423,-0.0049,0.0010,0.0149,-0.0018,0.0083,0.0016,-0.0011,0.0003,$\\
& &$-0.0065,0.1155,0.3076,0.1148,-0.0139,0.0030,0.0021,-0.0485,0.2750,0.1689,-0.0283,0.0067,$\\
& &$0.0037,0.0011,0.0136,0.1510,-0.0104,0.0026,-0.0013,0.0005,-0.0131,-0.0266,0.0080,-0.0015]$\\
\toprule[0.7pt]
\multirow{3}*{\shortstack{$\Lambda_c^{*}\left(\frac{1}{2}^{+}\right)$}}        &\multirow{3}*{$\Lambda_c(2765)^+$}
& $[-0.0244,-0.0944,-0.127630,0.1982,-0.0127,0.0034,-0.0375,-0.0588,0.0207,0.0079,-0.0028,0.0004,$\\
& &$0.0206,-0.4044,-0.4802,0.5686,-0.0447,0.0110,0.0099,0.0386,-0.2691,0.5855,-0.0219,0.0051,$\\
& &$-0.0080,-0.0099,-0.1144,0.3734,0.0859,-0.0156,0.0014,0.0089,0.0311,-0.0673,0.0101,0.0002]$\\
\toprule[0.7pt]
\multirow{3}*{\shortstack{$\Lambda\left(\frac{1}{2}^{+}\right)$}}    &\multirow{3}*{$\Lambda$}
& $[0.0171,0.0239,0.0847,0.0412,0.0035,-0.0005,0.0373,-0.0047,0.0106,0.0004,-0.0013,0.0006,$\\
& &$-0.0192,0.1969,0.2244,0.1544,-0.0024,0.0017,0.0014,-0.0693,0.2616,0.1231,0.0116,-0.0003,$\\
& &$0.0041,0.0115,-0.0091,0.1469,-0.0115,0.0038,-0.0016,-0.0021,-0.0071,-0.0248,0.0090,-0.0022]$\\
\toprule[0.7pt]
\multirow{3}*{\shortstack{$\Lambda\left(\frac{1}{2}^{-}\right)$}}    &\multirow{3}*{$\Lambda(1405)$}
& $[0.0058,0.0145,0.0715,0.0639,0.0019,0.0002,0.0193,-0.0075,0.0074,0.0028,-0.0015,0.0004,$\\
& &$-0.0101,0.1087,0.2125,0.1904,0.0020,0.0013,0.0027,-0.0469,0.2260,0.2215,-0.0009,0.0029,$\\
& &$0.0022,0.0061,-0.0113,0.1742,0.0127,-0.0007,-0.0010,-0.0005,-0.0054,-0.0329,0.0063,-0.0012]$\\
\toprule[0.7pt]
\multirow{3}*{\shortstack{$\Lambda^{*}\left(\frac{1}{2}^{+}\right)$}}    &\multirow{3}*{$\Lambda(1600)$}
& $[0.0310,0.0424,0.1590,-0.1262,-0.0378,0.0031,0.0429,0.0480,-0.0085,-0.0117,0.0011,0.0000,$\\
& &$-0.0119,0.2645,0.5786,-0.2742,-0.1134,0.0120,-0.0001,-0.0639,0.2197,-0.4162,-0.1299,0.0128,$\\
& &$0.0063,0.0065,0.1069,-0.3538,-0.2064,0.0247,-0.0012,-0.0062,-0.0280,0.0514,0.0050,-0.0022]$\\
\toprule[0.7pt]
\toprule[1pt]
\end{tabular*}
}
\end{table*}

Now we illustrate the construction of the baryon wave function. The total wave function is composed of color, flavor, space, and spin wave functions as
\begin{equation}
\Psi_{\mathbf{J},\mathbf{M_J}}=\chi^{c}\left\{\chi^{s}_{\mathbf{S},\mathbf{M_S}}\psi^{p}_{\mathbf{L},\mathbf{M_L}}\right\}_{\mathbf{J},\mathbf{M_J}}\psi^{f},
\end{equation}
where $\chi^{c}=\frac{1}{\sqrt{6}}(rgb-rbg+gbr-grb+brg-bgr)$ is the color wave function and is universal. The subscripts $\textbf{S}$, $\textbf{L}$ and $\textbf{J}$ are the total spin angular momentum, the total orbital angular momentum, and the total angular momentum quantum numbers, respectively. $\psi^{p}_{\mathbf{L},\mathbf{M_L}}$ denotes the spatial wave function consisting of the $\rho$-mode and $\lambda$-mode as
\begin{equation}
\psi^{p}_{\mathbf{L},\mathbf{M_L}}(\vec{\rho},\vec{\lambda})=\left\{\phi_{\pmb{l_{\rho}},\pmb{ml_{\rho}}}(\vec{\rho})
\phi_{\pmb{l_{\lambda}},\pmb{ml_{\lambda}}}(\vec{\lambda})\right\}_{\mathbf{L},\mathbf{M_L}},
\end{equation}
where the subscripts $\pmb{l_{\rho}}$ and $\pmb{l_{\lambda}}$ are the orbital angular momentum for the $\rho$ and $\lambda$-mode, respectively. In this paper, the internal Jacobi coordinates are chosen as
\begin{equation}
\vec{\rho}=\vec{r}_1-\vec{r}_2,~~\vec{\lambda}=\vec{r}_3-\frac{m_1 \vec{r}_1+m_2 \vec{r}_2}{m_1+m_2}.
\end{equation}
Note that we only use one channel here. It is suitable for singly heavy baryons while it is an approximation for light baryons.

The Rayleigh-Ritz variational principle is used in this work to solve the three-body Schr\"odinger equation
\begin{equation}
\mathcal{H} \Psi_{\mathbf{J},\mathbf{M_J}}=E \Psi_{\mathbf{J},\mathbf{M_J}},
\end{equation}
and thus we can obtain the eigenvalues $E$, and the spatial wave functions. The Gaussian bases \cite{Hiyama:2003cu,Yoshida:2015tia,Yang:2019lsg}
\begin{equation}
\phi_{n,l,m}^{G}(\vec{r})=\phi^{G}_{n,l}(r)Y_{l,m}(\hat{r}),
\label{Gaussian basis}
\end{equation}
are used to expand the spatial wave function, where $n=1,2,\cdots,n_{max}$ with a freedom parameter $n_{max}$ which should be chosen from positive integers, and the Gaussian size parameters $\nu_{n}$ are settled as a geometric progression as
\begin{equation}
\nu_{n}=1/r^2_{n}, ~r_{n}=r_{min}~a^{n-1}
\end{equation}
with $$a=\left(\frac{r_{max}}{r_{min}}\right)^{\frac{1}{n_{max}-1}}.$$

Using the parameters collected in Table \ref{parametersofGI} as input, we can obtain the masses \footnote{ For the ground states, the estimated masses are 1131 MeV, 2286 MeV and 5621 MeV for $\Lambda$, $\Lambda_c$ and $\Lambda_b$, respectively, which are close to the experimental data $1116$ MeV, $2286.46\pm0.14$ MeV and $5619.60\pm0.17$ MeV \cite{Zyla:2020zbs}. additionally, the estimated masses for $\Lambda_{c}(1/2^-)$ and $\Lambda_{c}^{*}(1/2^+)$ are 2615 MeV and 2765 MeV, respectively, which can also reproduce the experimental data $2592.25\pm0.28$ and $2766.6\pm2.4$ MeV \cite{Zyla:2020zbs}. For the $(uds)$-type $\Lambda^{*}$ resonances the estimations for $\Lambda(1/2^-)$ and $\Lambda^{*}(1/2^+)$ are 1517 MeV and 1679 MeV, respectively, which is markedly different compared to $\Lambda(1405)$'s $1405.1^{+1.3}_{-1.0}$ Mev and $\Lambda(1600)$'s $(1570\sim1630)$ Mev \cite{Zyla:2020zbs}. It is worth mentioning mention that there is only one channel \cite{Yoshida:2015tia,Yang:2019lsg} used in this work. The big differences indicate more complicated structures for light baryons. For $\Lambda(1405)$, our estimation is roughly consistent with the constituent quark model \cite{Pervin:2005ve,Yoshida:2015tia} although its structure is still a mystery  and needs to be understood. More discussions on this resonance can be found in Refs. \cite{Hall:2014uca,Liu:2016wxq,Azizi:2017xyx,Mai:2020ltx} and the references therein.} and spatial wave functions of baryons. Additionally, in our calculation the values of $\rho_{min}$ and $\rho_{max}$ are chosen as $0.2$~fm and $2.0$~fm, respectively, and $n_{\rho_{max}}=6$. For $\lambda$-mode, we also use the same Gaussian sized parameters. In Table \ref{Wave functions}, we collect the numerical spatial wave functions of bottom and charmed baryons, and the ground and excited states of $\Lambda$ involved in these discussed semileptonic decays. It should be noted that the spatial wave functions obtained in this work are based on a systematic analysis of all the related $\Lambda_b$, $\Lambda_c^{(*)}$ and $\Lambda^{(*)}$ baryons.

\section{numerical results}
\label{sec3}

\subsection{The calculated form factors}
\label{Numerical results of the form factors}

\begin{figure*}
  \centering
  \begin{tabular}{ccc}
  \includegraphics[width=56mm]{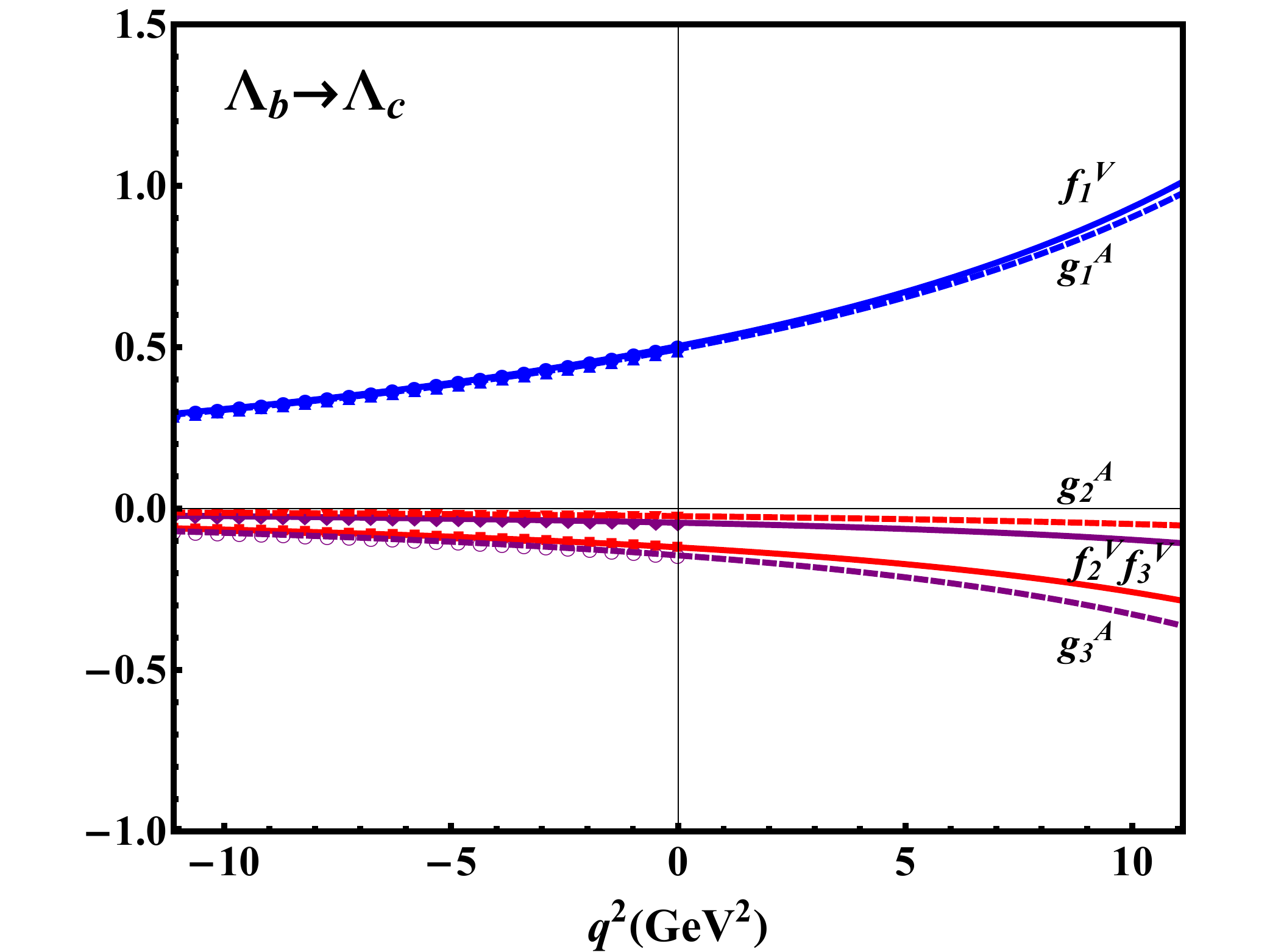}
  \includegraphics[width=56mm]{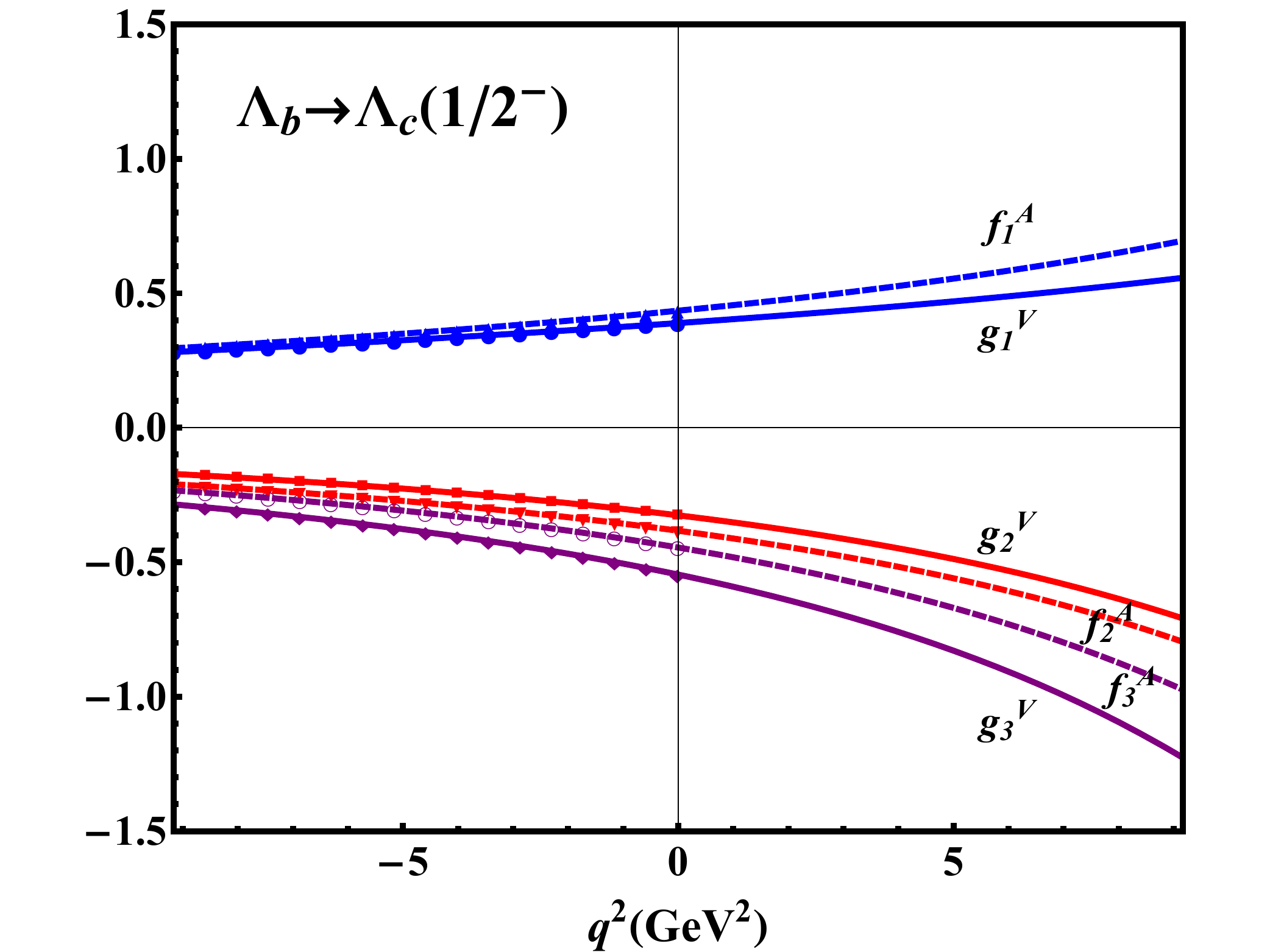}
  \includegraphics[width=56mm]{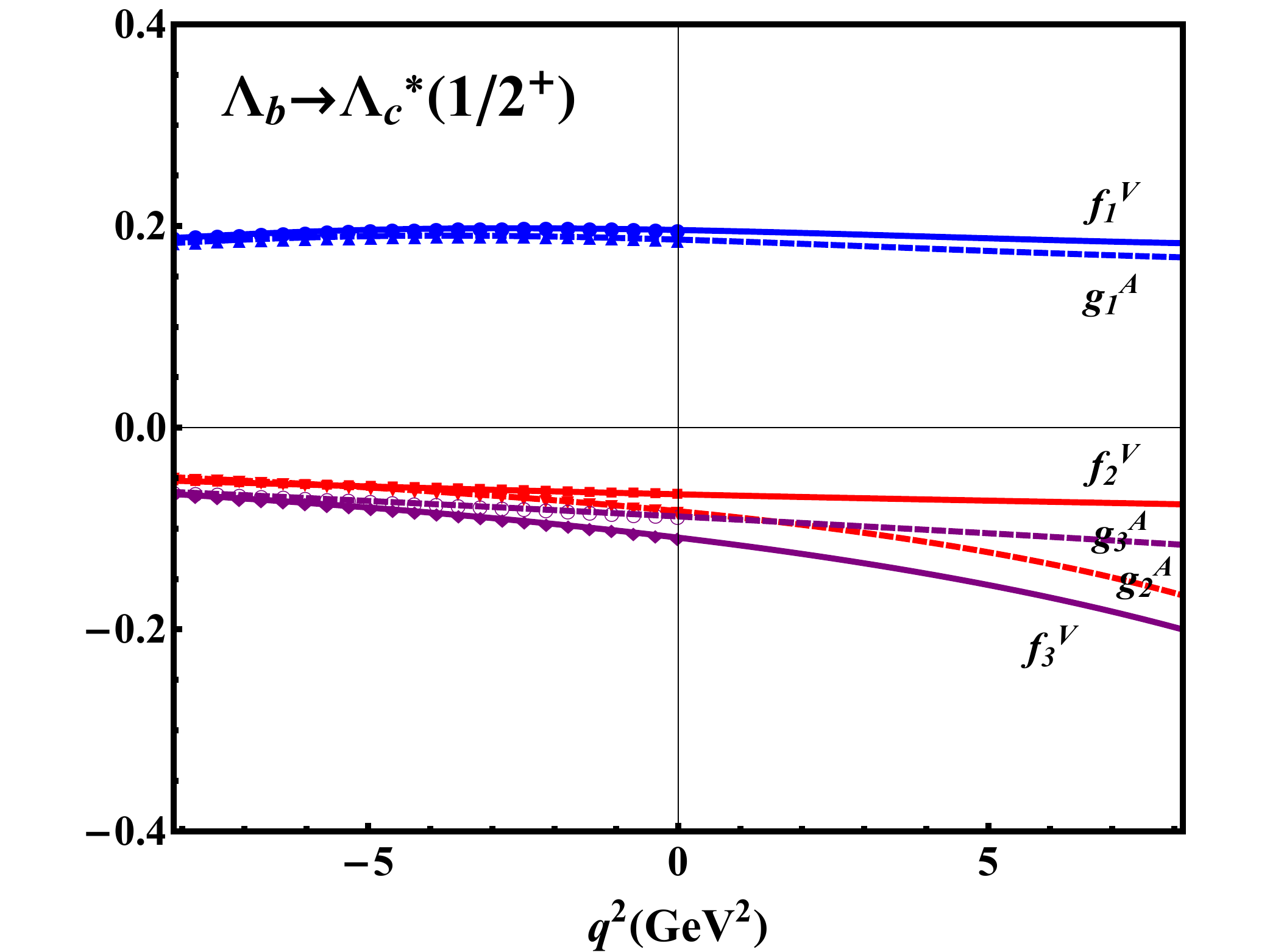}\\
  \includegraphics[width=56mm]{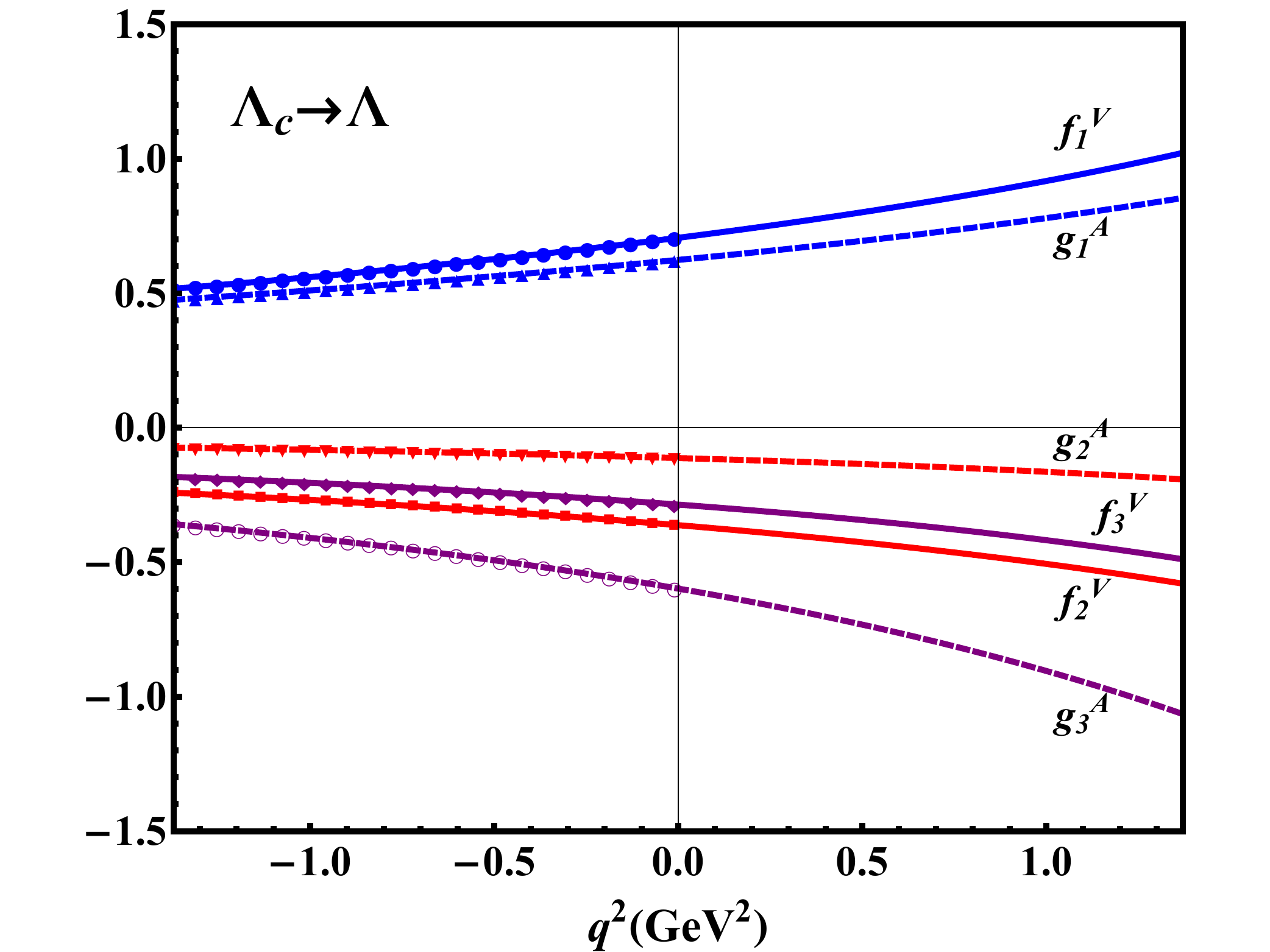}
  \includegraphics[width=56mm]{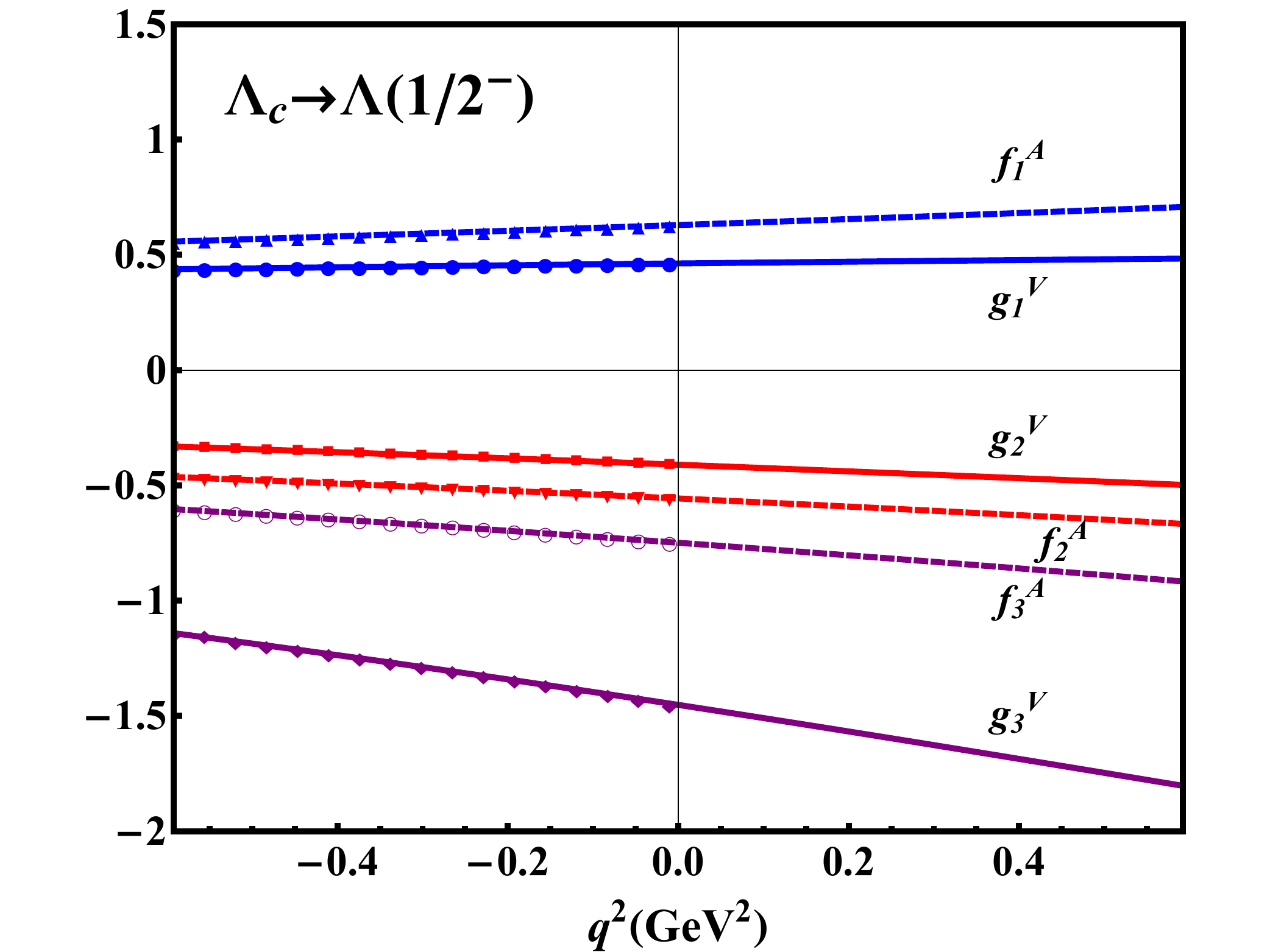}
  \includegraphics[width=56mm]{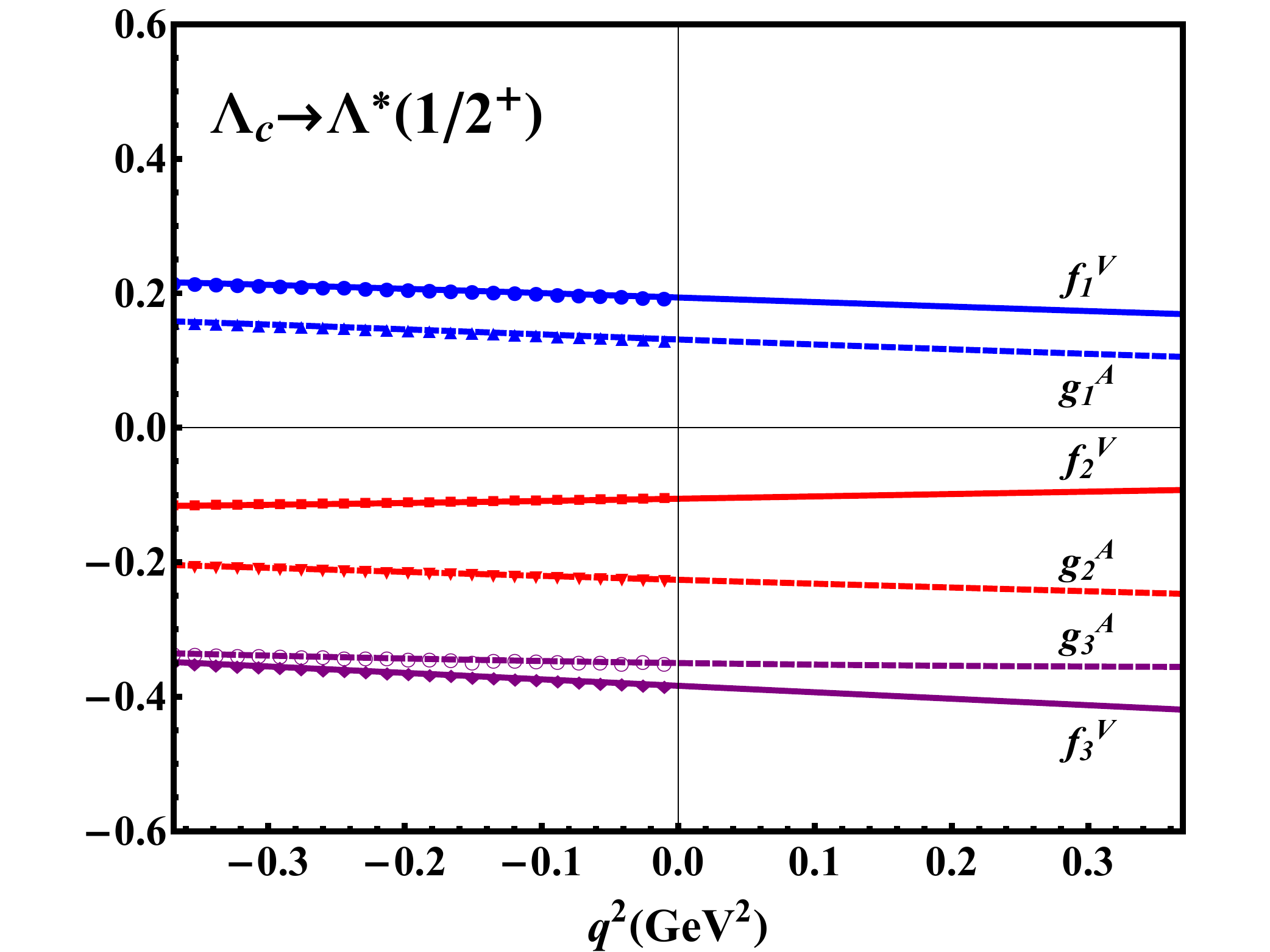}\\
  \end{tabular}
  \caption{The $q^2$ dependence of form factors of $\Lambda_b\to\Lambda_c^{(*)}(1/2^{\pm})$ and $\Lambda_c\to\Lambda^{(*)}(1/2^{\pm})$ transitions, in which the solid and dashed lines represent the vector or pseudoscalar-types form factors respectively, while the blue, red and purple lines (both solid and dashed) represent the $i$th form factors denoted by the subscripts respectively for each types. {In addition, the symbols {\it e.g.}, point, triangle and diamond in $q^2<0$ regions denote the fitted points for each of the form factors.}}
\label{formfactor}
\end{figure*}

In this subsection, we present our results for the form factors of $\Lambda_{b}\to\Lambda_{c}^{(*)}(1/2^{\pm})$ and $\Lambda_{c}\to\Lambda^{(*)}(1/2^{\pm})$ transitions. The mass of baryons we used in our calculation are cited from Ref. \cite{Zyla:2020zbs}, and the spatial wave functions are obtained from GEM. The form factors obtained in Sec. II are calculated out in a spacelike region, so we need to expand them to a timelike region. We would use the dipole parametrization to obtain our form factors in physical region. The concrete expression reads as \cite{Chua:2019yqh,Chua:2018lfa}
\begin{equation}
F(q^2)=\frac{F(0)}{\left(1-q^2/M^2\right)\left(1-b_{1}q^2/M^2+b_{2}(q^2/M^2)^2\right)},
\label{Forms}
\end{equation}
{where $F(0)$ is the form factor at $q^2=0$, $b_1$, and $b_2$ are parameters needing to be fitted.}
{We numerically compute 24 points for each form factor from $q^2=-q^2_{\text{max}}$ to $q^2=0$ and fit them with MINUIT program. The fitting of form factor is very good, which can be seen in Fig. \ref{formfactor}. Since it is difficult to estimate the uncertainty from the model itself, we roughly take an error of ten percent for the fitting points.}

With the spatial wave functions for $\Lambda_Q^{(*)}$, we can obtain the form factors numerically in the framework of the light-front quark model. In this way, we avoid the effective $\beta$ parameters \cite{Ke:2019smy,Chua:2019yqh,Zhu:2018jet,Ke:2007tg,Chen:2017vgi,Xu:2014mqa}, and all the free parameters can be fixed by fitting the mass spectra of the baryons. The parameters for the extended form factors are collected in Table \ref{b2cformfactor} and \ref{c2sformfactor} for $\Lambda_{b}\to\Lambda_{c}^{(*)}(1/2^{\pm})$ and $\Lambda_{c}\to\Lambda^{(*)}(1/2^{\pm})$ transitions, respectively.

\begin{table}
\centering \caption{The form factors for $\Lambda_{b}\to\Lambda_{c}^{(*)}$ in this work. We use a three-parameter form as illustrated in Eq. (\ref{Forms}) for these form factors.}
\label{b2cformfactor}
\renewcommand\arraystretch{1.05}
\begin{tabular*}{86mm}{c@{\extracolsep{\fill}}ccccc}
\toprule[1pt]
\toprule[0.7pt]
        &$F(0)$   &$F(q^2_{max})$   &$b_1$   &$b_2$  \\
\hline
&\multicolumn{4}{c}{$\Lambda_b\to\Lambda_c\left(\frac{1}{2}^+\right)$}  \\
\hline
$f^V_1$   &$0.50\pm0.05$      &$1.01\pm0.10$     &$0.71\pm0.01$     &$0.14\pm0.05$\\
$f^V_2$   &$-0.12\pm0.01$     &$-0.29\pm0.03$    &$1.14\pm0.02$     &$0.40\pm0.06$\\
$f^V_3$   &$-0.04\pm0.00$     &$-0.11\pm0.01$    &$1.16\pm0.02$     &$0.32\pm0.06$\\
$g^A_1$   &$0.49\pm0.05$      &$0.98\pm0.10$     &$0.68\pm0.01$     &$0.14\pm0.05$\\
$g^A_2$   &$-0.02\pm0.00$     &$-0.05\pm0.00$    &$0.98\pm0.02$     &$0.34\pm0.06$\\
$g^A_3$   &$-0.15\pm0.02$     &$-0.37\pm0.04$    &$1.29\pm0.02$     &$0.56\pm0.06$\\
\toprule[0.7pt]
&\multicolumn{4}{c}{$\Lambda_b\to\Lambda_c\left(\frac{1}{2}^-\right)$}\\
\toprule[0.7pt]
$g^V_1$   &$0.39\pm0.04$       &$0.56\pm0.06$       &$0.15\pm0.01$     &$0.33\pm0.06$\\
$g^V_2$   &$-0.33\pm0.03$      &$-0.71\pm0.07$      &$1.42\pm0.02$     &$0.69\pm0.07$\\
$g^V_3$   &$-0.55\pm0.06$      &$-1.23\pm0.12$      &$1.48\pm0.02$     &$0.66\pm0.07$\\
$f^A_1$   &$0.44\pm0.04$       &$0.70\pm0.07$       &$0.45\pm0.01$     &$0.14\pm0.06$\\
$f^A_2$   &$-0.38\pm0.04$      &$-0.80\pm0.08$      &$1.27\pm0.02$     &$0.53\pm0.07$\\
$f^A_3$   &$-0.45\pm0.05$      &$-0.98\pm0.10$      &$1.45\pm0.02$     &$0.74\pm0.07$\\
\toprule[0.7pt]
&\multicolumn{4}{c}{$\Lambda_b\to\Lambda_c^{*}\left(\frac{1}{2}^+\right)$}\\
\toprule[0.7pt]
$f^V_1$   &$0.20\pm0.02$      &$0.18\pm0.02$     &$-1.19\pm0.02$    &$2.05\pm0.07$\\
$f^V_2$   &$-0.07\pm0.01$     &$-0.08\pm0.01$    &$-0.33\pm0.02$    &$1.26\pm0.08$\\
$f^V_3$   &$-0.11\pm0.01$     &$-0.20\pm0.02$    &$1.13\pm0.02$     &$0.40\pm0.10$\\
$g^A_1$   &$0.19\pm0.02$      &$0.17\pm0.02$     &$-1.31\pm0.02$    &$2.24\pm0.07$\\
$g^A_2$   &$-0.08\pm0.01$     &$-0.17\pm0.02$    &$1.29\pm0.01$     &$0.12\pm0.07$\\
$g^A_3$   &$-0.09\pm0.01$     &$-0.12\pm0.01$    &$0.15\pm0.01$     &$0.89\pm0.06$\\
\bottomrule[0.7pt]
\bottomrule[1pt]
\end{tabular*}
\end{table}

\begin{table}
\centering \caption{The form factors for $\Lambda_{c}\to\Lambda^{(*)}$ in this work. We use a three-parameter form as illustrated in Eq. (\ref{Forms}) for these form factors.}
\label{c2sformfactor}
\renewcommand\arraystretch{1.05}
\begin{tabular*}{86mm}{c@{\extracolsep{\fill}}cccc}
\toprule[1pt]
\toprule[0.7pt]
        &$F(0)$   &$F(q^2_{max})$   &$b_1$   &$b_2$   \\
\toprule[0.7pt]
&\multicolumn{4}{c}{$\Lambda_c\to\Lambda\left(\frac{1}{2}^+\right)$}    \\
\toprule[0.7pt]
$f^V_1$ &$0.71\pm0.07$       &$1.02\pm0.10$      &$0.28\pm0.01$    &$0.13\pm0.06$\\
$f^V_2$ &$-0.36\pm0.04$      &$-0.58\pm0.06$     &$0.65\pm0.01$    &$0.23\pm0.06$\\
$f^V_3$ &$-0.29\pm0.03$      &$-0.49\pm0.05$     &$0.86\pm0.01$    &$0.21\pm0.07$\\
$g^A_1$ &$0.62\pm0.06$       &$0.86\pm0.09$      &$0.10\pm0.01$    &$0.21\pm0.06$\\
$g^A_2$ &$-0.11\pm0.01$      &$-0.19\pm0.02$     &$0.79\pm0.01$    &$0.05\pm0.06$\\
$g^A_3$ &$-0.60\pm0.06$      &$-1.07\pm0.11$     &$1.08\pm0.01$    &$0.56\pm0.07$\\
\toprule[0.7pt]
&\multicolumn{4}{c}{$\Lambda_c\to\Lambda\left(\frac{1}{2}^-\right)$}\\
\toprule[0.7pt]
$g^V_1$ &$0.46\pm0.05$        &$0.49\pm0.05$      &$-0.56\pm0.03$     &$1.13\pm0.35$\\
$g^V_2$ &$-0.41\pm0.04$       &$-0.53\pm0.05$     &$0.81\pm0.03$      &$1.54\pm0.38$\\
$g^V_3$ &$-1.45\pm0.15$       &$-1.92\pm0.19$     &$1.04\pm0.04$      &$1.94\pm0.39$\\
$f^A_1$ &$0.63\pm0.06$        &$0.73\pm0.07$      &$0.05\pm0.03$      &$0.64\pm0.37$\\
$f^A_2$ &$-0.56\pm0.06$       &$-0.71\pm0.07$     &$0.61\pm0.03$      &$0.75\pm0.38$\\
$f^A_3$ &$-0.75\pm0.08$       &$-0.98\pm0.10$     &$0.86\pm0.03$      &$1.38\pm0.38$\\
\toprule[0.7pt]
&\multicolumn{4}{c}{$\Lambda_c\to\Lambda^{*}\left(\frac{1}{2}^+\right)$}\\
\toprule[0.7pt]
$f^V_1$ &$0.19\pm0.02$       &$0.17\pm0.02$      &$-2.81\pm0.04$     &$6.92\pm0.70$\\
$f^V_2$ &$-0.11\pm0.01$      &$-0.09\pm0.01$     &$-2.68\pm0.04$     &$7.53\pm0.71$\\
$f^V_3$ &$-0.38\pm0.04$      &$-0.42\pm0.04$     &$0.31\pm0.04$      &$1.33\pm0.80$\\
$g^A_1$ &$0.13\pm0.01$       &$0.11\pm0.01$      &$-3.99\pm0.04$     &$11.42\pm0.73$\\
$g^A_2$ &$-0.23\pm0.02$      &$-0.25\pm0.03$     &$0.35\pm0.05$      &$1.89\pm0.81$\\
$g^A_3$ &$-0.35\pm0.04$      &$-0.36\pm0.04$     &$-0.60\pm0.04$     &$3.27\pm0.77$\\
\bottomrule[0.7pt]
\bottomrule[1pt]
\end{tabular*}
\end{table}

The form factors for $\Lambda_{b}\to\Lambda_{c}(1/2^{+}, 2286)$ transition can be worked out with the help of Eq. (\ref{general expression of 05+}). The $q^2$ dependence of $f^V_{1,2,3}$ and $g^A_{1,2,3}$ for this transition are plotted in the first panel of Fig. \ref{formfactor}, and the fitting parameters in Eq. (\ref{Forms}) are listed in Table \ref{b2cformfactor}. In the heavy quark limit (HQL), the corresponding transition matrix element can be rewritten as \cite{Bernlochner:2018bfn,Bernlochner:2018kxh,Georgi:1990ei}
\begin{equation}
\left<\Lambda_c(\nu',s')|\bar{c}_{\nu'}\Gamma b_{\nu}|\Lambda_b(\nu,s)\right>=\zeta(\omega)\bar{u}(\nu',s')\Gamma u(\nu,s),
\label{HQET form factors}
\end{equation}
at the leading order in the heavy quark expansion, so that the form factors have more simple behaviors as
\begin{equation}
\begin{split}
&f^V_1(q^2)=g^A_1(q^2)=\zeta(\omega),\\
&f^V_2=f^V_3=g^A_2=g^A_3=0,
\end{split}
\end{equation}
where $\omega=\nu\cdot\nu'=(M^2+M'^2-q^2)/(2MM')$ while $\nu'=p'/M'$ and $\nu=p/M$ are the four-velocities for $\Lambda_c$ and $\Lambda_b$, respectively. $\zeta(\omega)$ is the so-called Isgur-Wise function (IWF) and usually expressed as a Taylor series expansion as \cite{Aaij:2017svr}
\begin{equation}
\zeta(\omega)=1-\rho^2(\omega-1)+\frac{\sigma^2}{2}(\omega-1)^2+\cdots,
\label{Isgur-Wise Function}
\end{equation}
where $\rho^2=-\frac{d\zeta(\omega)}{d\omega}|_{\omega=1}$ and $\sigma^2=\frac{d^2\zeta(\omega)}{d\omega^2}|_{\omega=1}$ are the two important parameters for depicting the IWF. The most obvious character is in the point $q^2=q_{max}^2$ (or $\omega=1$),
\begin{equation*}
f^V_1(q_{max}^2)=g^A_1(q_{max}^2)=\zeta(1)=1.
\end{equation*}
It provides one strong restriction for our results. Either $f^V_1(q_{max}^2)=1.010\pm0.100$ or $g^A_1(q_{max}^2)=0.980\pm0.100$ in our results satisfies the above restriction. Besides, we can easily see that $f^V_1$ and $g^A_1$ are very close to each other, and dominate over $f^V_{2,3}$ and $g^A_{2,3}$. These results satisfy the predictions of heavy quark effective theory. At the mean time, we obtain the $\rho^2$ and $\sigma^2$ in Eq. (\ref{Isgur-Wise Function}), by fitting $\zeta(\omega)$ from $f^V_1(q^2)$ and $g^A_1(q^2)$. The comparison with other theoretical predictions and the fitting of experimental data are listed in Table \ref{IWF}. The slope parameter $\rho^2$, either from fitting $f^V_1$ or $g^A_1$, is close to the results from DELPHI \cite{Abdallah:2003gn} and LHCb \cite{Aaij:2017svr}. The parameter $\sigma^2$ from $g^A_1(q^2)$ is slightly greater than the measurement of LHCb Collaboration, while the one from $f^V_1(q^2)$ is agrees well with the measurement.

\begin{table}
\centering \caption{Our results for the IWF's shape parameters of the $\Lambda_b\rightarrow\Lambda_c$ transition. The superscripts $[a]$ and $[b]$ in the second and third rows represent the fitting of $f^V_1$ and $g^A_1$ respectively. In Ref. \cite{Thakkar:2020vpv}, they list various theoretical predictions of $\rho^2$ (in Table 3) and we do not list those here.}
\label{IWF}
\renewcommand\arraystretch{1.05}
\begin{tabular*}{86mm}{c@{\extracolsep{\fill}}ccc}
\toprule[1pt]
\toprule[0.5pt]
                                     &$\rho^2$                  &$\sigma^2$\\
\toprule[0.5pt]
This work$^{[a]}$                    &$1.67\pm0.11$             &$2.45\pm0.63$\\
This work$^{[b]}$                    &$1.85\pm0.11$             &$3.25\pm0.61$\\
RQM \cite{Ebert:2006rp}              &$1.51$                    &$4.06$\\
QCDSR \cite{Huang:2005mea}           &$1.35\pm0.13$             &\\
LQCD \cite{Bowler:1997ej}            &$1.2^{+0.8}_{-1.1}$       &\\
DELPHI \cite{Abdallah:2003gn}        &$2.03\pm0.46^{+0.72}_{-1.00}$ &\\
LHCb \cite{Aaij:2017svr}             &$1.63\pm0.07$             &$2.16\pm0.34$\\
\bottomrule[0.5pt]
\bottomrule[1pt]
\end{tabular*}
\end{table}

\begin{table*}
\centering \caption{Theoretical predictions for the form factors $f_{1,2}^{V}(0)$ and $g_{1,2}^{A}(0)$ of $\Lambda_{b}\to\Lambda_{c}$ and $\Lambda_{c}\to\Lambda$ at $q^2=0$ using different approaches.}
\label{Comparationofformfactors}
\renewcommand\arraystretch{1.05}
\begin{tabular*}{165mm}{c@{\extracolsep{\fill}}ccccc}
\toprule[1pt]
\toprule[0.5pt]
$$     & $f^V_1(0)$   & $f^V_2(0)$      & $g^A_1(0)$    & $g^A_2(0)$  \\
\toprule[0.5pt]
$\Lambda_{b}\to\Lambda_{c}$\\
This work &$0.50\pm0.05$  &$-0.12\pm0.01$  &$0.49\pm0.05$  &$-0.02\pm0.00$ \\
LFQM \cite{Chua:2019yqh}& $0.474^{+0.069}_{-0.072}$ & $-0.109^{+0.019}_{-0.021}$  & $0.468^{+0.067}_{-0.007}$ & $0.051^{+0.008}_{-0.012}$ \\
LFQM \cite{Zhu:2018jet}& $0.500^{+0.028}_{-0.031}$ & $-0.098^{+0.005}_{-0.004}$  & $0.509^{+0.029}_{-0.031}$ & $-0.014^{+0.001}_{-0.001}$ \\
LFQM \cite{Ke:2019smy}& $0.488$ & $-0.180$  & $0.470$ & $-0.0479$\\
RQM \cite{Faustov:2016pal}& $0.526$ & $-0.137$  & $0.505$ & $0.027$\\
CCQM \cite{Gutsche:2015mxa}& $0.549$ & $-0.110$  & $0.542$ & $-0.018$\\
LQCD \cite{Detmold:2015aaa}& $0.418\pm0.161$ & $-0.100\pm0.055$  & $0.378\pm0.102$ & $-0.004\pm0.002$\\
\toprule[0.5pt]
$\Lambda_c\to\Lambda$\\
This work  &$0.71\pm0.07$  &$-0.36\pm0.04$  &$0.62\pm0.06$  &$-0.11\pm0.01$\\
LFQM \cite{Zhao:2018zcb}& $0.468$ & $-0.222$  & $0.407$ & $-0.035$ \\
\multirow{2}*{{LFQM} \cite{Geng:2020fng,Geng:2020gjh}} &\multirow{2}*{$0.67\pm0.01$} &\multirow{2}*{$-0.76\pm0.02$} &\multirow{2}*{$0.59\pm0.01$} &$(1.59\pm0.05)\times10^{-3}$ \cite{Geng:2020fng} \\
                                                                   & & & &$(3.8\pm1.2)\times10^{-3}$ \cite{Geng:2020gjh}\\
MBM \cite{Geng:2020fng}& $0.54$ & $-0.22$  & $0.52$ & $0.06$\\
CCQM \cite{Gutsche:2015rrt}& $0.511$ & $-0.289$  & $0.466$ & $0.025$\\
RQM \cite{Faustov:2016yza}& $0.700$ & $-0.295$  & $0.488$ & $0.135$\\
LCSR\cite{Liu:2009sn}& $0.517$ & $-0.123$  & $0.517$ & $0.123$\\
LCSR \cite{Huang:2006ny}& $0.449$ & $-0.193$  & $0.449$ & $0.193$\\
LQCD \cite{Meinel:2016dqj}& $0.647\pm0.122$ & $-0.310\pm0.075$  & $0.577\pm0.076$ & $0.001\pm0.048$\\
\toprule[0.5pt]
\toprule[1pt]
\end{tabular*}
\end{table*}

For $\Lambda_b\to\Lambda_c(1/2^{-},2595)$ transition, the $q^2$ dependence of $g^V_{1,2,3}$ and $f^A_{1,2,3}$ are plotted in the second panel of Fig. \ref{formfactor}. Similarly, the corresponding transition matrix element can be simplified and be rewritten as \cite{Chua:2019yqh}
\begin{equation}
\left<\Lambda_c^{*}(1/2^{-})(\nu')|\bar{c}_{\nu'}\Gamma b_{\nu}|\Lambda_b(\nu)\right>\\
=\frac{\sigma(\omega)}{\sqrt{3}}\bar{u}(\nu')\gamma_{5}(\slashed{\nu}+\nu\cdot \nu)\Gamma u(\nu),
\end{equation}
in HQL. At the mean time, the form factors can be simplified as
\begin{equation}
\begin{split}
g_1^V(q^2)=f_1^A(q^2)&=\left(\omega-\frac{M'}{M}\right)\frac{\sigma(\omega)}{\sqrt{3}},\\
g_2^V(q^2)=g_3^V(q^2)=&f_2^A(q^2)=f_3^A(q^2)=-\frac{\sigma(\omega)}{\sqrt{3}}.
\end{split}
\end{equation}
When comparing our results with the predictions of the heavy quark limit, we can draw the conclusion that our results match the main requirements of the above analysis with: 1. $g^V_1$ is close to $f^A_1$, while $g^V_{2,3}$ and $f^A_{2,3}$ are close to each others. 2. $f^{A}_{1}/f^{A}_{2,3}$ and $f^{A}_{1}/f^{A}_{2,3}$ roughly close to $(M'-M)/M$ at $q^2=q_{max}^{2}$.

For the $\Lambda_b\to\Lambda_c^{*}(1/2^{+}, 2765)$ transition, the $q^2$ dependence of $f^V_{1,2,3}$ and $g^A_{1,2,3}$ are plotted in the third panel of Fig. \ref{formfactor}. The HQL expects $f_1^{V}=g_1^{A}=0$ at $q^2=q^2_{max}$, because the wave functions of the low-lying $\Lambda_b$ and the radial excited state $\Lambda_c^{*}(2765)$ are orthogonal \cite{Chua:2019yqh}. Evidently our results embody this character according to Fig. \ref{formfactor} and Table \ref{b2cformfactor}.

For $\Lambda_c\to\Lambda(1/2^{+}, 1116)$ form factors, The $q^2$ dependence of $f^V_{1,2,3}$ and $g^A_{1,2,3}$ are plotted in the fourth panel of Fig. \ref{formfactor}. One interesting prediction from heavy quark symmetry at $q^{2}=0$ is that the daughter $\Lambda$ baryon is predicted to emerge $100\%$ polarized, with the decay asymmetry parameter $\alpha_{q^2=0}=-1$. Our result of $-0.99$ is very close to the prediction. Besides, the mean value of the polarization $\langle\alpha\rangle$ has been measured as $-0.86\pm0.03\pm0.02$ by CLEO with high precision \cite{Hinson:2004pj}. Our result of $-0.87$ is very consistent with that  experiment.

As the form factors depict $\Lambda_c\to\Lambda(1/2^{-}, 1P)$ and $\Lambda_c\to\Lambda^{*}(1/2^{+}, 2S)$ transitions, we also plot the $q^2$ dependence of $f_{1,2,3}$ and $g_{1,2,3}$ in the fifth and sixth panels of Fig. \ref{formfactor}. Presently, we do not have enough information to test these channels, we expect more experiments and more theories (especially the LQCD) may bring us the crucial breakthroughs in the future.

Lastly, we list our results for form factors of the $\Lambda_{b}\to\Lambda_{c}$ and $\Lambda_{c}\to\Lambda$ transitions at $q^2=0$ in Table \ref{Comparationofformfactors} and compare them with other approaches. {The calculations in Refs. \cite{Zhu:2018jet,Ke:2019smy,Chua:2019yqh,Geng:2020fng,Geng:2020gjh} were based on the light-front quark model. Reference. \cite{Chua:2019yqh} worked in the diquark picture while Ref. \cite{Ke:2019smy,Geng:2020fng,Geng:2020gjh} worked in the triquark picture. In addition, the MBM was also used in Ref. \cite{Geng:2020fng}, in which the inputs are based on the baryon spectroscopy.} Refs. \cite{Zhu:2018jet} used covariant light-front quark model. A covariant consistent quark model was used in Ref. \cite{Gutsche:2015mxa} and a RQM was used in Ref. \cite{Faustov:2016pal}. (The result from LQCD \cite{Meinel:2016dqj,Detmold:2015aaa} are also listed here.) {It can be seen from Table \ref{Comparationofformfactors} that our results are consistent with most of the other theoretical predictions including  LQCD.}
These form factors are extremely important to help us to understand the weak decays and search for the undiscovered semileptonic channels. We expect more theoretical and experimental data for testing these results ulteriorly.

These transition form factors would be useful as inputs not only in the studies of semileptonic decays as to be shown in the following subsection, but also in the nonleptonic decays of $\Lambda_b$ and $\Lambda_c$ \cite{Lu:2016ogy,Geng:2017esc,Geng:2017mxn,Wang:2017gxe,Geng:2018plk,Cheng:2018hwl,Jiang:2018iqa,Geng:2018bow,Geng:2018upx,Zhao:2018mov,Jia:2019zxi,Zou:2019kzq,Niu:2020gjw,Hu:2020nkg,Meng:2020euv}.

\subsection{Numerical results of the semileptonic decays observables}
\label{Numerical results of the semileptonic decays}

\begin{table*}
\centering
\caption{Our results for the absolute branching fractions $\mathcal{B}(\Lambda_{b}\to\Lambda_{c}^{(*)}\ell^{-}\nu_{\ell})$ with $\ell^{-}=e^{-}, \mu^{-}, \tau^{-}$ and $\mathcal{B}(\Lambda_{c}\to\Lambda^{(*)}\ell^{+}\nu_{\ell})$ with $\ell^{+}=e^{+}, \mu^{+}$. The numbers out of the brackets are our predictions, while the ones in the brackets are the referential numbers with the superscript $\text{[exp]}$ and $\text{[LQCD]}$ denoting the data which comes from experiments \cite{Zyla:2020zbs, Ablikim:2015prg, Ablikim:2016vqd, Hinson:2004pj} or LQCD \cite{Meinel:2016dqj, Detmold:2015aaa}, respectively. All the values are given as a percent ($\%$).}
\label{branchingfractions}
\renewcommand\arraystretch{1.25}
\begin{tabular*}{165mm}{c@{\extracolsep{\fill}}ccccccc}
\toprule[1pt]
\toprule[0.5pt]
&\multicolumn{3}{c}{$\Lambda_b\to\Lambda_c^{(*)}\ell\nu_\ell$} &\multicolumn{3}{c}{$\Lambda_c\to\Lambda^{(*)}\ell\nu_\ell$}  \\
\toprule[0.5pt]
               &$\Lambda_c\left(\frac{1}{2}^{+}\right)$       &$\Lambda_c\left(\frac{1}{2}^{-}\right)$   &$\Lambda_c^{*}\left(\frac{1}{2}^{+}\right)$   &$\Lambda\left(\frac{1}{2}^{+}\right)$         &$\Lambda\left(\frac{1}{2}^{-}\right)$     &$\Lambda^{*}\left(\frac{1}{2}^{+}\right)$\\
\toprule[0.5pt]
\multirow{2}*{\shortstack{$\ell=e$}}       &$6.47\pm0.96$ (${6.2^{+1.4}_{-1.3}}^{\text{[exp]}},$          &$1.73\pm0.59$      &$0.21\pm0.05$
                                           &$4.04\pm0.75$ (${3.63\pm0.43}^{\text{[exp]}},$     &$0.31\pm0.08$      &$0.007\pm0.002$ \\
                                           &${5.49\pm0.34}^{\text{[LQCD]}}$)                   &          &
                                           &$3.80\pm0.22^{\text{[LQCD]}}$)                     &          &\\
\bottomrule[0.5pt]
\multirow{2}*{\shortstack{$\ell=\mu$}}     &$6.45\pm0.95$ (${6.2^{+1.4}_{-1.3}}^{\text{[exp]}},$          &$1.72\pm0.58$      &$0.21\pm0.05$
                                           &$3.90\pm0.73$ (${3.49\pm0.46}^{\text{[exp]}},$     &$0.28\pm0.07$      &$0.005\pm0.002$\\
                                           &${5.49\pm0.34}^{\text{[LQCD]}}$)                   &(${0.79^{+0.40}_{-0.35}}^{\text{[exp]}}$)    & &$3.69\pm0.22^{\text{[LQCD]}}$)                     &          &\\
\bottomrule[0.5pt]
\multirow{1}*{\shortstack{$\ell=\tau$}}    &$1.97\pm0.29$ (${1.57\pm0.08}^{\text{[LQCD]}}$)               &$0.24\pm0.11$      &$0.02\pm0.005$  &-  &-  &-\\
\bottomrule[0.5pt]
\bottomrule[1pt]
\end{tabular*}
\end{table*}

\begin{figure*}
  \centering
  \begin{tabular}{ccc}
  \includegraphics[width=50mm]{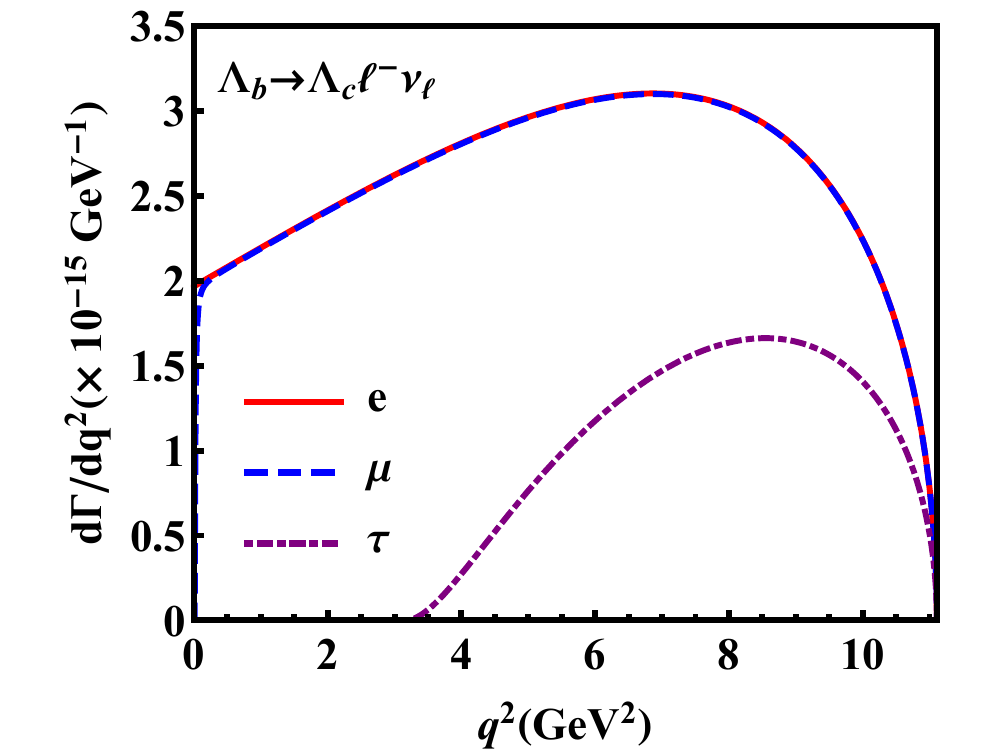}
  \includegraphics[width=50mm]{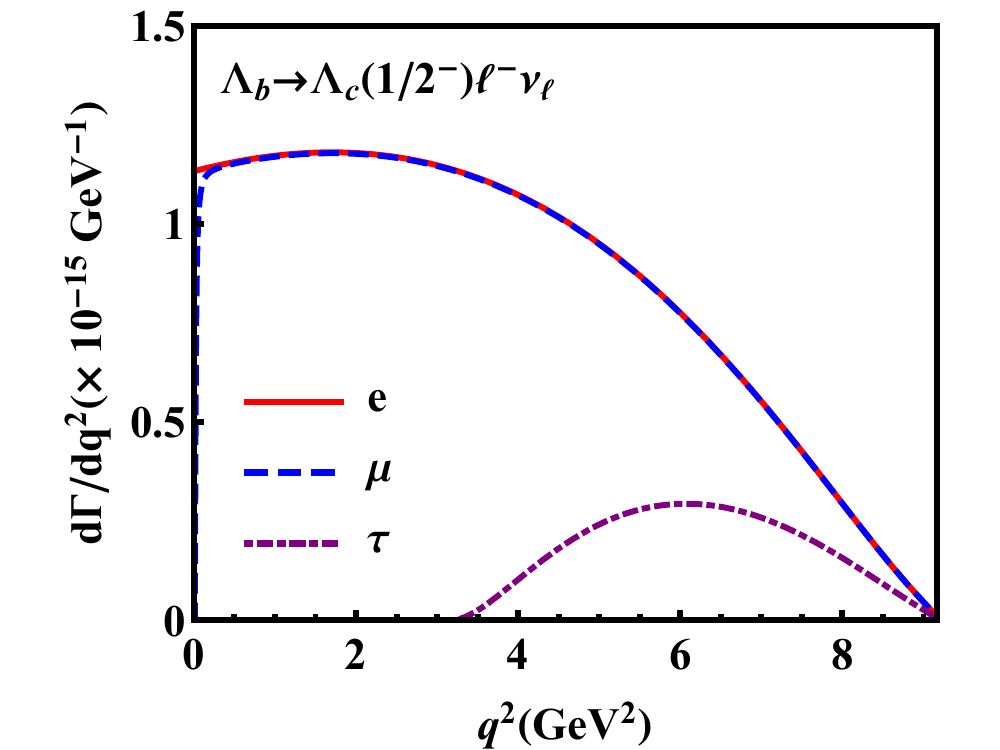}
  \includegraphics[width=50mm]{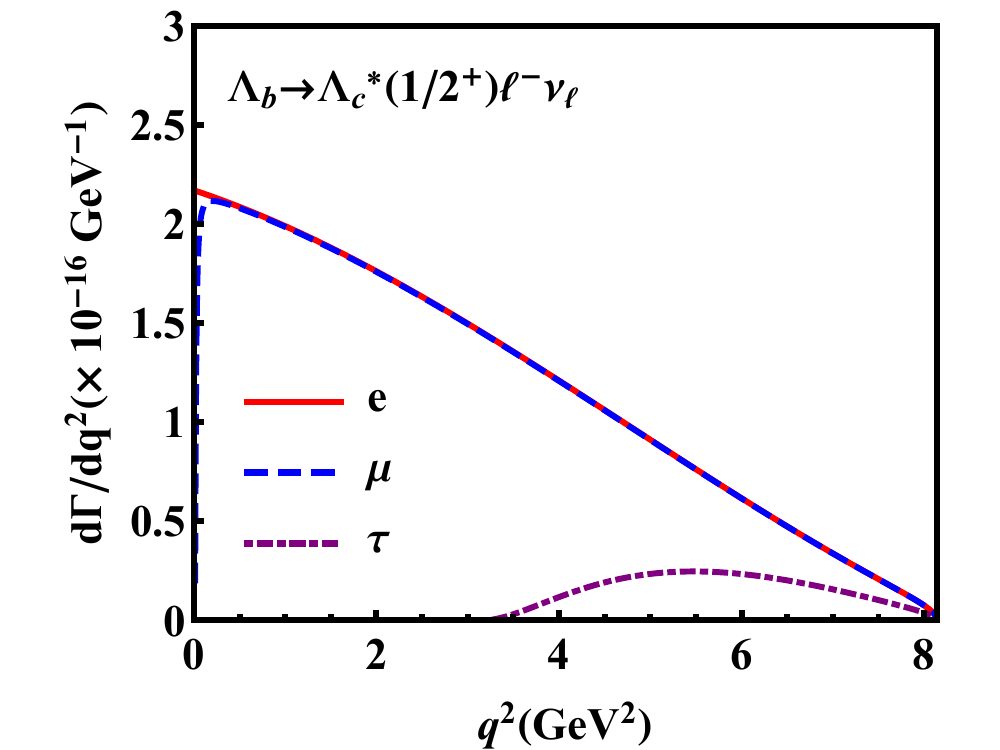}\\
  \includegraphics[width=50mm]{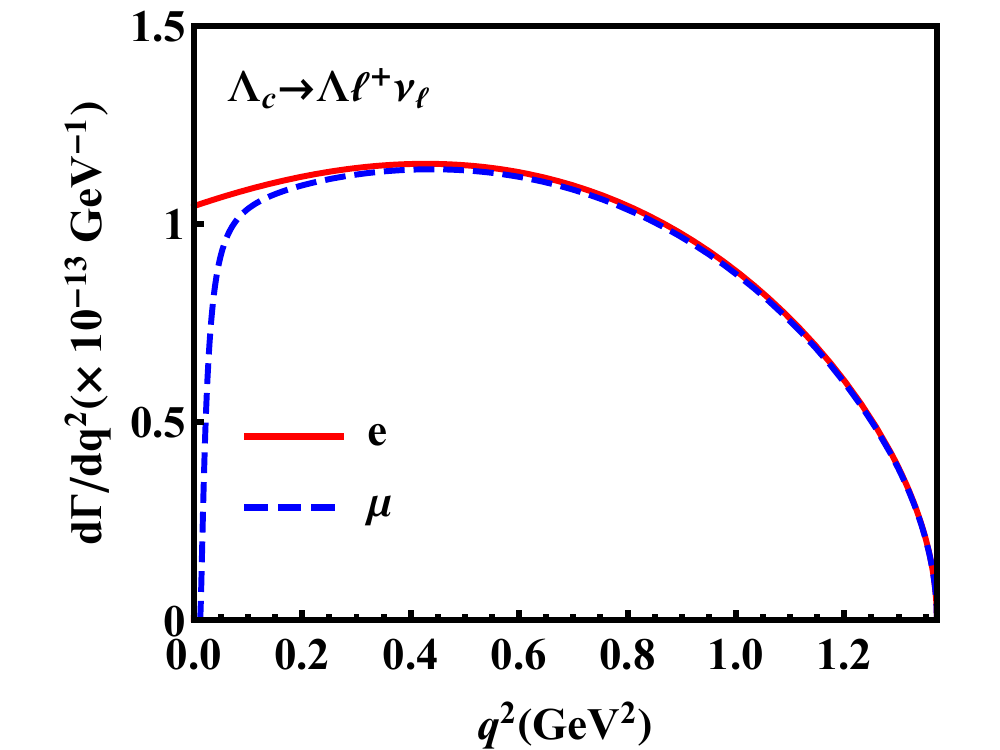}
  \includegraphics[width=50mm]{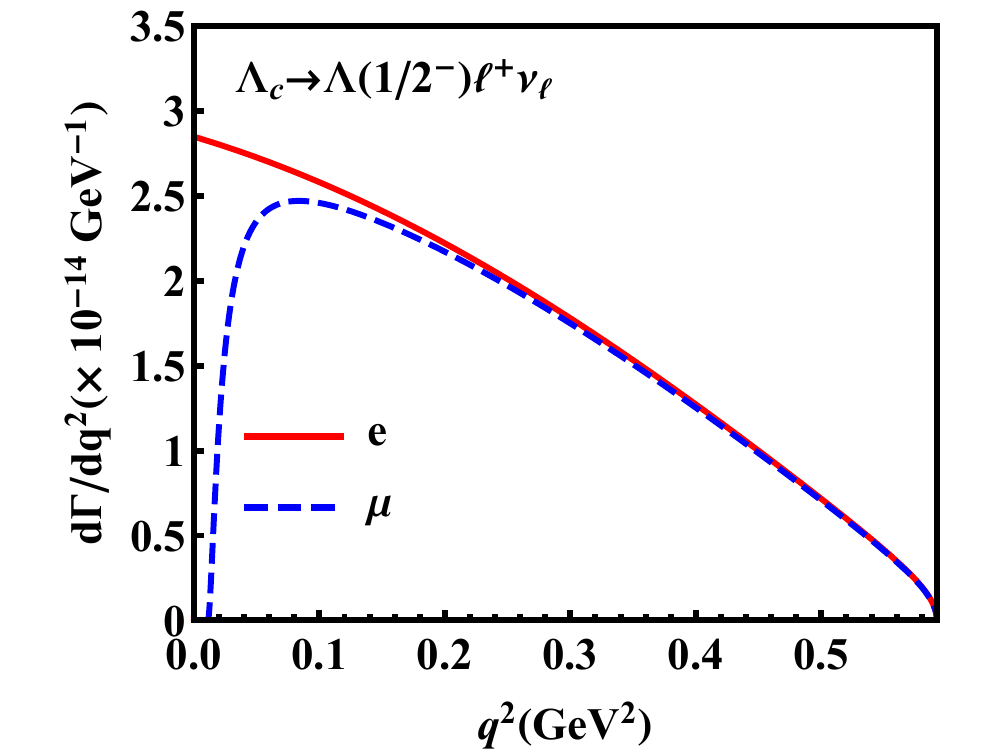}
  \includegraphics[width=50mm]{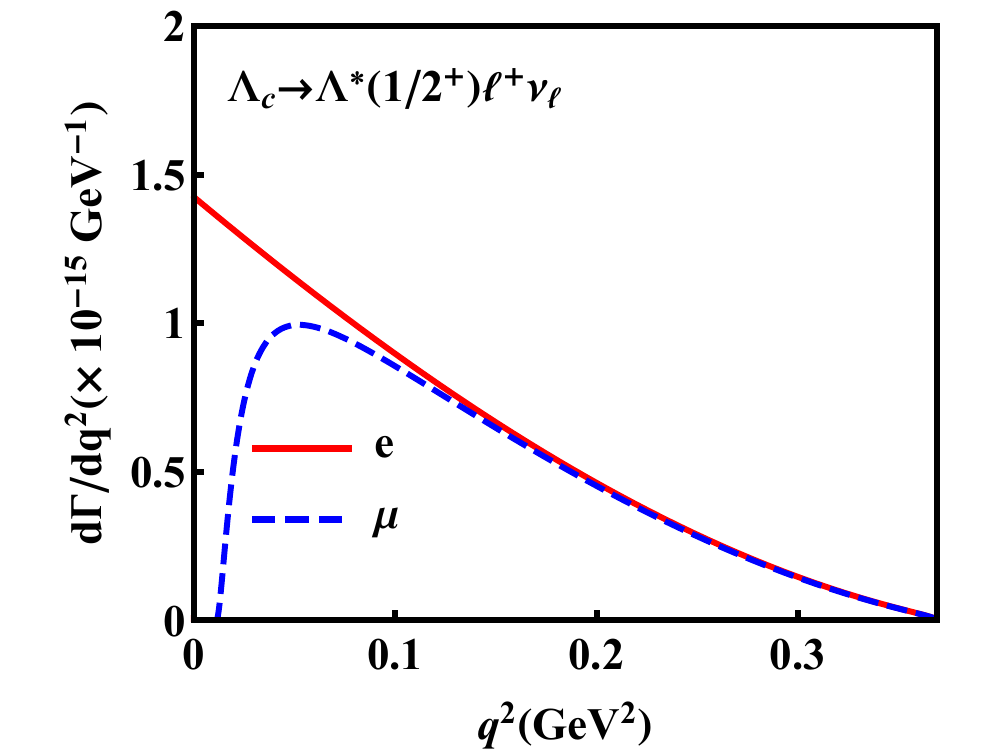}\\
  \end{tabular}
  \caption{The $q^2$ dependence of differential decay widths for $\Lambda_b\to\Lambda_c^{(*)}\ell^{-}\nu_{\ell}$ semileptonic decays with $\ell^{-}= e^{-}$ (red solid line), $\mu^{-}$ (blue dashed line), $\tau^{-}$ (purple dot dashed line) and for $\Lambda_c\to\Lambda^{(*)}\ell^{+}\nu_{\ell}$ semileptonic decays with $\ell^{+}= e^{+}$ (red solid line), $\mu^{+}$ (blue dashed line).}
\label{width}
\end{figure*}

\begin{table}
\centering
\caption{The predictions for averaged leptonic forward-backward asymmetry $\langle A_{FB}\rangle$, the averaged final hadron polarization $\langle P_{B}\rangle$ and the averaged lepton polarization $\langle P_{\ell}\rangle$. The values in the same column for $\Lambda_{b}\to\Lambda_{c}^{(*)}\ell^{-}\nu_{\ell}$ processes are the results for $\ell^{-}=e^-,\mu^-,\tau^-$ (top to bottom), while the ones for $\Lambda_{c}\to\Lambda^{(*)}\ell^{+}\nu_{\ell}$ processes are the results for $\ell^{+}=e^{+},\mu^{+}$ (top to bottom).}
\label{PhysicalObservables}
\renewcommand\arraystretch{1.05}
\begin{tabular*}{86mm}{c@{\extracolsep{\fill}}ccccc}
\toprule[1pt]
\toprule[0.5pt]
Channels   &$\ell=$  &$\langle A_{FB}\rangle$   &$\langle P_{B}\rangle$ &$\langle P_{\ell}\rangle$\\
\toprule[0.5pt]
\multirow{3}*{\shortstack{$\Lambda_{b}\to\Lambda_c(1/2^{+})\ell^{-}\nu_{\ell}$}}      &$e$     &$0.18\pm0.05$ &$-0.81\pm0.12$  &$-1.00\pm0.00$\\
                                                                                      &$\mu$   &$0.17\pm0.05$
&$-0.81\pm0.12$  &$-0.98\pm0.00$\\
                                                                                      &$\tau$  &$-0.08\pm0.03$
&$-0.77\pm0.11$  &$-0.24\pm0.03$\\
\bottomrule[0.5pt]
\multirow{3}*{\shortstack{$\Lambda_{b}\to\Lambda_c(1/2^{-})\ell^{-}\nu_{\ell}$}}      &$e$     &$0.23\pm0.04$  &$-0.95\pm0.08$  &$-1.00\pm0.00$\\
                                                                                      &$\mu$   &$0.22\pm0.04$
&$-0.95\pm0.08$  &$-0.98\pm0.00$\\
                                                                                      &$\tau$  &$-0.07\pm0.07$
&$-0.95\pm0.06$  &$-0.18\pm0.04$\\
\bottomrule[0.5pt]
\multirow{3}*{\shortstack{$\Lambda_{b}\to\Lambda_c^{*}(1/2^{+})\ell^{-}\nu_{\ell}$}}  &$e$     &$0.18\pm0.07$  &$-0.91\pm0.12$  &$-1.00\pm0.00$\\
                                                                                      &$\mu$   &$0.17\pm0.07$
&$-0.91\pm0.12$  &$-0.97\pm0.00$\\
                                                                                      &$\tau$  &$-0.13\pm0.05$
&$-0.88\pm0.13$  &$-0.13\pm0.04$\\
\bottomrule[0.5pt]
\multirow{2}*{\shortstack{$\Lambda_{c}\to\Lambda(1/2^{+})\ell^{+}\nu_{\ell}$}}        &$e$     &$0.20\pm0.05$  &$-0.87\pm0.09$  &$-1.00\pm0.00$\\
                                                                                      &$\mu$   &$0.16\pm0.04$
&$-0.87\pm0.09$  &$-0.89\pm0.04$\\
\bottomrule[0.5pt]
\multirow{2}*{\shortstack{$\Lambda_{c}\to\Lambda(1/2^{-})\ell^{+}\nu_{\ell}$}}        &$e$     &$0.18\pm0.22$  &$-0.91\pm0.05$  &$-1.00\pm0.00$\\
                                                                                      &$\mu$   &$0.10\pm0.13$
&$-0.91\pm0.05$  &$-0.80\pm0.07$\\
\bottomrule[0.5pt]
\multirow{2}*{\shortstack{$\Lambda_{c}\to\Lambda^{*}(1/2^{+})\ell^{+}\nu_{\ell}$}}    &$e$     &$0.11\pm0.002$     &$-0.90\pm0.06$  &$-1.00\pm0.00$\\
                                                                                      &$\mu$   &$-0.001\pm0.001$
&$-0.90\pm0.06$  &$-0.71\pm0.09$\\
\bottomrule[0.5pt]
\bottomrule[1pt]
\end{tabular*}
\end{table}

\begin{figure*}
  \centering
  \begin{tabular}{ccc}
  \includegraphics[width=50mm]{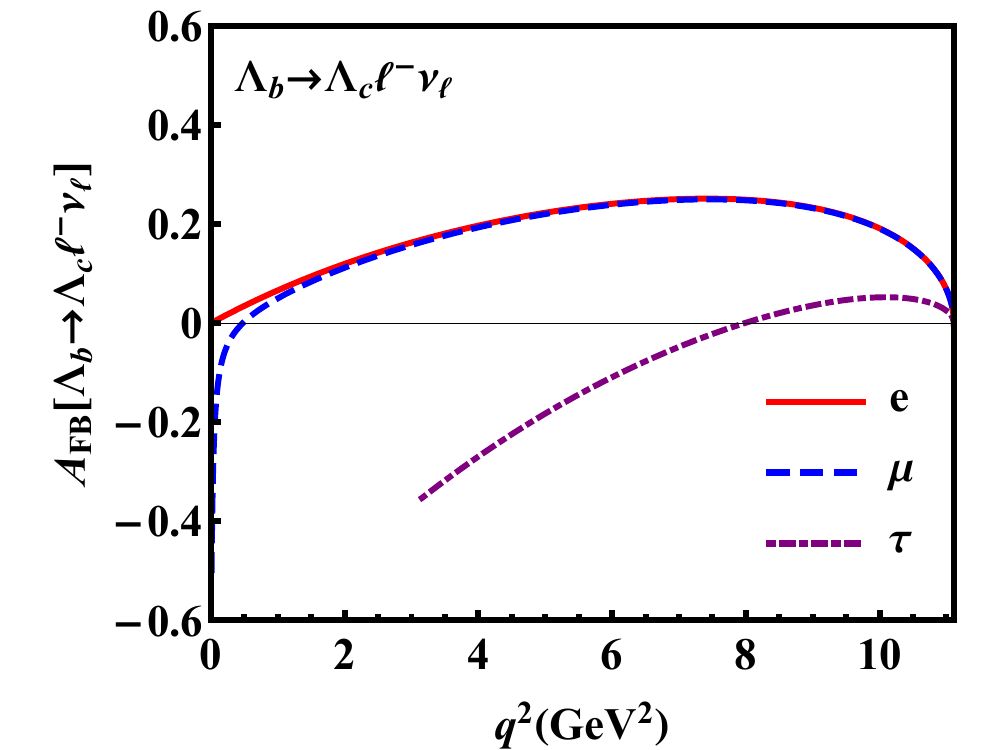}
  \includegraphics[width=50mm]{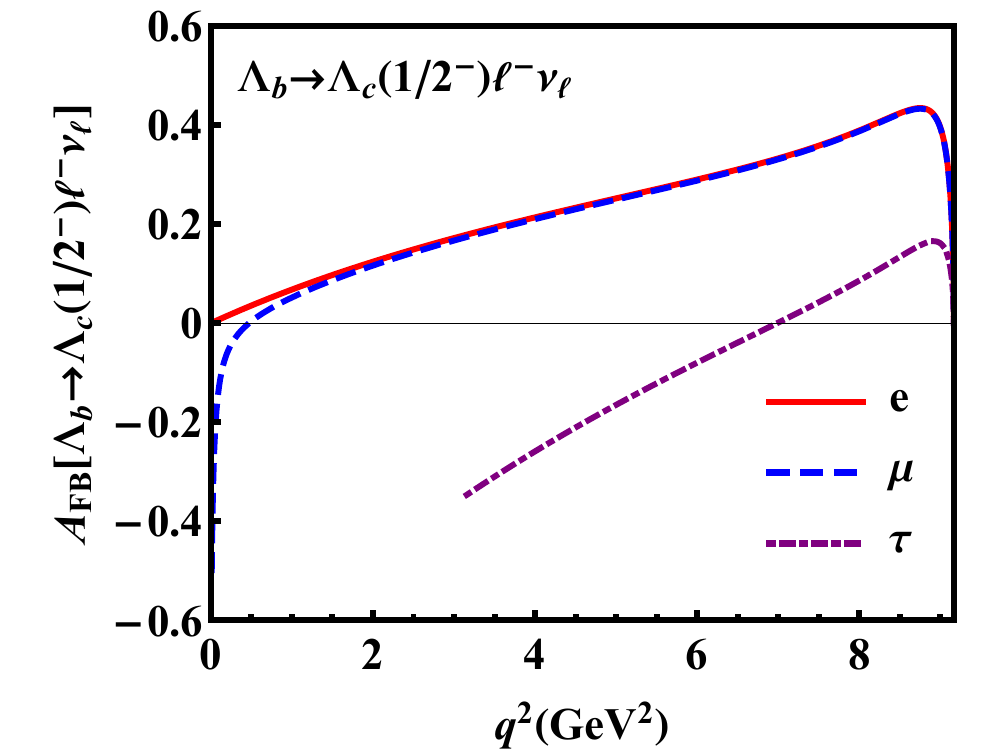}
  \includegraphics[width=50mm]{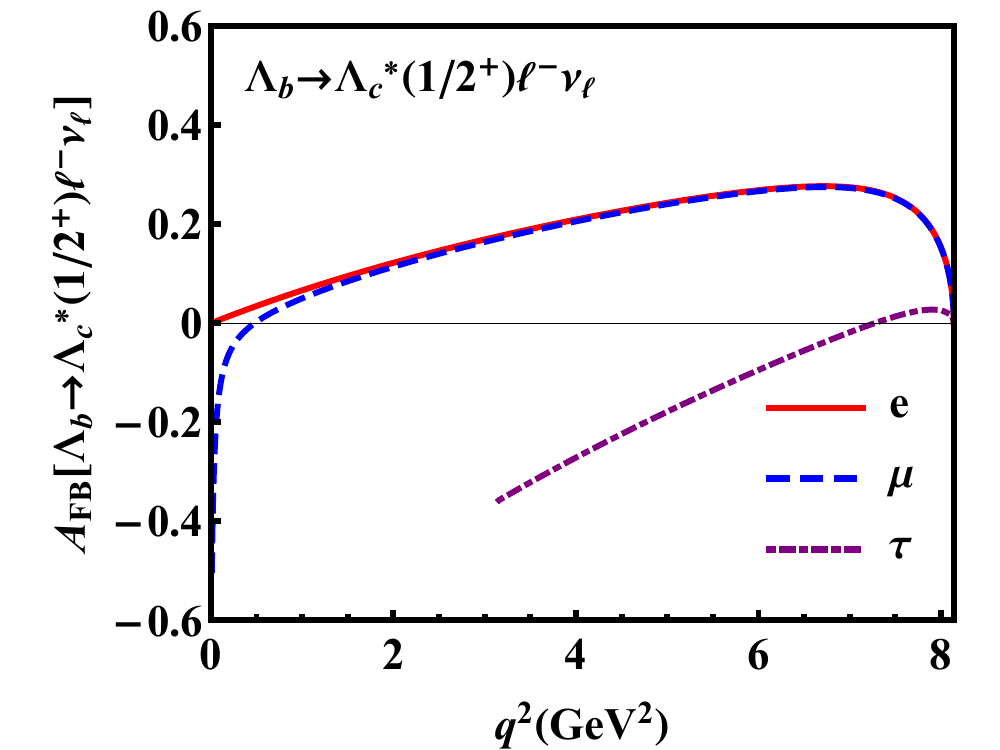}\\
  \includegraphics[width=50mm]{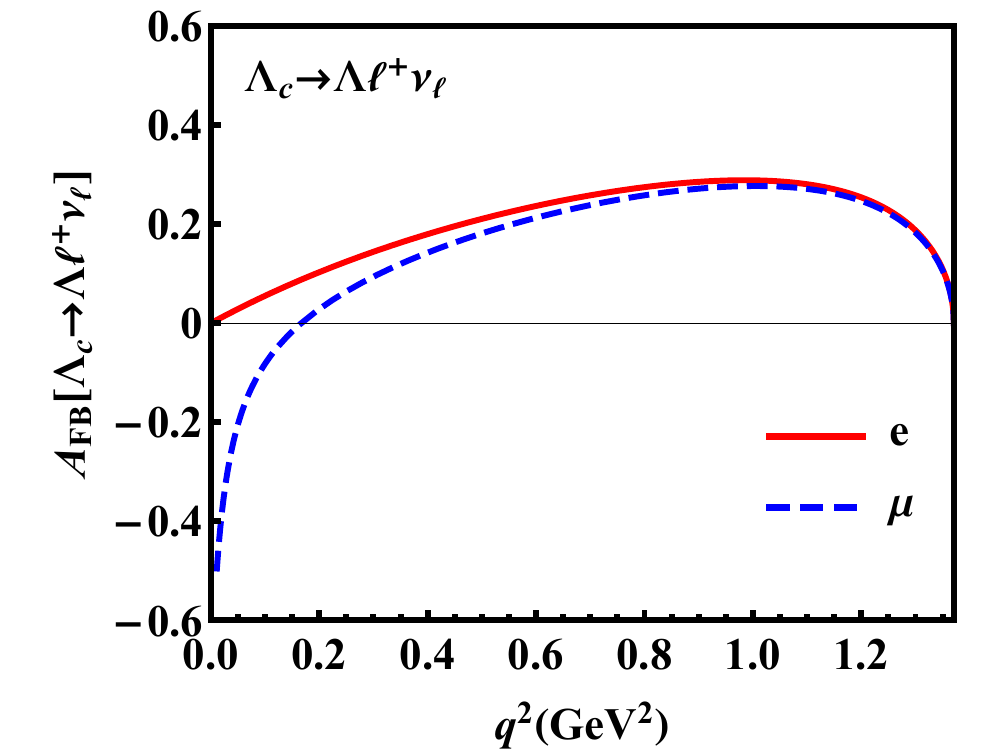}
  \includegraphics[width=50mm]{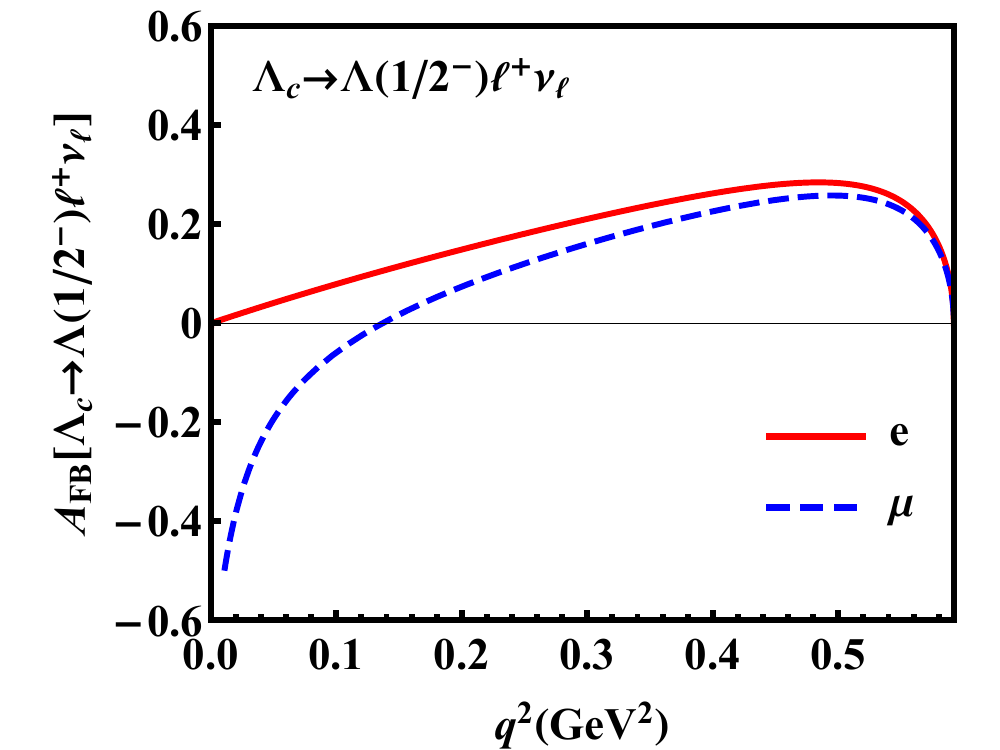}
  \includegraphics[width=50mm]{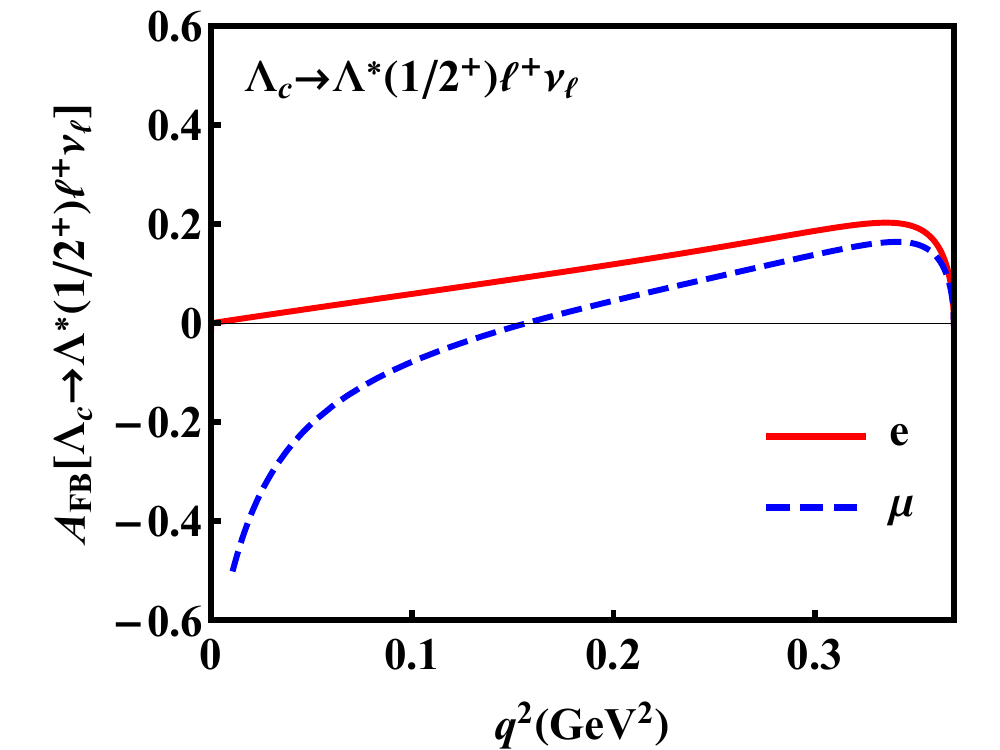}\\
  \end{tabular}
  \caption{The $q^2$ dependence of lepton forward-backward asymmetry ($A_{FB}$) for $\Lambda_b\to\Lambda_c^{(*)}\ell^{-}\nu_{\ell}$ semileptonic decays with $\ell^{-}= e^{-}$ (red solid line), $\mu^{-}$ (blue dashed line), $\tau^{-}$ (purple dot dashed line) and for $\Lambda_c\to\Lambda^{(*)}\ell^{+}\nu_{\ell}$ semileptonic decays with $\ell^{+}= e^{+}$ (red solid line), $\mu^{+}$ (blue dashed line).}
\label{AFB}
\end{figure*}

\begin{figure*}
  \centering
  \begin{tabular}{ccc}
  \includegraphics[width=50mm]{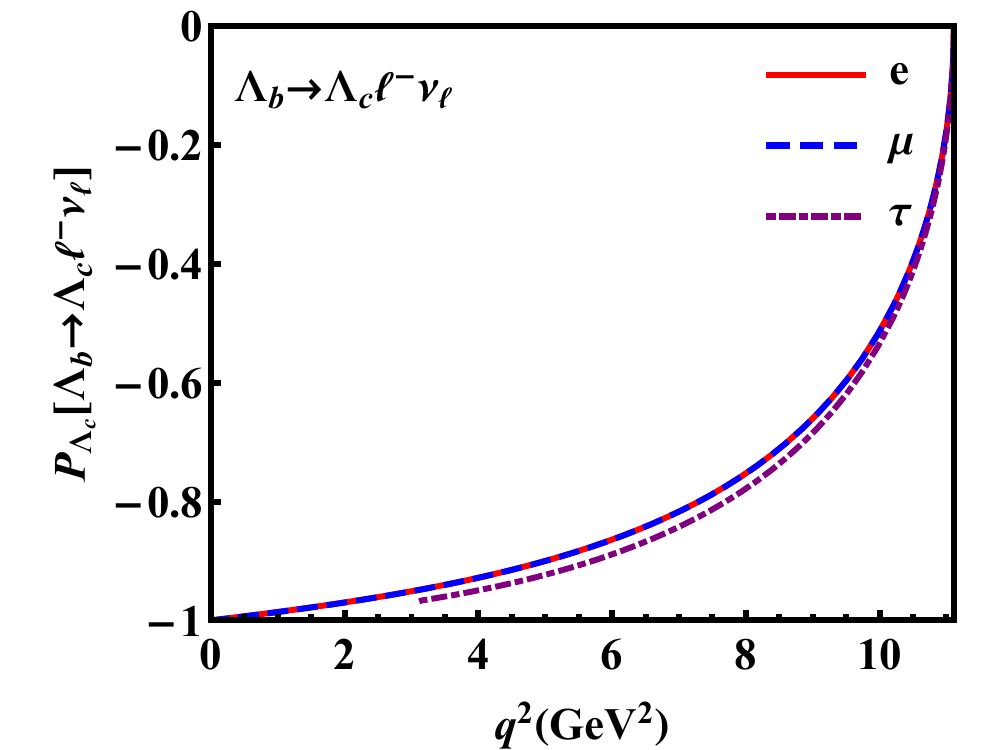}
  \includegraphics[width=50mm]{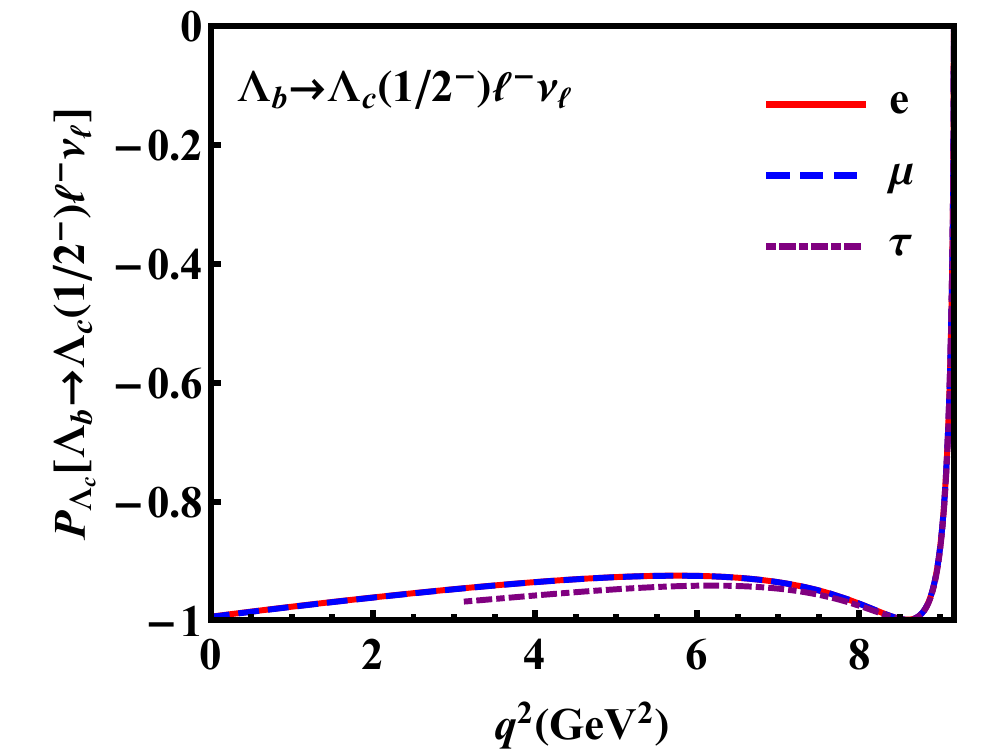}
  \includegraphics[width=50mm]{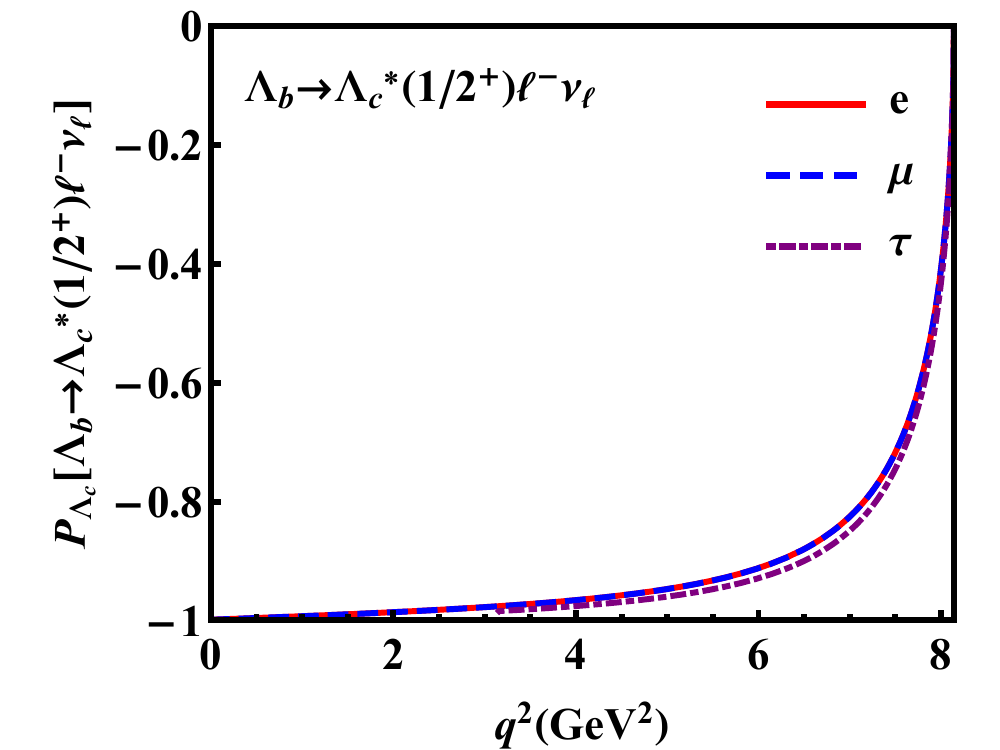}\\
  \includegraphics[width=50mm]{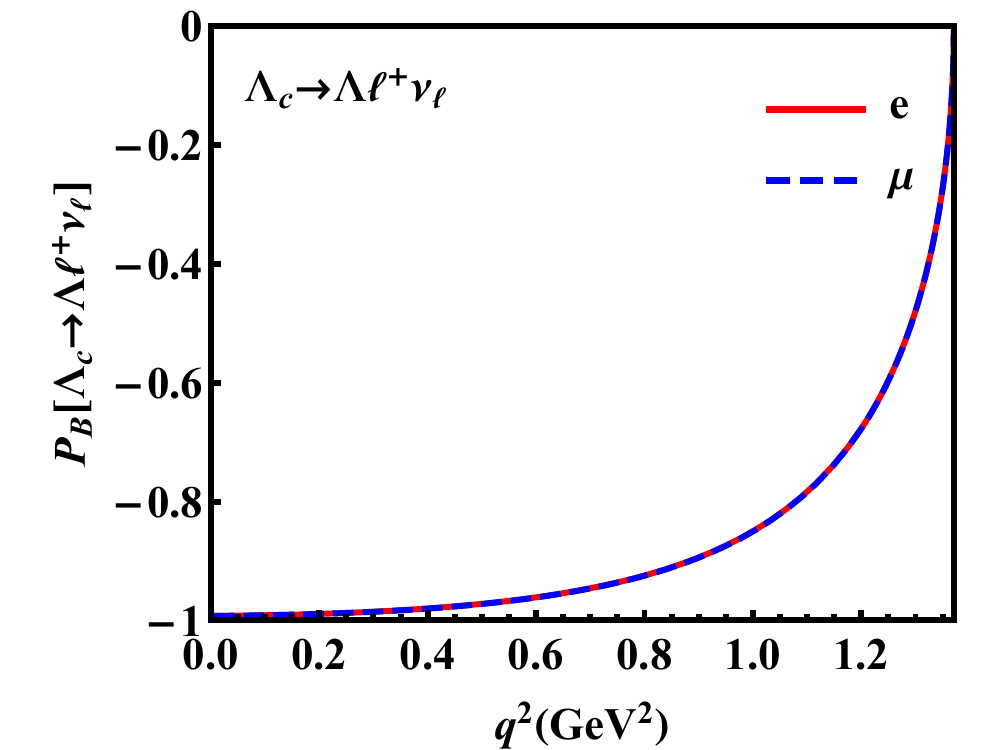}
  \includegraphics[width=50mm]{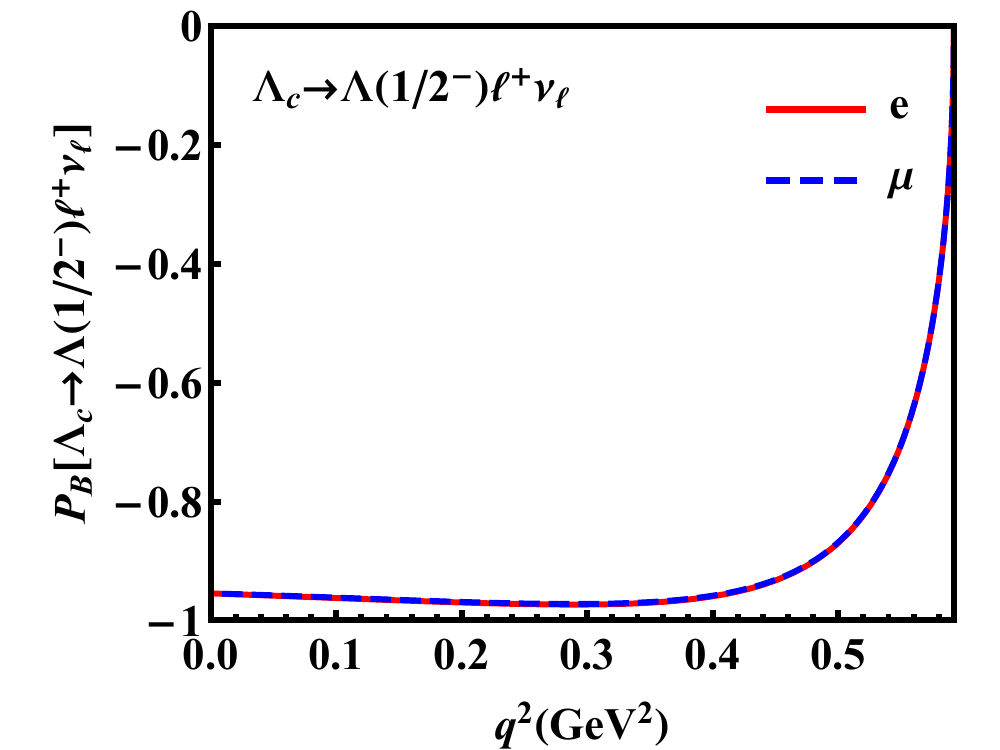}
  \includegraphics[width=50mm]{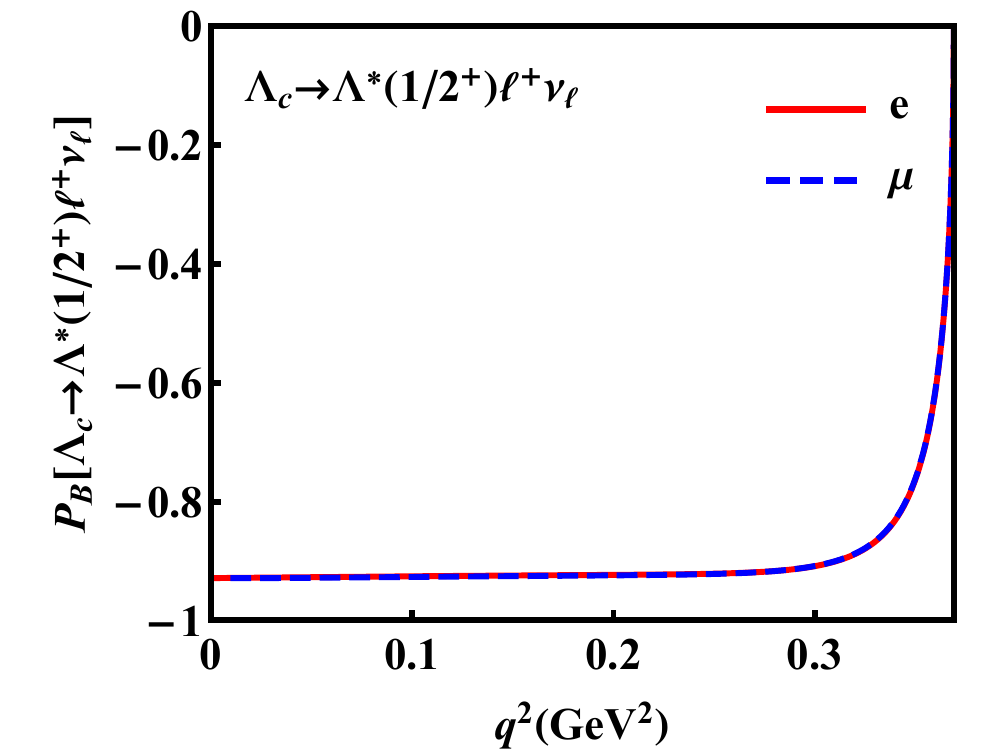}\\
  \end{tabular}
  \caption{The $q^2$ dependence of final hadron longitudinal polarization ($P_{B}$)for $\Lambda_b\to\Lambda_c^{(*)}\ell^{-}\nu_{\ell}$ semileptonic decays with $\ell^{-}= e^{-}$ (red solid line), $\mu^{-}$ (blue dashed line), $\tau^{-}$ (purple dot dashed line) and for $\Lambda_c\to\Lambda^{(*)}\ell^{+}\nu_{\ell}$ semileptonic decays with $\ell^{+}= e^{+}$ (red solid line), $\mu^{+}$ (blue dashed line).}
\label{PB}
\end{figure*}

\begin{figure*}
  \centering
  \begin{tabular}{ccc}
  \includegraphics[width=50mm]{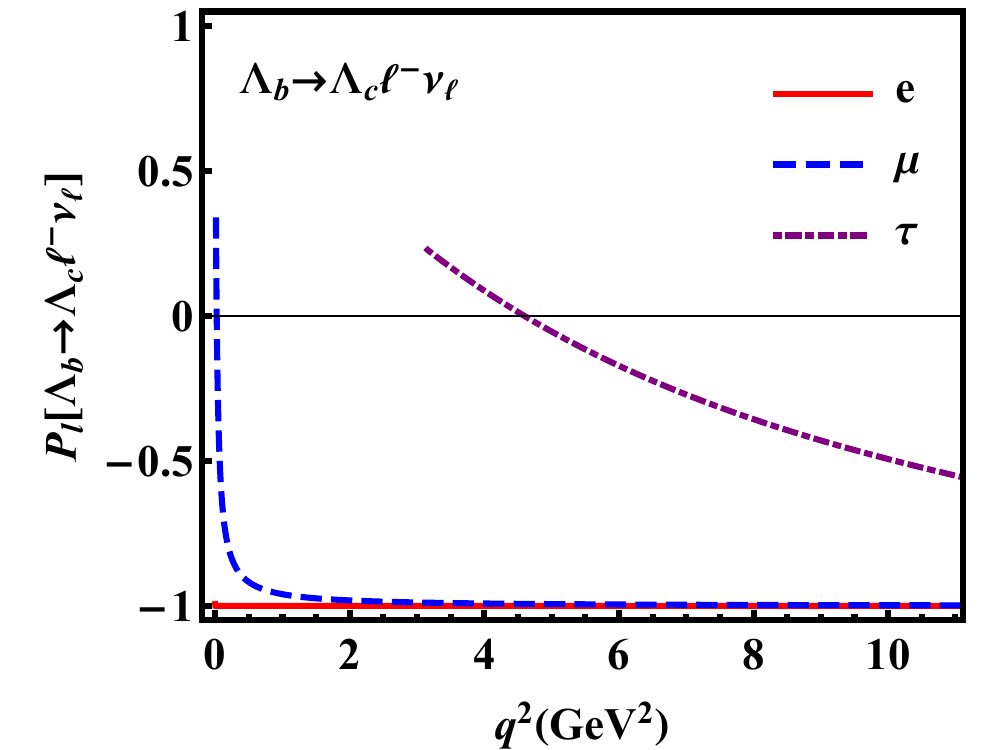}
  \includegraphics[width=50mm]{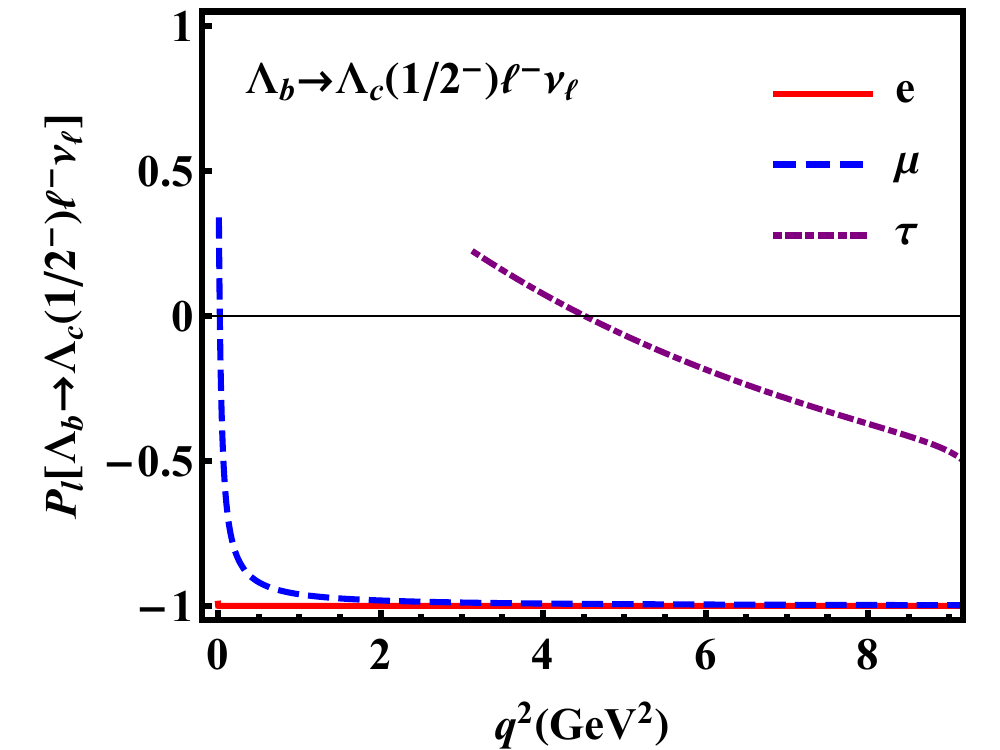}
  \includegraphics[width=50mm]{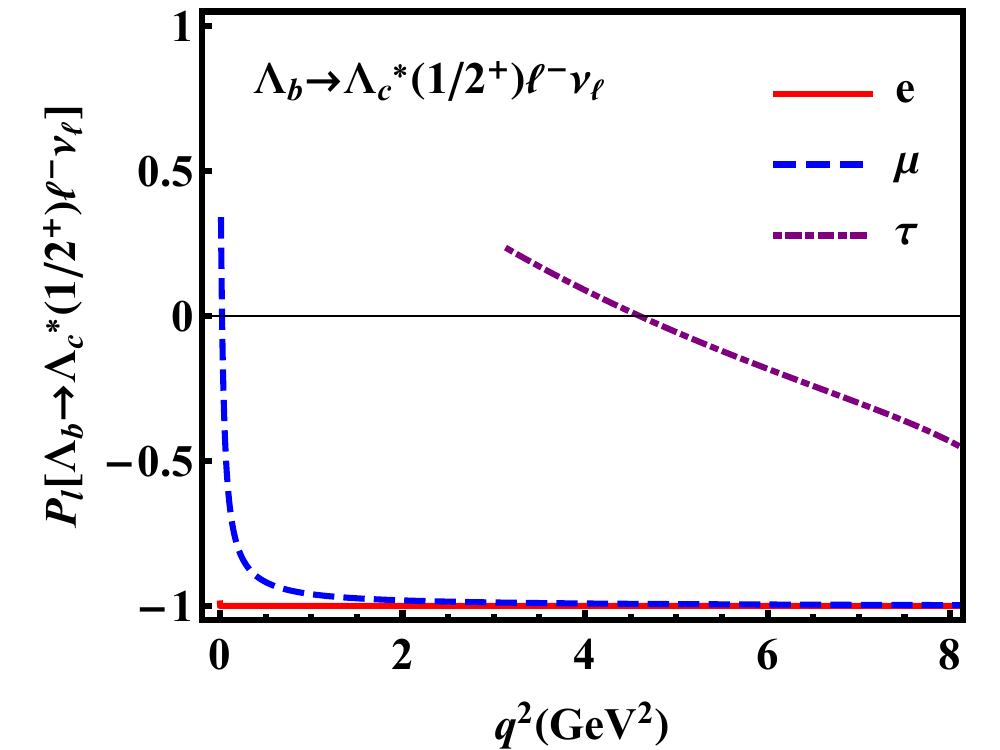}\\
  \includegraphics[width=50mm]{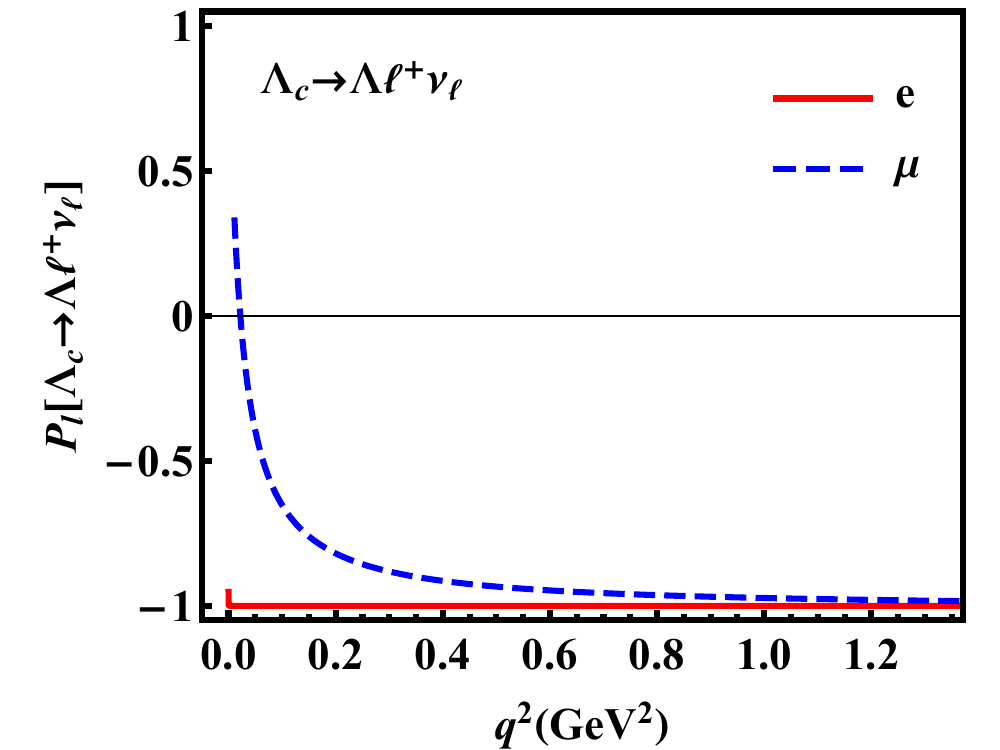}
  \includegraphics[width=50mm]{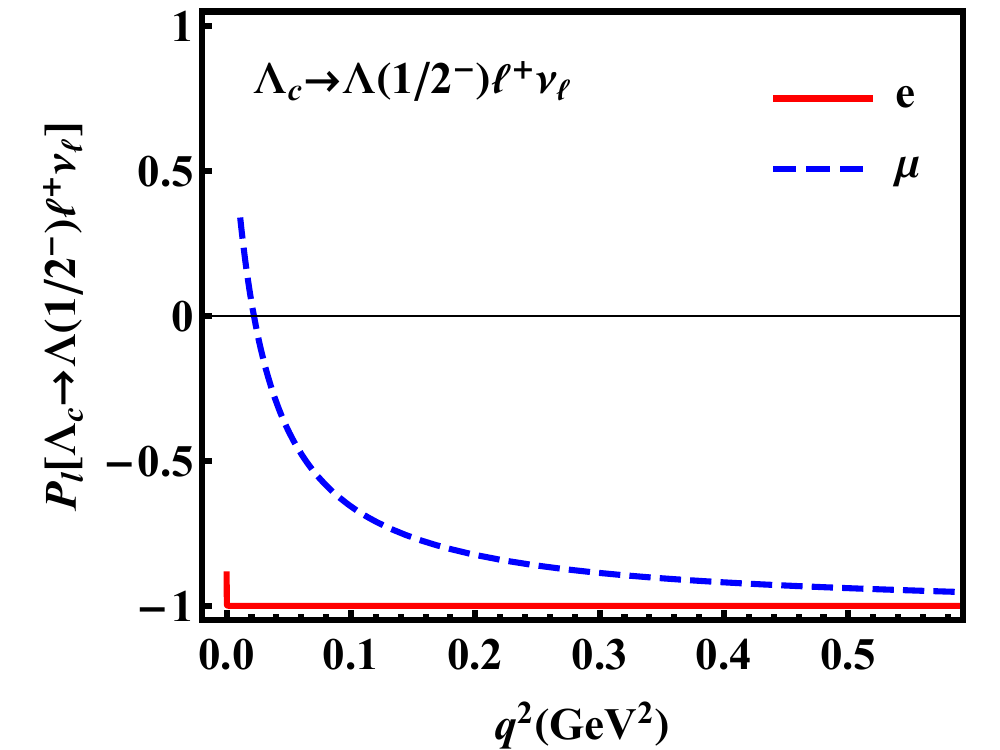}
  \includegraphics[width=50mm]{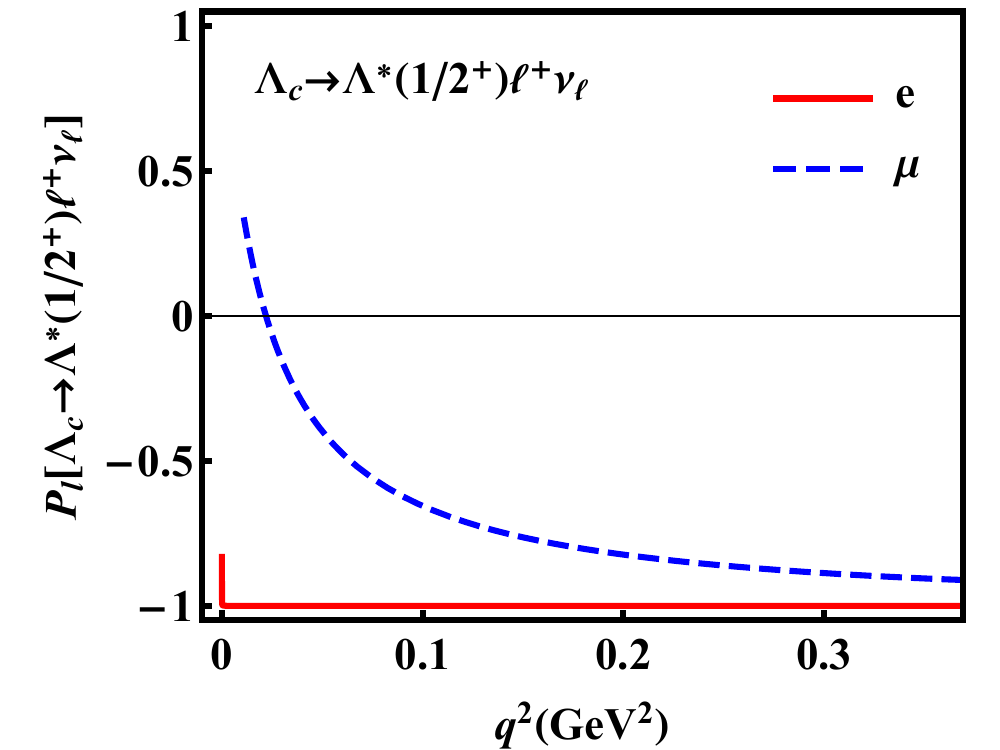}\\
  \end{tabular}
  \caption{The $q^2$ dependence of final lepton longitudinal polarization ($P_{l}$) for $\Lambda_b\to\Lambda_c^{(*)}\ell^{-}\nu_{\ell}$ semileptonic decays with $\ell^{-}= e^{-}$ (red solid line), $\mu^{-}$ (blue dashed line), $\tau^{-}$ (purple dot dashed line) and for $\Lambda_c\to\Lambda^{(*)}\ell^{+}\nu_{\ell}$ semileptonic decays with $\ell^{+}= e^{+}$ (red solid line), $\mu^{+}$ (blue dashed line).}
\label{Pl}
\end{figure*}

In order to calculate the physical observables in semileptonic precess, we adopt the helicity formalism.  The baryon and lepton masses used in this paper come from  from the PDG \cite{Zyla:2020zbs}, as well as $\tau_{\Lambda_b}=1.470~\rm{ps}$, and $\tau_{\Lambda_c^+}=203.5~\rm{fs}$ \cite{Aaij:2019lwg}. The CKM matrix elements are chosen as $V_{cs}=0.987$ and $V_{cb}=0.041$ \cite{Zyla:2020zbs}.

The $q^2$ dependence of the differential decay widths of $\Lambda_{b}\to\Lambda_{c}^{(*)}(1/2^{\pm})\ell^{-}\nu_{\ell}$ and $\Lambda_{c}\to\Lambda^{(*)}(1/2^{\pm})\ell^{-}\nu_{\ell}$ processes, which can be obtained by integrating $\theta_{\ell}$, are shown in Fig. \ref{width}.In the meantime we list the branching fractions in Table \ref{branchingfractions}, and compare our results with the data from experiments or LQCD.

It is obvious that our predictions for $\mathcal{B}(\Lambda_b\to\Lambda_ce^{-}\nu_e(\mu^{-}\nu_{\mu}))$ are consistent with the experimental data although slightly greater than the LQCDs' results. As for the $\tau$ one, it is slightly over the upper limit of LQCD but tallies with Ref. \cite{Gutsche:2015mxa} and Ref. \cite{Faustov:2016pal}.
$\mathcal{B}(\Lambda_c\to\Lambda e^{+}\nu_e(\mu^{+}\nu_{\mu}))$ is consistent with either the LQCD's or the BESIII's results.
In 2009, the CDF Collaboration reported their estimation for $\mathcal{B}(\Lambda_b\to\Lambda_c(2595)\mu^{-}\nu_{\mu})=0.9\%$ with almost $50\%$ uncertainty. This is the only experimental data for excited channels. Our prediction for this one is near $1.8\%$, which is about two times better than the limit for the current data. This leads us to search for more information, in particular the form factors and the absolute branching rates (by further experiment), on these channels. While for the $\Lambda_{c}^{*}(1/2^{+})$ and $\Lambda(1/2^{-})$ channels, considering that our predictions for the $e$ and $\mu$-modes are up to a few thousand of magnitude-we would like to take a positive attitude to discovering them in future experiments. These decays can provide a unique platform for studying the nature of the involved excited hadron states ulteriorly.

We briefly estimate the contributions from ground states, and obtain
\begin{equation}
\begin{split}
\frac{\mathcal{B}(\Lambda_{b}\to\Lambda_{c}\ell^-\nu_{\ell})}{\sum\mathcal{B}(\Lambda_{b}\to\Lambda_{c}^{(*)}\ell^-\nu_{\ell})}
= & \ 0.77\pm0.21,\\
\frac{\mathcal{B}(\Lambda_{c}\to\Lambda\ell^+\nu_{\ell})}{\sum\mathcal{B}(\Lambda_{c}\to\Lambda^{(*)}\ell^+\nu_{\ell})}
= & \ 0.93\pm0.34,
\end{split}
\end{equation}
with $\ell=e$ or $\mu$. It indicates the decays to ground state account for a large proportion of the decays in the corresponding $\Lambda_{b}$ and $\Lambda_{c}$ semileptonic channels, which is consistent with the BESIII measurements of $\mathcal{B}(\Lambda_{c}\to\Lambda e^+\nu_{e})/\mathcal{B}(\Lambda_{c}\to e^+ X)=(91.9 \pm 13.6)\%$.

Additionally, other important physical observables, e.g., the leptonic forward-backward asymmetry ($A_{FB}$), the final hadron polarization ($P_{B}$), and the lepton polarization ($P_l$) are also explored. Their characters dependant on $q^2$ are shown in Figs. \ref{AFB}, \ref{PB}, and \ref{Pl}, respectively. Their mean values are also listed in Table \ref{PhysicalObservables}. These parameters are defined as \cite{Azizi:2019tcn}
\begin{equation}
A_{FB}(q^2)=\frac{\int_0^1\frac{d\Gamma}{dq^2d\cos\theta_{\ell}}d\cos\theta_{\ell}-\int_{-1}^{0}\frac{d\Gamma}{dq^2d\cos\theta_{\ell}}d\cos\theta_{\ell}}
{\int_0^1\frac{d\Gamma}{dq^2d\cos\theta_{\ell}}d\cos\theta_{\ell}+\int_{-1}^{0}\frac{d\Gamma}{dq^2d\cos\theta_{\ell}}d\cos\theta_{\ell}},
\end{equation}
\begin{equation}
P_{B}(q^2)=\frac{d\Gamma^{\lambda_{2}=1/2}/dq^2-d\Gamma^{\lambda_{2}=-1/2}/dq^2}{d\Gamma/dq^2},
\end{equation}
\begin{equation}
P_{\ell}(q^2)=\frac{d\Gamma^{\lambda_{\ell}=1/2}/dq^2-d\Gamma^{\lambda_{\ell}=-1/2}/dq^2}{d\Gamma/dq^2},
\end{equation}
respectively, where
\begin{align}
\frac{d\Gamma^{\lambda_{2}=1/2}}{dq^2}=&\frac{4m_{l}^2}{3q^2}\left(H^2_{1/2,1}+H^2_{1/2,0}+3H^2_{1/2,t}\right)\nonumber\\
&+\frac{8}{3}\left(H^2_{1/2,0}+H^2_{1/2,1}\right),
\end{align}
\begin{align}
\frac{d\Gamma^{\lambda_{2}=-1/2}}{dq^2}=&\frac{4m_{l}^2}{3q^2}\left(H^2_{-1/2,-1}+H^2_{-1/2,0}+3H^2_{-1/2,t}\right)\nonumber\\
&+\frac{8}{3}\left(H^2_{-1/2,-1}+H^2_{-1/2,0}\right),
\end{align}
\begin{align}
\frac{d\Gamma^{\lambda_{\ell}=1/2}}{dq^2}=&\frac{m_{l}^2}{q^2}\left[\frac{4}{3}\left(H^2_{1/2,1}+H^2_{1/2,0}+H^2_{-1/2,-1}+H^2_{-1/2,0}\right)\right.\nonumber\\
&\left.+4\left(H^2_{1/2,t}+H^2_{-1/2,t}\right)\right],
\end{align}
\begin{align}
\frac{d\Gamma^{\lambda_{\ell}=-1/2}}{dq^2}=&\frac{8}{3}\left(H^2_{1/2,1}+H^2_{1/2,0}+H^2_{-1/2,-1}+H^2_{-1/2,0}\right).
\end{align}

In Fig. \ref{AFB}, the leptonic forward-backward asymmetry $F_{FB}$ always changes its sign at a $q^2$ value which is associated with the mass of a lepton. This character is consistent with Refs. \cite{Faustov:2016pal,Gutsche:2015rrt,Becirevic:2020nmb}. In Fig. \ref{PB}, the hadron polarization always varies from $P_{B}=-1$ to $P_{B}=0$ as the $q^2$ increases from zero to $q^2_{\rm max}$. The nonmonotone behavior of $\Lambda_b\to\Lambda_c(1/2^-)\ell^-\nu_\ell$ is consistent with \cite{Becirevic:2020nmb}. The dependence on the lepton mass in the hadron polarization is less than the other physical observables. Our estimation of $\langle P_{B}\rangle_{\Lambda_c\to\Lambda e^{+}\nu_{e}}=-0.87\pm0.09$ agrees with $-0.86\pm0.03\pm0.02$ \cite{Hinson:2004pj} reported by CLEO Collaboration. In Fig. \ref{Pl}, the lepton polarization is close to $P_{\ell}=-1$, especially for electron modes. It stems from the weak interaction which is purely left handed. With the negligible mass, the electrons are approximately absolutely polarized.  The line shapes of $\mu$ and $\tau$ modes can be understood by their masses which change the chirality. As the $q^2$ increases, the leptons are highly boosted, so that their helicity are more left-handed with the polarization approaching  $-1$. The future experiments on the values of these observables and their comparison with the predictions of the present study would help us understand the corresponding channels and the internal structures of the baryons. These observables are also important to investigate the new physics effects beyond the SM \cite{Azizi:2019tcn}.

In conclusion, we would like to focus on the ratios of branching fractions which reflect the leptonic flavor universality
\begin{equation}
R(\Lambda_c^{(*)})=\frac{Br(\Lambda_b\to\Lambda_c^{(*)}\tau^-\nu_{\tau})}{Br(\Lambda_b\to\Lambda_c^{(*)}\ell^{-}\nu_{\ell})},
\label{R1}
\end{equation}
with $\ell^{-}=e^{-}$ or $\mu^{-}$. These ratios, obtained by our and other approaches, e.g. HQET \cite{Bernlochner:2018kxh,Bernlochner:2018bfn,Han:2020sag}, QCDSR \cite{Azizi:2018axf}, various quark models \cite{Pervin:2005ve,Faustov:2016pal,Faustov:2016yza,Gutsche:2018nks}, the SU(3) flavor symmetry \cite{Geng:2019bfz}, as well as LQCD \cite{Bernlochner:2018bfn,Mu:2019bin}, are displayed in Table \ref{Ratios}. Our estimation for $R(\Lambda_c)$ is consistent with other models. The $\tau$-channel for $\Lambda_{b}\to\Lambda_{c}^{*}$ decays is smaller than the $\Lambda_{b}\to\Lambda_{c}$ process, due to the smaller phase spaces.

\begin{table}
\centering
\caption{Our predictions for the ratios of branching fractions $R(\Lambda_c^{(*)})$.}
\label{Ratios}
\renewcommand\arraystretch{1.05}
\begin{tabular*}{86mm}{c@{\extracolsep{\fill}}ccc}
\toprule[1pt]
\toprule[0.5pt]
                 &This work     &Others    \\
\toprule[0.5pt]
$R(\Lambda_c)$   &$0.30\pm0.09$         &about 0.3 \cite{Azizi:2018axf,Bernlochner:2018bfn,Faustov:2016pal,Gutsche:2018nks,Mu:2019bin,Han:2020sag}\\
$R\left(\Lambda_c\left(1/2^{-}\right)\right)$  &$0.14\pm0.01$     &0.13 \cite{Gutsche:2018nks}, 0.21-0.31 \cite{Pervin:2005ve}\\
$R\left(\Lambda_c^{*}\left(1/2^{+}\right)\right)$  &$0.10\pm0.05$     &\\
\bottomrule[0.5pt]
\bottomrule[1pt]
\end{tabular*}
\end{table}

\section{conclusion and discussion}
\label{sec4}
In this work, we study the weak transition form factors for $\Lambda_{b}\to\Lambda_{c}^{(*)}\left(1/2^{\pm}\right)$ and $\Lambda_{c}\to\Lambda^{(*)}\left(1/2^{\pm}\right)$ in the framework of light-front quark model, and investigate the involved semileptonic decays. A semirelativistic three-body potential model and GEM are used to extract the baryon wave functions, which are the input of our calculation. This treatment of baryon wave functions is different from former ways of taking a simple harmonic oscillator wave function with a $\beta$ value in the calculation. To some extent, we can avoid the $\beta$ parameter dependence of the results based on the support from baryon spectroscopy.

Our results of form factors for the ground-state modes are comparable to other theoretical studies, especially to the LQCD or HQL. The ones for the excited states involving modes need to be tested by more theories. These form factors will be useful for studying the weak decays of $\Lambda_{b}$ and $\Lambda_{b}$.

With the obtained form factors, we predict the branching fractions of the considered channels. Our results are consistent with the experimental data and the LQCD calculations. The excited-state modes have much smaller branching fractions compared to the ground-state modes, which can explain the measured result of $\mathcal{B}(\Lambda_{c}\to\Lambda e^+\nu_{e})/\mathcal{B}(\Lambda_{c}\to e^+ X)=(91.9 \pm 13.6) \%$. The ratios of $R(\Lambda_c^{(*)})=\mathcal{B}(\Lambda_b\to\Lambda_c^{(*)}\tau^-\nu_{\tau})/\mathcal{B}(\Lambda_b\to\Lambda_c^{(*)}\ell^{-}\nu_{\ell})$ are predicted to be tested for the lepton flavor universality. Moreover, the leptonic forward-backward asymmetry ($A_{FB}$), the final hadron polarization ($P_{B}$) and the lepton polarization ($P_{\ell}$) are also investigated. The future BESIII, LHCb and Belle II experiments can measure these observables and test our results.

\section*{ACKNOWLEDGMENTS}

We would like to thank Chun-Khiang Chua, Si-Qiang Luo and Jian-Peng Wang for useful discussions. This work is supported by the China National Funds for Distinguished Young Scientists under Grant No. 11825503, National Key Research and Development Program of China under Contract No. 2020YFA0406400, the 111 Project under Grant No. B20063, and the National Natural Science Foundation of China under Grant No. 11975112 and 12047501.


\begin{thebibliography}{99}

\bibitem{Aaij:2017ueg}
R.~Aaij \textit{et al.} [LHCb],
Observation of the doubly charmed baryon $\Xi_{cc}^{++}$,
Phys. Rev. Lett. \textbf{119} (2017) no.11, 112001.

\bibitem{Aaij:2018gfl}
R.~Aaij \textit{et al.} [LHCb],
First Observation of the Doubly Charmed Baryon Decay $\Xi_{cc}^{++}\rightarrow \Xi_{c}^{+}\pi^{+}$,
Phys. Rev. Lett. \textbf{121} (2018) no.16, 162002.

\bibitem{Zupanc:2013iki}
A.~Zupanc \textit{et al.} [Belle],
Measurement of the Branching Fraction $\mathcal B(\Lambda_c^+ \to p K^- \pi^+)$,
Phys. Rev. Lett. \textbf{113} (2014) no.4, 042002.

\bibitem{Ablikim:2015prg}
M.~Ablikim \textit{et al.} [BESIII],
Measurement of the absolute branching fraction for $\Lambda^+_{c}\to \Lambda e^+\nu_e$,
Phys. Rev. Lett. \textbf{115} (2015) no.22, 221805.

\bibitem{Ablikim:2015flg}
M.~Ablikim \textit{et al.} [BESIII],
Measurements of absolute hadronic branching fractions of $\Lambda_{c}^{+}$ baryon,
Phys. Rev. Lett. \textbf{116} (2016) no.5, 052001.

\bibitem{Ablikim:2016vqd}
M.~Ablikim \textit{et al.} [BESIII],
Measurement of the absolute branching fraction for $\Lambda_c^+\rightarrow \Lambda \mu^+\nu_{\mu}$,
Phys. Lett. B \textbf{767} (2017), 42-47.

\bibitem{Ablikim:2018woi}
M.~Ablikim \textit{et al.} [BESIII],
Measurement of the absolute branching fraction of the inclusive semileptonic $\Lambda_c^+$ decay,
Phys. Rev. Lett. \textbf{121} (2018) no.25, 251801.

\bibitem{Li:2018qak}
Y.~B.~Li \textit{et al.} [Belle],
First Measurements of Absolute Branching Fractions of the $\Xi_c^0$ Baryon at Belle,
Phys. Rev. Lett. \textbf{122} (2019) no.8, 082001.

\bibitem{Aaij:2016cla}
R.~Aaij \textit{et al.} [LHCb],
Measurement of matter-antimatter differences in beauty baryon decays,
Nature Phys. \textbf{13} (2017), 391-396.

\bibitem{Amhis:2019ckw}
Y.~S.~Amhis \textit{et al.} [HFLAV],
Averages of $b$-hadron, $c$-hadron, and $\tau$-lepton properties as of 2018.

\bibitem{Bernlochner:2018bfn}
F.~U.~Bernlochner, Z.~Ligeti, D.~J.~Robinson and W.~L.~Sutcliffe,
Precise predictions for $\Lambda_b \to \Lambda_c$ semileptonic decays,
Phys. Rev. D \textbf{99} (2019) no.5, 055008.

\bibitem{Bernlochner:2018kxh}
F.~U.~Bernlochner, Z.~Ligeti, D.~J.~Robinson and W.~L.~Sutcliffe,
New predictions for $\Lambda_b\to\Lambda_c$ semileptonic decays and tests of heavy quark symmetry,
Phys. Rev. Lett. \textbf{121} (2018) no.20, 202001.

\bibitem{Bernlochner:2021vlv}
F.~U.~Bernlochner, M.~F.~Sevilla, D.~J.~Robinson and G.~Wormser,
Semitauonic $b$-hadron decays: A lepton flavor universality laboratory,
[arXiv:2101.08326 [hep-ex]].

\bibitem{Zyla:2020zbs}
P.~A.~Zyla \textit{et al.} [Particle Data Group],
Review of Particle Physics,
PTEP \textbf{2020} (2020) no.8, 083C01.

\bibitem{Yu:2017zst}
F.~S.~Yu, H.~Y.~Jiang, R.~H.~Li, C.~D.~L\"u, W.~Wang and Z.~X.~Zhao,
Discovery Potentials of Doubly Charmed Baryons,
Chin. Phys. C \textbf{42} (2018) no.5, 051001.

\bibitem{Wang:2017mqp}
W.~Wang, F.~S.~Yu and Z.~X.~Zhao,
Weak decays of doubly heavy baryons: the $1/2\rightarrow 1/2$ case,
Eur. Phys. J. C \textbf{77} (2017) no.11, 781.

\bibitem{Han:2021azw}
J.~J.~Han, H.~Y.~Jiang, W.~Liu, Z.~J.~Xiao and F.~S.~Yu,
Rescattering mechanism of weak decays of double-charm baryons,
[arXiv:2101.12019 [hep-ph]].

\bibitem{Yu:2019lxw}
F.~S.~Yu,
Role of decay in the search for double-charm baryons,
Sci. China Phys. Mech. Astron. \textbf{63} (2020) no.2, 221065.

\bibitem{Guo:2005qa}
P.~Guo, H.~W.~Ke, X.~Q.~Li, C.~D.~Lu and Y.~M.~Wang,
Diquarks and the semi-leptonic decay of $\Lambda_b$ in the hyrid scheme,
Phys. Rev. D \textbf{75} (2007), 054017.

\bibitem{Chua:2018lfa}
C.~K.~Chua,
Color-allowed bottom baryon to charmed baryon nonleptonic decays,
Phys. Rev. D \textbf{99} (2019) no.1, 014023.

\bibitem{Zhao:2018zcb}
Z.~X.~Zhao,
Weak decays of heavy baryons in the light-front approach,
Chin. Phys. C \textbf{42} (2018) no.9, 093101.

\bibitem{Chua:2019yqh}
C.~K.~Chua,
Color-allowed bottom baryon to $s$-wave and $p$-wave charmed baryon nonleptonic decays,
Phys. Rev. D \textbf{100} (2019) no.3, 034025.

\bibitem{Zhu:2018jet}
J.~Zhu, Z.~T.~Wei and H.~W.~Ke,
Semileptonic and nonleptonic weak decays of $\Lambda_b^0$,
Phys. Rev. D \textbf{99} (2019) no.5, 054020.

\bibitem{Ke:2007tg}
H.~W.~Ke, X.~Q.~Li and Z.~T.~Wei,
Diquarks and $\Lambda_b\to\Lambda_c$ weak decays,
Phys. Rev. D \textbf{77} (2008), 014020.

\bibitem{Ke:2019smy}
H.~W.~Ke, N.~Hao and X.~Q.~Li,
Revisiting $\Lambda _{b}\rightarrow \Lambda _{c}$ and $\Sigma _{b}\rightarrow \Sigma _{c}$ weak decays in the light-front quark model,
Eur. Phys. J. C \textbf{79} (2019) no.6, 540.

\bibitem{Geng:2020fng}
C.~Q.~Geng, C.~C.~Lih, C.~W.~Liu and T.~H.~Tsai,
Semileptonic decays of $\Lambda_c^+$ in dynamical approaches,
Phys. Rev. D \textbf{101} (2020) no.9, 094017.

\bibitem{Geng:2020gjh}
C.~Q.~Geng, C.~W.~Liu and T.~H.~Tsai,
Semileptonic weak decays of antitriplet charmed baryons in the light-front formalism,
Phys. Rev. D \textbf{103} (2021) no.5, 054018.

\bibitem{Ballagh:1981yh}
H.~C.~Ballagh, H.~H.~Bingham, T.~Lawry, G.~R.~Lynch, J.~Lys, J.~Orthel, M.~D.~Sokoloff, M.~L.~Stevenson, G.~P.~Yost and D.~Gee, \textit{et al.}
Dilepton Production by Neutrinos in the {Fermilab} 15-ft Bubble Chamber,
Phys. Rev. D \textbf{24} (1981), 7.

\bibitem{Vella:1982ei}
E.~Vella, G.~Trilling, G.~S.~Abrams, M.~S.~Alam, C.~A.~Blocker, A.~Blondel, A.~Boyarski, M.~Breidenbach, D.~L.~Burke and W.~C.~Carithers, \textit{et al.}
Observation of Semileptonic Decays of Charmed Baryons,
Phys. Rev. Lett. \textbf{48} (1982), 1515.

\bibitem{Klein:1989pu}
S.~Klein, T.~Himel, G.~S.~Abrams, D.~Amidei, A.~R.~Baden, T.~Barklow, A.~Boyarski, J.~Boyer, P.~Burchat and D.~L.~Burke, \textit{et al.}
$\Lambda_c^+$ Production and Semileptonic Decay in 29-{GeV} $e^+ e^-$ Annihilation,
Phys. Rev. Lett. \textbf{62} (1989), 2444.

\bibitem{Albrecht:1991bu}
H.~Albrecht \textit{et al.} [ARGUS],
Observations of $\Lambda_c^+$ semileptonic decay,
Phys. Lett. B \textbf{269} (1991), 234-242.

\bibitem{Bergfeld:1994gt}
T.~Bergfeld \textit{et al.} [CLEO],
Study of the decay $\Lambda_c^+\to\Lambda\ell^+\nu_{\ell}$,
Phys. Lett. B \textbf{323} (1994), 219-226.

\bibitem{PerezMarcial:1989yh}
R.~Perez-Marcial, R.~Huerta, A.~Garcia and M.~Avila-Aoki,
Predictions for Semileptonic Decays of Charm Baryons. 2. Nonrelativistic and {MIT} Bag Quark Models,
Phys. Rev. D \textbf{40} (1989), 2955
[erratum: Phys. Rev. D \textbf{44} (1991), 2203].

\bibitem{Hussain:1990ai}
F.~Hussain and J.~G.~Korner,
Semileptonic charm baryon decays in the relativistic spectator quark model,
Z. Phys. C \textbf{51} (1991), 607-614.

\bibitem{Efimov:1991ex}
G.~V.~Efimov, M.~A.~Ivanov and V.~E.~Lyubovitskij,
Predictions for semileptonic decay rates of charmed baryons in the quark confinement model,
Z. Phys. C \textbf{52} (1991), 149-158.

\bibitem{Cheng:1995fe}
H.~Y.~Cheng and B.~Tseng,
1/M corrections to baryonic form-factors in the quark model,
Phys. Rev. D \textbf{53} (1996), 1457
[erratum: Phys. Rev. D \textbf{55} (1997), 1697].

\bibitem{MarquesdeCarvalho:1999bqs}
R.~S.~Marques de Carvalho, F.~S.~Navarra, M.~Nielsen, E.~Ferreira and H.~G.~Dosch,
Form-factors and decay rates for heavy Lambda semileptonic decays from QCD sum rules,
Phys. Rev. D \textbf{60} (1999), 034009.

\bibitem{Liu:2009sn}
Y.~L.~Liu, M.~Q.~Huang and D.~W.~Wang,
Improved analysis on the semi-leptonic decay $\Lambda_c\to\Lambda \ell^+\nu$ from QCD light-cone sum rules,
Phys. Rev. D \textbf{80} (2009), 074011.

\bibitem{Zhao:2020mod}
Z.~X.~Zhao, R.~H.~Li, Y.~L.~Shen, Y.~J.~Shi and Y.~S.~Yang,
The semi-leptonic form factors of $\Lambda_{b}\to\Lambda_{c}$ and $\Xi_{b}\to\Xi_{c}$ in QCD sum rules,
Eur. Phys. J. C \textbf{80} (2020) no.12, 1181.

\bibitem{Pervin:2005ve}
M.~Pervin, W.~Roberts and S.~Capstick,
Semileptonic decays of heavy lambda baryons in a quark model,
Phys. Rev. C \textbf{72} (2005), 035201.

\bibitem{Migura:2006en}
S.~Migura, D.~Merten, B.~Metsch and H.~R.~Petry,
Semileptonic decays of baryons in a relativistic quark model,
Eur. Phys. J. A \textbf{28} (2006), 55.

\bibitem{Faustov:2020thr}
R.~N.~Faustov and V.~O.~Galkin,
Semileptonic Decays of Heavy Baryons in the Relativistic Quark Model,
Particles \textbf{3} (2020) no.1, 208-222.

\bibitem{Gutsche:2015rrt}
T.~Gutsche, M.~A.~Ivanov, J.~G.~Korner, V.~E.~Lyubovitskij and P.~Santorelli,
Semileptonic decays $\Lambda_c^+ \to \Lambda \ell^+ \nu_\ell\,\,(\ell=e,\mu)$ in the covariant quark model and comparison with the new absolute branching fraction measurements of Belle and BESIII,
Phys. Rev. D \textbf{93} (2016) no.3, 034008.

\bibitem{Singleton:1990ye}
R.~L.~Singleton,
Semileptonic baryon decays with a heavy quark,
Phys. Rev. D \textbf{43} (1991), 2939-2950.

\bibitem{Mannel:1991ii}
T.~Mannel, W.~Roberts and Z.~Ryzak,
$1/m_c$ suppressed semileptonic $\Lambda_b$ decays,
Phys. Lett. B \textbf{271} (1991), 421-424.

\bibitem{Korner:1991ph}
J.~G.~Korner and M.~Kramer,
Polarization effects in exclusive semileptonic $\Lambda_c$ and $\Lambda_b$ charm and bottom baryon decays,
Phys. Lett. B \textbf{275} (1992), 495-505.

\bibitem{Mannel:1991bs}
T.~Mannel and G.~A.~Schuler,
Semileptonic decays of bottom baryons at LEP,
Phys. Lett. B \textbf{279} (1992), 194-200.

\bibitem{Dai:1996xv}
Y.~B.~Dai, C.~S.~Huang, M.~Q.~Huang and C.~Liu,
QCD sum rule analysis for the $\Lambda_b\to\Lambda_c$ semileptonic decay,
Phys. Lett. B \textbf{387} (1996), 379-385.

\bibitem{Lee:1998bj}
J.~P.~Lee, C.~Liu and H.~S.~Song,
Analysis of $\Lambda_b\to\Lambda_c$ weak decays in heavy quark effective theory,
Phys. Rev. D \textbf{58} (1998), 014013.

\bibitem{Ivanov:1998ya}
M.~A.~Ivanov, J.~G.~Korner, V.~E.~Lyubovitskij and A.~G.~Rusetsky,
Charm and bottom baryon decays in the Bethe-Salpeter approach: Heavy to heavy semileptonic transitions,
Phys. Rev. D \textbf{59} (1999), 074016.

\bibitem{Cardarelli:1998tq}
F.~Cardarelli and S.~Simula,
Analysis of the $\Lambda_b\to\Lambda_c^+\ell\bar{\nu}_{\ell}$ decay within a light front constituent quark model,
Phys. Rev. D \textbf{60} (1999), 074018.

\bibitem{Guo:1999ss}
X.~H.~Guo, A.~W.~Thomas and A.~G.~Williams,
$1/m_Q$ corrections to the Bethe-Salpeter equation for $\Lambda_{Q}$ in the diquark picture,
Phys. Rev. D \textbf{61} (2000), 116015.

\bibitem{Lees:2012xj}
J.~P.~Lees \textit{et al.} [BaBar],
Evidence for an excess of $\bar{B} \to D^{(*)} \tau^-\bar{\nu}_\tau$ decays,
Phys. Rev. Lett. \textbf{109} (2012), 101802.

\bibitem{Belle:2019rba}
G.~Caria \textit{et al.} [Belle],
Measurement of $\mathcal{R}(D)$ and $\mathcal{R}(D^*)$ with a semileptonic tagging method,
Phys. Rev. Lett. \textbf{124} (2020) no.16, 161803.

\bibitem{Faustov:2016pal}
R.~N.~Faustov and V.~O.~Galkin,
Semileptonic decays of $\Lambda_b$ baryons in the relativistic quark model,
Phys. Rev. D \textbf{94} (2016) no.7, 073008.

\bibitem{Ebert:2006rp}
D.~Ebert, R.~N.~Faustov and V.~O.~Galkin,
Semileptonic decays of heavy baryons in the relativistic quark model,
Phys. Rev. D \textbf{73} (2006), 094002.

\bibitem{Gutsche:2015mxa}
T.~Gutsche, M.~A.~Ivanov, J.~G.~K\"orner, V.~E.~Lyubovitskij, P.~Santorelli and N.~Habyl,
Semileptonic decay $\Lambda_b \to \Lambda_c + \tau^- + \bar{\nu_\tau}$ in the covariant confined quark model,
Phys. Rev. D \textbf{91} (2015) no.7, 074001
[erratum: Phys. Rev. D \textbf{91} (2015) no.11, 119907].

\bibitem{Gutsche:2018nks}
T.~Gutsche, M.~A.~Ivanov, J.~G.~K\"orner, V.~E.~Lyubovitskij, P.~Santorelli and C.~T.~Tran,
Analyzing lepton flavor universality in the decays $\Lambda_b\to\Lambda_c^{(\ast)}(\frac12^\pm,\frac32^-) + \ell\,\bar\nu_\ell$,
Phys. Rev. D \textbf{98} (2018) no.5, 053003.

\bibitem{Rahmani:2020kjd}
S.~Rahmani, H.~Hassanabadi and J.~K\v{r}\'\i{}\v{z},
Nonleptonic and semileptonic ${\Lambda _b} \rightarrow {\Lambda _c}$ transitions in a potential quark model,
Eur. Phys. J. C \textbf{80} (2020) no.7, 636.

\bibitem{Thakkar:2020vpv}
K.~Thakkar,
Semileptonic transition of $\varLambda _{b}$ baryon,
Eur. Phys. J. C \textbf{80} (2020) no.10, 926.

\bibitem{Wang:2003it}
D.~W.~Wang and M.~Q.~Huang,
Choice of heavy baryon currents in QCD sum rules,
Phys. Rev. D \textbf{67} (2003), 074025.

\bibitem{Huang:2005mea}
M.~Q.~Huang, H.~Y.~Jin, J.~G.~Korner and C.~Liu,
Note on the slope parameter of the baryonic $\Lambda_b\to\Lambda_c$ Isgur-Wise function,
Phys. Lett. B \textbf{629} (2005), 27-32.

\bibitem{Wang:2009yma}
Z.~G.~Wang,
Analysis of the Isgur-Wise function of the $\Lambda_b\to\Lambda_c$ transition with light-cone QCD sum rules,
[arXiv:0906.4206 [hep-ph]].

\bibitem{Azizi:2018axf}
K.~Azizi and J.~Y.~S\"ung\"u,
Semileptonic $\Lambda_{b}\rightarrow \Lambda_{c}{\ell}\bar\nu_{\ell}$ Transition in Full QCD,
Phys. Rev. D \textbf{97} (2018) no.7, 074007.

\bibitem{Hussain:2017lir}
M.~M.~Hussain and W.~Roberts,
$\Lambda_c$ Semileptonic Decays in a Quark Model,
Phys. Rev. D \textbf{95} (2017) no.5, 053005.

\bibitem{Nieves:2019kdh}
J.~Nieves, R.~Pavao and S.~Sakai,
$\Lambda _b$ decays into $\Lambda _c^*\ell \bar{\nu }_\ell $ and $\Lambda _c^*\pi ^-$ $[\Lambda _c^*=\Lambda _c(2595)$ and $\Lambda _c(2625)]$ and heavy quark spin symmetry,
Eur. Phys. J. C \textbf{79} (2019) no.5, 417.

\bibitem{Becirevic:2020nmb}
D.~Be\v{c}irevi\'c, A.~Le Yaouanc, V.~Mor\'enas and L.~Oliver,
Heavy baryon wave functions, Bakamjian-Thomas approach to form factors, and observables in ${\Lambda_b \to \Lambda_c\left({1 \over 2}^\pm \right) \ell \overline{\nu}}$ transitions,
Phys. Rev. D \textbf{102} (2020) no.9, 094023.

\bibitem{Ablikim:2019hff}
M.~Ablikim \textit{et al.} [BESIII],
Future Physics Programme of BESIII,
Chin. Phys. C \textbf{44} (2020) no.4, 040001.

\bibitem{Bowler:1997ej}
K.~C.~Bowler \textit{et al.} [UKQCD],
First lattice study of semileptonic decays of $\Lambda_b$ and $\Xi_b$ baryons,
Phys. Rev. D \textbf{57} (1998), 6948-6974.

\bibitem{Gottlieb:2003yb}
S.~A.~Gottlieb and S.~Tamhankar,
A Lattice study of $\Lambda_b$ semileptonic decay,
Nucl. Phys. B Proc. Suppl. \textbf{119} (2003), 644-646.

\bibitem{Meinel:2016dqj}
S.~Meinel,
$\Lambda_c \to \Lambda l^+ \nu_l$ form factors and decay rates from lattice QCD with physical quark masses,
Phys. Rev. Lett. \textbf{118} (2017) no.8, 082001.

\bibitem{Meinel:2021rbm}
S.~Meinel and G.~Rendon,
$\Lambda_b \to \Lambda_c^*(2595,2625)\ell^-\bar{\nu}$ form factors from lattice QCD,
[arXiv:2103.08775 [hep-lat]].

\bibitem{Detmold:2015aaa}
W.~Detmold, C.~Lehner and S.~Meinel,
$\Lambda_b \to p \ell^- \bar{\nu}_\ell$ and $\Lambda_b \to \Lambda_c \ell^- \bar{\nu}_\ell$ form factors from lattice QCD with relativistic heavy quarks,
Phys. Rev. D \textbf{92} (2015) no.3, 034503.

\bibitem{Cheng:2004cc}
H.~Y.~Cheng, C.~K.~Chua and C.~W.~Hwang,
Light front approach for heavy pentaquark transitions,
Phys. Rev. D \textbf{70} (2004), 034007.

\bibitem{Wei:2009np}
Z.~T.~Wei, H.~W.~Ke and X.~Q.~Li,
Evaluating decay Rates and Asymmetries of $\Lambda_b$ into Light Baryons in LFQM,
Phys. Rev. D \textbf{80} (2009), 094016.

\bibitem{Chang:2019obq}
Q.~Chang, L.~T.~Wang and X.~N.~Li,
Form factors of $V'\to V''$ transition within the light-front quark models,
JHEP \textbf{12} (2019), 102.

\bibitem{Chang:2020wvs}
Q.~Chang, X.~L.~Wang and L.~T.~Wang,
Tensor form factors of $P\to P,\,S,\,V$ and $A$ transitions within standard and covariant light-front approaches,
Chin. Phys. C \textbf{44} (2020) no.8, 083105.

\bibitem{Ke:2019lcf}
H.~W.~Ke, F.~Lu, X.~H.~Liu and X.~Q.~Li,
Study on $\Xi_{cc}\to\Xi_c$ and $\Xi_{cc}\to\Xi'_c$ weak decays in the light-front quark model,
Eur. Phys. J. C \textbf{80} (2020) no.2, 140.

\bibitem{Ke:2012wa}
H.~W.~Ke, X.~H.~Yuan, X.~Q.~Li, Z.~T.~Wei and Y.~X.~Zhang,
$\Sigma_{b}\to\Sigma_c$ and $\Omega_b\to\Omega_c$ weak decays in the light-front quark model,
Phys. Rev. D \textbf{86} (2012), 114005.

\bibitem{Godfrey:1985xj}
S.~Godfrey and N.~Isgur,
Mesons in a Relativized Quark Model with Chromodynamics,
Phys. Rev. D \textbf{32} (1985), 189-231.

\bibitem{Capstick:1986bm}
S.~Capstick and N.~Isgur,
Baryons in a Relativized Quark Model with Chromodynamics,
AIP Conf. Proc. \textbf{132} (1985), 267-271.

\bibitem{Song:2015nia}
Q.~T.~Song, D.~Y.~Chen, X.~Liu and T.~Matsuki,
Charmed-strange mesons revisited: mass spectra and strong decays,
Phys. Rev. D \textbf{91} (2015), 054031.

\bibitem{Pang:2017dlw}
C.~Q.~Pang, J.~Z.~Wang, X.~Liu and T.~Matsuki,
A systematic study of mass spectra and strong decay of strange mesons,
Eur. Phys. J. C \textbf{77} (2017) no.12, 861.

\bibitem{Wang:2018rjg}
J.~Z.~Wang, Z.~F.~Sun, X.~Liu and T.~Matsuki,
Higher bottomonium zoo,
Eur. Phys. J. C \textbf{78} (2018) no.11, 915.

\bibitem{Hiyama:2003cu}
E.~Hiyama, Y.~Kino and M.~Kamimura,
Gaussian expansion method for few-body systems,
Prog. Part. Nucl. Phys. \textbf{51} (2003), 223-307.

\bibitem{Yoshida:2015tia}
T.~Yoshida, E.~Hiyama, A.~Hosaka, M.~Oka and K.~Sadato,
Spectrum of heavy baryons in the quark model,
Phys. Rev. D \textbf{92} (2015) no.11, 114029.

\bibitem{Yang:2019lsg}
G.~Yang, J.~Ping, P.~G.~Ortega and J.~Segovia,
Triply heavy baryons in the constituent quark model,
Chin. Phys. C \textbf{44} (2020) no.2, 023102.

\bibitem{Hall:2014uca}
J.~M.~M.~Hall, W.~Kamleh, D.~B.~Leinweber, B.~J.~Menadue, B.~J.~Owen, A.~W.~Thomas and R.~D.~Young,
Lattice QCD Evidence that the $\ensuremath{\Lambda}(1405)$ Resonance is an Antikaon-Nucleon Molecule,
Phys. Rev. Lett. \textbf{114} (2015) no.13, 132002.

\bibitem{Liu:2016wxq}
Z.~W.~Liu, J.~M.~M.~Hall, D.~B.~Leinweber, A.~W.~Thomas and J.~J.~Wu,
Structure of the $\mathbf{\Lambda(1405)}$ from Hamiltonian effective field theory,
Phys. Rev. D \textbf{95} (2017) no.1, 014506.

\bibitem{Azizi:2017xyx}
K.~Azizi, B.~Barsbay and H.~Sundu,
Mass and residue of $\Lambda (1405)$ as hybrid and excited ordinary baryon,
Eur. Phys. J. Plus \textbf{133} (2018) no.3, 121.

\bibitem{Mai:2020ltx}
M.~Mai,
Review of the ${\mathbf \Lambda}$(1405): A curious case of a strange-ness resonance,
[arXiv:2010.00056 [nucl-th]].

\bibitem{Chen:2017vgi}
K.~Chen, H.~W.~Ke, X.~Liu and T.~Matsuki,
Estimating the production rates of $D$-wave charmed mesons via the semileptonic decays of bottom mesons,
Chin. Phys. C \textbf{43} (2019) no.2, 023106.

\bibitem{Xu:2014mqa}
H.~Xu, Q.~Huang, H.~W.~Ke and X.~Liu,
Numerical analysis of the production of $D^{(*)}$(3000), $D_{sJ}$(3040) and their partners through the semileptonic decays of $B_{(s)}$ mesons in terms of the light front quark model,
Phys. Rev. D \textbf{90} (2014) no.9, 094017.

\bibitem{Georgi:1990ei}
H.~Georgi, B.~Grinstein and M.~B.~Wise,
$\Lambda_b$ semileptonic decay form-factors for $m_c$ does not equal infinity,
Phys. Lett. B \textbf{252} (1990), 456-460.

\bibitem{Abdallah:2003gn}
J.~Abdallah {\it et al.} [DELPHI Collaboration],
Measurement of the $\Lambda^0_b$ decay form-factor,
Phys.\ Lett.\ B {\bf 585}, 63 (2004).

\bibitem{Aaij:2017svr}
R.~Aaij {\it et al.} [LHCb Collaboration],
Measurement of the shape of the $\Lambda_b^0\to\Lambda_c^+ \mu^- \overline{\nu}_{\mu}$ differential decay rate,
Phys.\ Rev.\ D {\bf 96}, no. 11, 112005 (2017).

\bibitem{Hinson:2004pj}
J.~W.~Hinson \textit{et al.} [CLEO],
Improved measurement of the form-factors in the decay $\Lambda^+_{c}\to\Lambda e^+\nu_{e},$
Phys. Rev. Lett. \textbf{94} (2005), 191801.

\bibitem{Huang:2006ny}
M.~Q.~Huang and D.~W.~Wang,
semi-leptonic decay $\Lambda_c\to\Lambda\ell^+\nu$ from QCD light-cone sum rules,
[arXiv:hep-ph/0608170 [hep-ph]].

\bibitem{Faustov:2016yza}
R.~N.~Faustov and V.~O.~Galkin,
semi-leptonic decays of $\Lambda _c$ baryons in the relativistic quark model,
Eur.\ Phys.\ J.\ C {\bf 76}, no. 11, 628 (2016).



\bibitem{Lu:2016ogy}
C.~D.~L\"u, W.~Wang and F.~S.~Yu,
{\it Test flavor SU(3) symmetry in exclusive $\Lambda_c$ decays},
Phys. Rev. D \textbf{93}, no.5, 056008 (2016)
[arXiv:1601.04241 [hep-ph]].

\bibitem{Geng:2017esc}
C.~Q.~Geng, Y.~K.~Hsiao, Y.~H.~Lin and L.~L.~Liu,
{\it Non-leptonic two-body weak decays of $\Lambda_c(2286)$},
Phys. Lett. B \textbf{776}, 265-269 (2018)
[arXiv:1708.02460 [hep-ph]].

\bibitem{Geng:2017mxn}
C.~Q.~Geng, Y.~K.~Hsiao, C.~W.~Liu and T.~H.~Tsai,
{\it Charmed Baryon Weak Decays with SU(3) Flavor Symmetry},
JHEP \textbf{11}, 147 (2017)
[arXiv:1709.00808 [hep-ph]].


\bibitem{Wang:2017gxe}
D.~Wang, P.~F.~Guo, W.~H.~Long and F.~S.~Yu,
{\it K$_{S}^{0}$  \ensuremath{-} K$_{L}^{0}$ asymmetries and CP violation in charmed baryon decays into neutral kaons},
JHEP \textbf{03}, 066 (2018)
[arXiv:1709.09873 [hep-ph]].

\bibitem{Geng:2018plk}
C.~Q.~Geng, Y.~K.~Hsiao, C.~W.~Liu and T.~H.~Tsai,
{\it Antitriplet charmed baryon decays with SU(3) flavor symmetry},
Phys. Rev. D \textbf{97}, no.7, 073006 (2018)
[arXiv:1801.03276 [hep-ph]].

\bibitem{Cheng:2018hwl}
H.~Y.~Cheng, X.~W.~Kang and F.~Xu,
{\it Singly Cabibbo-suppressed hadronic decays of $\Lambda_c^+$},
Phys. Rev. D \textbf{97}, no.7, 074028 (2018)
[arXiv:1801.08625 [hep-ph]].

\bibitem{Jiang:2018iqa}
H.~Y.~Jiang and F.~S.~Yu,
{\it Fragmentation-fraction ratio $f_{\Xi_b}/f_{\Lambda_b}$ in $b$- and $c$-baryon decays},
Eur. Phys. J. C \textbf{78}, no.3, 224 (2018)
[arXiv:1802.02948 [hep-ph]].

\bibitem{Geng:2018bow}
C.~Q.~Geng, Y.~K.~Hsiao, C.~W.~Liu and T.~H.~Tsai,
{\it SU(3) symmetry breaking in charmed baryon decays},
Eur. Phys. J. C \textbf{78}, no.7, 593 (2018)
[arXiv:1804.01666 [hep-ph]].

\bibitem{Geng:2018upx}
C.~Q.~Geng, Y.~K.~Hsiao, C.~W.~Liu and T.~H.~Tsai,
{\it Three-body charmed baryon Decays with SU(3) flavor symmetry},
Phys. Rev. D \textbf{99}, no.7, 073003 (2019)
[arXiv:1810.01079 [hep-ph]].

\bibitem{Zhao:2018mov}
H.~J.~Zhao, Y.~L.~Wang, Y.~K.~Hsiao and Y.~Yu,
{\it A diagrammatic analysis of two-body charmed baryon decays with flavor symmetry},
JHEP \textbf{02}, 165 (2020)
[arXiv:1811.07265 [hep-ph]].

\bibitem{Jia:2019zxi}
C.~P.~Jia, D.~Wang and F.~S.~Yu,
{\it Charmed baryon decays in $SU(3)_F$ symmetry},
Nucl. Phys. B \textbf{956}, 115048 (2020)
[arXiv:1910.00876 [hep-ph]].

\bibitem{Zou:2019kzq}
J.~Zou, F.~Xu, G.~Meng and H.~Y.~Cheng,
{\it Two-body hadronic weak decays of antitriplet charmed baryons},
Phys. Rev. D \textbf{101}, no.1, 014011 (2020)
[arXiv:1910.13626 [hep-ph]].

%
\bibitem{Niu:2020gjw}
P.~Y.~Niu, J.~M.~Richard, Q.~Wang and Q.~Zhao,
{\it Hadronic weak decays of $\Lambda_c$ in the quark model},
Phys. Rev. D \textbf{102}, no.7, 073005 (2020)
[arXiv:2003.09323 [hep-ph]].

\bibitem{Meng:2020euv}
G.~Meng, S.~M.~Y.~Wong and F.~Xu,
{\it Doubly Cabibbo-suppressed decays of antitriplet charmed baryons},
JHEP \textbf{11}, 126 (2020)
[arXiv:2005.12111 [hep-ph]].

\bibitem{Hu:2020nkg}
S.~Hu, G.~Meng and F.~Xu,
{\it Hadronic weak decays of the charmed baryon $\Omega_c$},
Phys. Rev. D \textbf{101}, no.9, 094033 (2020)
[arXiv:2003.04705 [hep-ph]].

\bibitem{Aaij:2019lwg}
R.~Aaij \textit{et al.} [LHCb],
Precision measurement of the $\Lambda_c^+$, $\Xi_c^+$ and $\Xi_c^0$ baryon lifetimes,
Phys. Rev. D \textbf{100} (2019) no.3, 032001.

\bibitem{Azizi:2019tcn}
K.~Azizi, A.~T.~Olgun and Z.~Tavuko\u{g}lu,
``Effects of vector leptoquarks on $\Lambda_b \to \Lambda_c\ell\bar \nu_\ell$ decay,''
Chin. Phys. C \textbf{45} (2021) no.1, 013113.

\bibitem{Han:2020sag}
C.~Han and C.~Liu,
b-baryon semi-tauonic decays in the Standard Model,
Nucl. Phys. B \textbf{961} (2020), 115262

\bibitem{Geng:2019bfz}
C.~Geng, C.~Liu, T.~Tsai and S.~Yeh,
semi-leptonic decays of anti-triplet charmed baryons,
Phys.\ Lett.\ B \textbf{792} (2019), 214-218.

\bibitem{Mu:2019bin}
X.~Mu, Y.~Li, Z.~Zou and B.~Zhu,
Investigation of effects of new physics in $\Lambda_b\to\Lambda_c \tau\bar\nu_\tau$ decay,
Phys.\ Rev.\ D \textbf{100} (2019) no.11, 113004.

\end{thebibliography}
\end{document}